\documentclass[journal]{IEEEtran}

\usepackage{subcaption}
\usepackage[table,xcdraw]{xcolor}
\usepackage{multirow}
\usepackage{url}
\usepackage{makecell}
 
\usepackage{caption}
\usepackage{verbatim}

\usepackage{cite}
\usepackage{amsmath,amssymb,amsfonts}
\usepackage{graphicx}
\usepackage{textcomp}
\usepackage[noend]{algpseudocode}

\usepackage[linesnumbered,ruled]{algorithm2e}

\def\BibTeX{{\rm B\kern-.05em{\sc i\kern-.025em b}\kern-.08em
T\kern-.1667em\lower.7ex\hbox{E}\kern-.125emX}}

\definecolor{light-gray}{gray}{0.9}
\definecolor{dark-gray}{gray}{0.5}

 \newcommand{\black}[1]{\textcolor{black}{#1}}

\begin{document}

\title{SCDP: Systematic Rateless Coding for Efficient Data Transport in Data Centres (Complete Version)}

 

\author{\IEEEauthorblockN{Mohammed Alasmar\IEEEauthorrefmark{1}, George Parisis\IEEEauthorrefmark{1}, Jon Crowcroft \IEEEauthorrefmark{2}} 
	
	\IEEEauthorblockA{\IEEEauthorrefmark{1}School of Engineering and Informatics, University of Sussex, UK, Email: \{m.alasmar, g.parisis\}@sussex.ac.uk}
	\IEEEauthorblockA{\IEEEauthorrefmark{2}Computer Laboratory, University of Cambridge, UK, Email: Jon.Crowcroft@cl.cam.ac.uk}
 \vspace{-5mm}
}

	\markboth{Accepted  for publication  at IEEE/ACM TRANSACTIONS ON NETWORKING}
{SCDP: Systematic Rateless Coding for Efficient Data Transport in Data Centres (Complete Version)}
\maketitle

\begin{abstract}
	
	In this paper we propose SCDP, a general-purpose data transport protocol for data centres that, in contrast to all other protocols proposed to date, supports efficient one-to-many and many-to-one  communication, which is extremely common in modern data centres. SCDP does so without compromising on efficiency for short and long unicast flows. SCDP achieves this by integrating RaptorQ codes with receiver-driven data transport, packet trimming and Multi-Level Feedback Queuing (MLFQ); (1) RaptorQ codes enable efficient one-to-many and many-to-one data transport; (2) on top of RaptorQ codes, receiver-driven flow control, in combination with in-network packet trimming, enable efficient usage of network resources as well as multi-path transport and packet spraying for all transport modes. Incast and Outcast are eliminated; (3) the systematic nature of RaptorQ codes, in combination with MLFQ, enable fast, decoding-free completion of short flows. We extensively evaluate SCDP in a wide range of simulated scenarios with realistic data centre workloads. For one-to-many and many-to-one transport sessions, SCDP performs significantly better compared to NDP and PIAS. For short and long unicast flows, SCDP performs equally well or better compared to NDP and PIAS.
	
\end{abstract}

 
 \section{Introduction}
 \label{introduction}
 Data centres support the provision of core Internet services and it is therefore crucial to have in place data transport mechanisms that ensure high performance for the diverse set of supported services. Data centres consist of a large number of commodity servers and switches, support multiple paths among servers, which can be multi-homed, very large aggregate bandwidth and very low latency communication with shallow buffers at the switches.
 
 \noindent\textbf{One-to-many and many-to-one communication.} \black{Modern data centres support a plethora of services that produce one-to-many and many-to-one traffic workloads. Distributed storage systems, such as GFS/HDFS~\cite{GFS, HDFSlink} and Ceph \cite{Ceph}, replicate data blocks across the data centre (with or without daisy chaining\footnote{\url{https://patents.google.com/patent/US20140215257}}). Partition-aggregate \cite{MapReduce,SparkRDD}, streaming telemetry~\cite{sflow-streaming,gangliaDistributed}, distributed messaging~\cite{Akka,JGroups}, publish-subscribe systems \cite{ApacheKafka, GooglePubSub}, high frequency trading~\cite{tradingex1, tradingex2} and replicated state machines~\cite{statemachines1, statemachines2} also produce similar workloads. Multicast has already been deployed in data centres (e.g. to support virtualised workloads \cite{virtualisedNet1} and financial services~\cite{tradingex3}). With the advent of P4, multicasting in data centres is becoming practical \cite{ElmoMulticast}. As a result, much research on scalable network-layer multicasting in data centres has recently emerged \cite{rfcMulticast2, reliablemulticast, DualStructure, ScalingIP,infocom-multicast}, including approaches for optimising multicast flows in reconfigurable data centre networks \cite{SplitCast} and programming interfaces for applications requesting data multicast \cite{republic}.}
 
 Existing data centre transport protocols are suboptimal in terms of network and server utilisation for these workloads. One-to-many data transport is implemented through multi-unicasting or daisy chaining for distributed storage. As a result, copies of the same data are transmitted multiple times, wasting network bandwidth and creating hotspots that severely impair the performance of short, latency-sensitive flows. In many application scenarios, multiple copies of the same data can be found in the network at the same time (e.g. in replicated distributed storage) but only one replica server is used to fetch it. Fetching data from all servers, in parallel, from all available replica servers (many-to-one data transport) would provide significant benefits in terms of eliminating hotspots and naturally balancing load among servers.
 
\begin{figure}[!h]
 	\setlength{\belowcaptionskip}{-3pt}
 	\centering
 	\subcaptionbox{One-to-many (write)}[.48\linewidth][c]{%
 		\includegraphics[scale=0.22]{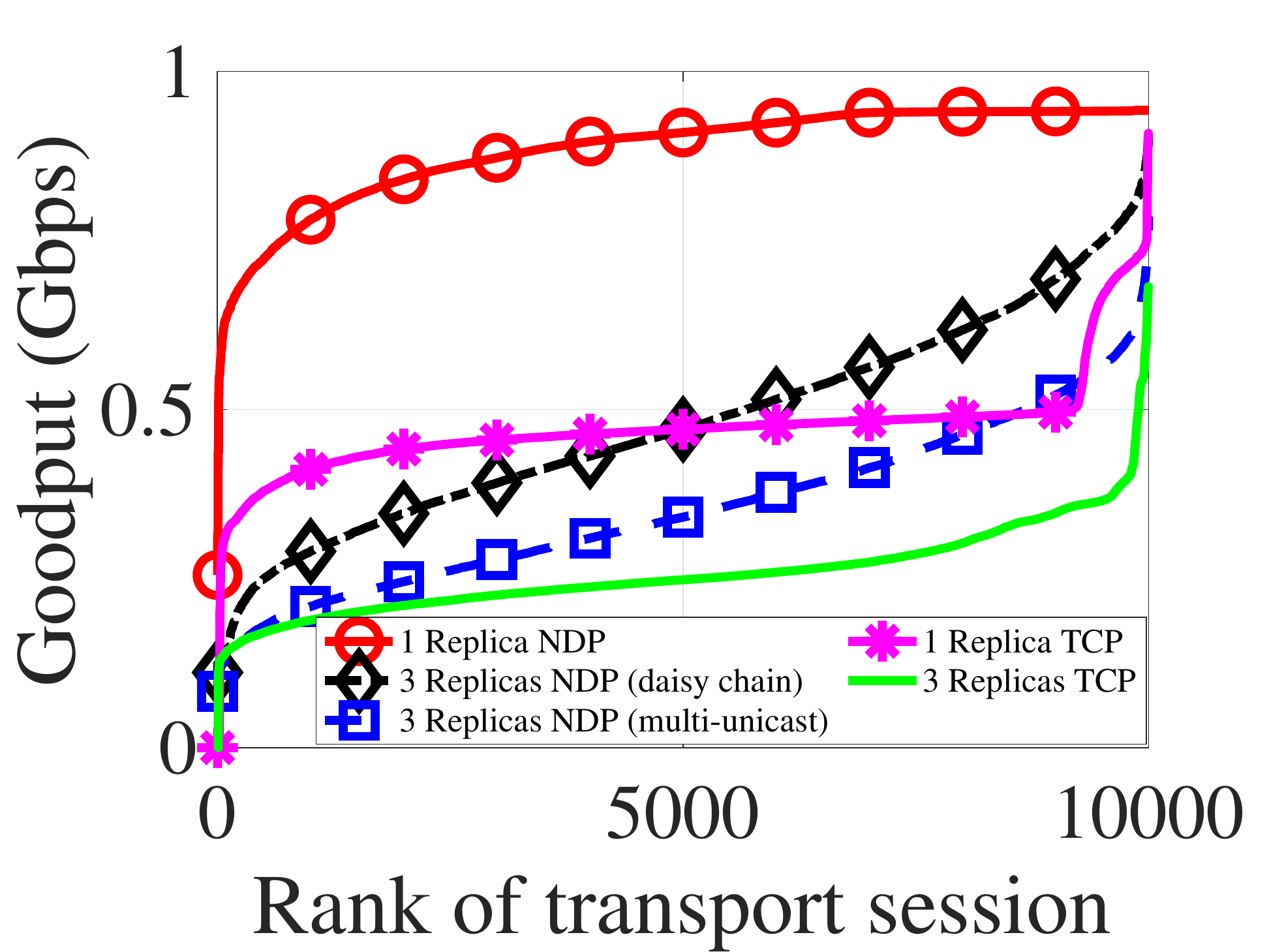}}\quad
 	\subcaptionbox{Many-to-one (read)}[.48\linewidth][c]{%
 		\includegraphics[scale=0.22]{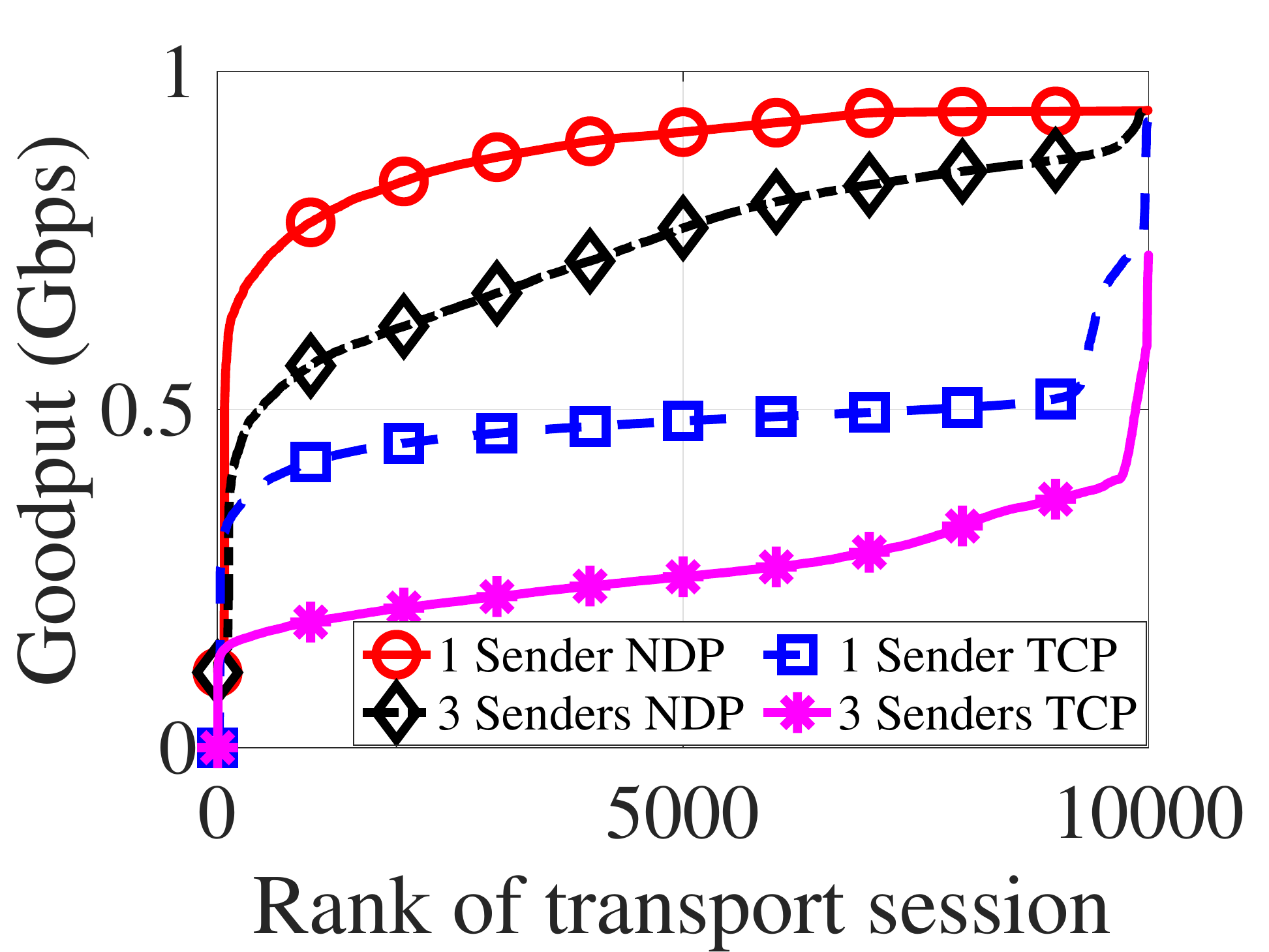}}\quad
 	\caption{Goodput in a 250-server FatTree topology with 1GB link speed \& 10$\mu$s link delay. Background traffic is present to simulate congestion. Results are for 10,000 (a) write and (b) read block requests (2MB each). Each I/O request is `assigned' to a host in the network, which is selected uniformly at random and acts as the client. Requests' arrival times follow a Poisson process with an inter-arrival rate $\lambda=1000$. Replica selection and placement is based on HDFS' default policy.}
 	\label{tcpndp} 
 	\vspace{-2mm}
 \end{figure}

 These performance limitations are illustrated in Figure \ref{tcpndp}, where we plot the application goodput for TCP and NDP (Novel Datacenter transport Protocol) \cite{NDP} in a distributed storage scenario with $1$ and $3$ replicas. When a single replica is stored in the data centre, NDP performs very well, as also demonstrated in \cite{NDP}. TCP performs poorly\footnote{It is well-established that TCP is ill-suited for meeting throughput and latency requirements of applications in data centre networks, therefore we will be using NDP and  PIAS~\cite{PIAS} as the baseline protocols in this paper.}. On the other hand, when three replicas are stored in the network, both NDP and TCP perform poorly in both write and read workloads. Writing data involves either multi-unicasting replicas to all three servers (blue and green lines in Figure \ref{tcpndp}a) or daisy chaining replica servers (black line); although daisy chaining performs better, avoiding the bottleneck at the client's uplink, they both consume excessive bandwidth by moving multiple copies of the same block in the data centre. Fetching a data block from a single server when it is stored in two more servers creates hotspots at servers' uplinks due to collisions from randomly selecting a replica server for each read request (see black and purple lines in Figure \ref{tcpndp}b).

 \noindent\textbf{Long and short flows. }Modern cloud applications commonly have strict latency requirements \cite{DCTCP,minimizingfct,RepFlow,OneMoreQueue,HOMA,pfabric}. At the same time, background services require high network utilisation \cite{infocom-morteza, Improving-Datacenter-MPTCP, packet-spraying, Hedera}. A plethora of mechanisms and protocols have been proposed to date to provide efficient access to network resources to data centre applications, by exploiting support for multiple equal-cost paths between any two servers \cite{Improving-Datacenter-MPTCP,NDP,packet-spraying,FMTCP2015} and hardware capable of low latency communication \cite{HOMA,AUTO,pHost-CoNEXT-2015} and eliminating Incast \cite{TCP-Incast-2012, LTTP , Jiang} and Outcast \cite{TCP-Outcast}.  Recent proposals commonly focus on a single dimension of the otherwise complex problem space; e.g. TIMELY\cite{TIMELY}, DCQCN\cite{DCQCN-RDMA}, QJUMP \cite{qjump} and RDMA over Converged Ethernet v2 \cite{rdma} focus on low latency communication but do not support multi-path routing. Other approaches \cite{Hedera, packet-spraying} do provide excellent performance for long flows but perform poorly for short flows \cite{Improving-Datacenter-MPTCP,infocom-morteza}. None of these protocols support efficient one-to-many and many-to-one communication.
 
 \noindent\textbf{Contribution. }In this paper we propose SCDP\footnote{\black{SCDP builds on our early work on integrating fountain coding in data transport protocols \cite{polyraptor, Trevi, scdp-arxiv}. In \cite{Trevi} we motivated the need for a novel data transport mechanism to efficiently support one-to-many and many-to-one communication and argued that rateless codes is the way forward in doing so. In \cite{polyraptor}, we introduced an early version of SCDP to the research community.}}, a general-purpose data transport protocol for data centres that, unlike any other protocol proposed to date, supports efficient one-to-many and many-to-one communication. This, in turn, results in significantly better overall network utilisation, minimising hotspots and providing more resources to long and short unicast flows. At the same time, SCDP supports fast completion of latency-sensitive flows and consistently high-bandwidth communication for long flows. SCDP eliminates Incast and Outcast.  All these are made possible by integrating RaptorQ codes \cite{RFC-6330-RQ, RaptorQ-book} with receiver-driven data transport \cite{NDP, HOMA}, in-network packet trimming \cite{cuttingPayload, NDP} and Multi-Level Feedback Queuing (MLFQ) \cite{PIAS}.
 
 \noindent\textbf{SCDP performance overview. }We found that SCDP improves goodput performance by up to $\sim$50\% compared to NDP and  $\sim$60\% compared to PIAS with different application workloads involving one-to-many and many-to-one communication (\textsection\ref{goodput-performance}). Equally importantly, it reduces the average FCT for short flows by up to $\sim$45\% compared to NDP and $\sim$70\% compared to PIAS under two realistic data centre traffic workloads (\textsection\ref{realistic-workloads}).  For short flows, decoding latency is minimised by the combination of the systematic nature of RaptorQ codes and MLFQ; even in a $70\%$ loaded network, decoding was needed for only $9.6\%$ of short flows. This percentage was less than $1\%$ in a $50\%$ congested network (\textsection\ref{network-overhead}). The network overhead induced by RaptorQ codes is negligible compared to the benefits of supporting one-to-many and many-to-one communication. Only 1\% network overhead was introduced when the network was very heavily congested (\textsection\ref{unnecessary-overhead}). RaptorQ codes have been shown to perform exceptionally well even on a single core, in terms of encoding/decoding rates. We therefore expect that with hardware offloading, in combination with SCDP's block pipelining mechanism {(\textsection\ref{pipelining})}, the required computational overhead will not be significant.
 
 \vspace{-1mm}
 \section{RaptorQ Encoding and Decoding}
 \label{raptorQ}
 
 \noindent\textbf{Encoding. }RaptorQ codes are \emph{rateless} and \emph{systematic}. The input to the encoder is one or more \emph{source blocks}; for each one of these source blocks, the encoder creates a potentially very large number of \emph{encoding symbols} (rateless coding). All  $K$ source symbols (i.e. the original fragments of a source block) are amongst the set of encoding symbols (systematic coding). All other symbols are called \emph{repair} symbols. Senders initially send source symbols, followed by repair symbols, if needed.
 
 \noindent\textbf{Decoding. }A source block can be decoded after receiving a number of symbols that must be equal to or larger than the number of source symbols; all symbols contribute to the decoding process equally. In a lossless communication scenario, decoding is not required, because all source symbols are available (systematic coding).
 
 \noindent\textbf{Performance. } In the absence of loss, RaptorQ codes do not incur any network or computational overhead. The trade-off associated with RaptorQ codes when loss occurs is with respect to some (1) minimal network overhead to enable successful decoding of the original fragments and (2) computational overhead for decoding the received symbols to the original fragments. RaptorQ codes behave exceptionally well in both respects.  With two extra encoding symbols (compared to the number of original fragments), the decoding failure probability is in the order of $10^{-6}$. It is important to note that decoding failure is not fatal; instead one or more encoding symbols can be requested in order to ensure that decoding is successful~\cite{RaptorQ-book}. The time complexity of RaptorQ encoding and decoding is linear to the number of source symbols. RaptorQ codes support excellent performance for all block sizes, including very small ones, which is very important for building a general-purpose data transport protocol that is able to handle efficiently a diverse set of workloads. In \cite{CodornicesRqNew, LiquidCloudStorage}, the authors report encoding and decoding speeds of over 10 Gbps using a RaptorQ software prototype running on a single core.  With hardware offloading, RaptorQ codes would be able to support data transport at line speeds in modern data centre deployments. On top of that, multiple blocks can be decoded in parallel, independently of each other (e.g. on different cores). Decoding small source blocks is even faster, as reported in~\cite{CodornicesRqNew}. The decoding performance does not depend on the sequence that symbols arrived nor on which ones do.
 
\begin{figure}[t]
 	\centering
 	\includegraphics[scale=0.073]{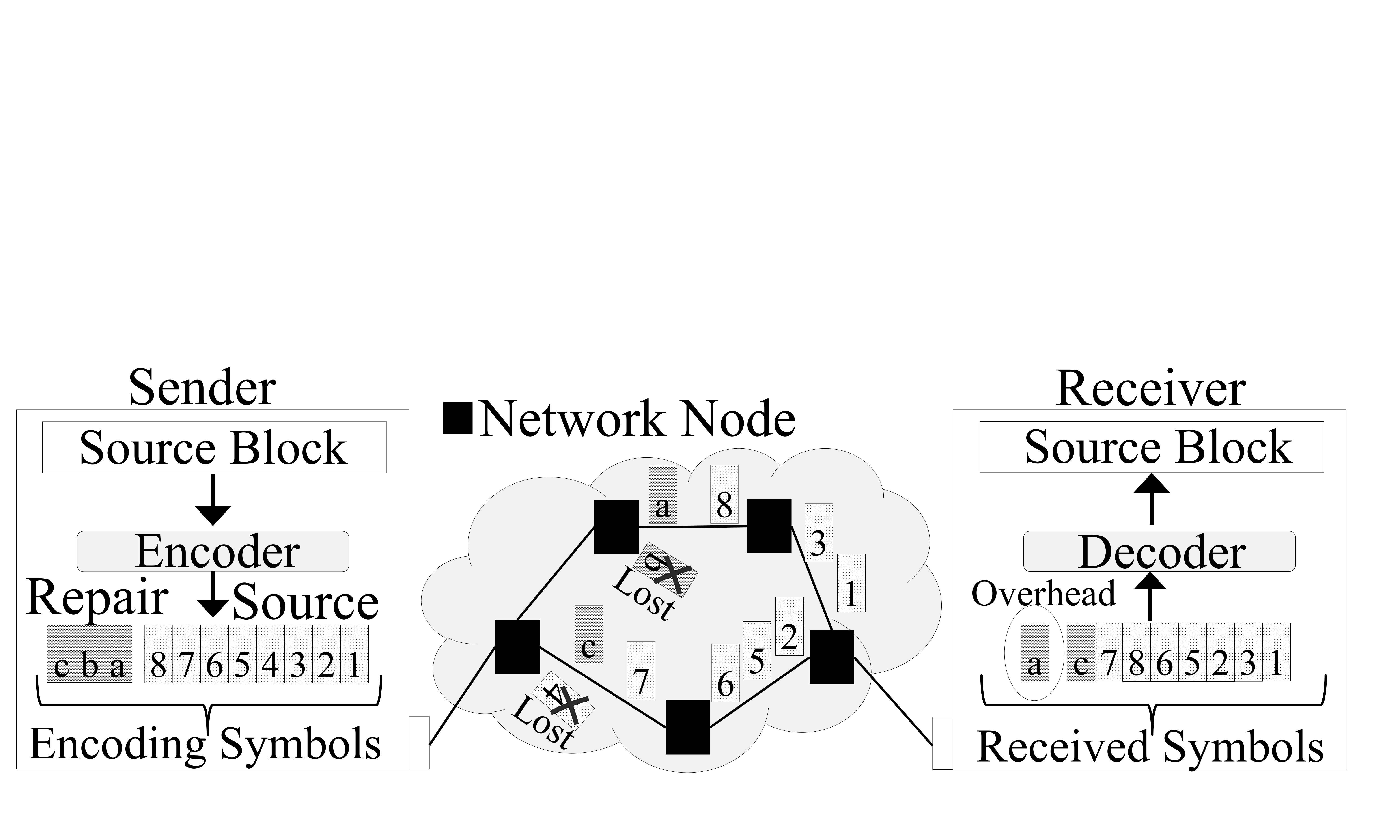}\quad
 	\caption{RaptorQ-based communication}
 	\label{RQ-block-diagram}
 	\vspace{-4mm}
 \end{figure}

 \noindent\textbf{Example. }Before explaining how RaptorQ codes are integrated in SCDP, we present a simple example of point-to-point communication between two hosts, which is illustrated in Figure \ref{RQ-block-diagram}\footnote{Note that Figure \ref{RQ-block-diagram} does not illustrate SCDP's underlying mechanisms. The design of SCDP is discussed extensively in Section \ref{design}.}. On the sender side, a single source block is passed to the encoder that fragments it into $8$ equal-sized source symbols $S\textsubscript{1},S\textsubscript{2}, ..., S\textsubscript{8}$. The encoder uses the source symbols to generate repair symbols $S\textsubscript{a},S\textsubscript{b},S\textsubscript{c}$ (here, the decision to encode $3$ repair symbols is arbitrary). Encoding symbols are transmitted to the network, along with the respective encoding symbol identifiers (ESI) and source block numbers (SBN) \cite{RFC-6330-RQ}. As shown in Figure \ref{RQ-block-diagram}, symbols $S\textsubscript{4}$ and $S\textsubscript{b}$ are lost. Symbols take different paths in the network but this is transparent to the receiver that only needs to collect a specific amount of encoding symbols (source and/or repair). The receiver can receive symbols from multiple senders from different network interfaces. In this example, the receiver attempts to decode the original source block upon receiving $9$ symbols, i.e. with one extra symbol which is network overhead (as shown in Figure \ref{RQ-block-diagram}). Decoding is successful and the source block is passed to the receiver application. As mentioned above, if no loss had occurred, there would be no need for decoding and the data would have been directly passed to the application.
 
 \noindent \textbf{Erasure coding in data transport. } There is a long and interesting trail of research that integrates erasure coding into data transport protocols. SCDP is unique compared to all these works, efficiently supporting one-to-many and many-to-one data transport sessions for distributed storage and numerous other workloads prevalent in modern data centres, without sacrificing performance for traditional short and long flows. In ~\cite{end-to-end-coding}, the authors explore the advantages and challenges of integrating end-to-end coding into TCP. Corrective~\cite{Reducing-Web-Latency} employs coding for faster loss recovery but it can only deal with one packet loss in one window as its coding redundancy is fixed. FMTCP~\cite{FMTCP2015} employs fountain coding to improve the performance of MPTCP~\cite{Improving-Datacenter-MPTCP} by recovering data over multiple subflows. LTTP~\cite{LTTP} is a UDP-based transport protocol that uses fountain codes to mitigate Incast in data centres. CAPS~\cite{CAPS-coding} deals with out of order data by integrating forward error correction on short flows, in order to reduce their flow completion time, and employs ECMP for achieving high throughput for long flows. RC-UDP~\cite{RCUDP} is a rateless coding data transport protocol that enables reliable data transfer over high bandwidth networks. It uses block-by-block flow control where the sender keeps sending encoded symbols until the receiver sends an acknowledgement indicating successful decoding. PPUSH~\cite{scdpcitedpaper} is a multi-source data delivery protocol that employs RaptorQ codes for sending multiple flows in parallel using all available replicas.

 \section{The Case for RaptorQ Coding in Data Transport for Data Centre Networks}
 \label{case-for-rq}
 
 The starting point in designing SCDP, which is also the key differentiator to the rest of the literature, is its efficient handling of one-to-many and many-to-one communication, without sacrificing performance for traditional unicast flows.
 
 \noindent\black{\textbf{One-to-many communication. }None of the existing data transport protocols for data centres can support communication beyond traditional unicast flows, even if network-level multicasting was deployed in the network. Congestion control in reliable multicasting is a challenging problem and traditional sender-driven, reliable multicasting approaches (e.g. as in~\cite{RFC2362, RizzoSigcomm}) would suffer from Incast \cite{TCP-Incast-2012}, and lack of support for multipath routing and multi-homed servers, as well as their inability to spray packets in the network. A receiver-driven approach would be more suitable. However, extending approaches, such as NDP \cite{NDP} or Homa \cite{HOMA}, is far from trivial as this would entail complications with flow control, when losses occur, because lost packets must be retransmitted. Senders would have to maintain state, enqueuing incoming pull requests by multiple receivers, while waiting to multicast a new packet or retransmit a lost packet. Equally importantly, the slowest receiver would slow down all other receivers.}\footnote{\black{How existing protocols for data centres could be extended to support one-to-many and many-to-one communication is beyond the scope of this paper.}}
 
 \black{With RaptorQ codes and receiver-driven flow control, one-to-many communication is simple and efficient: a sender multicasts a new symbol after receiving a \textit{pull} request from all receivers (see Section \ref{multicast} for a detailed description). A sender does not need to remember which symbols it has sent as there is no notion of retransmission. Instead, it only needs to count the number of pending pull requests from each receiver so it can `clock' symbol sending. A receiver can decode the original data and complete the session after it receives the necessary amount of symbols (see Section \ref{raptorQ}), independently of other receivers that may be behind in terms of receiving symbols because of network congestion (e.g. when they are connected to a congested ToR switch).}
 
 \noindent\black{\textbf{Many-to-one communication. }Existing protocols do not and could not support many-to-one communication in a way that benefits the overall performance. Even if senders were instructed to only send a subset of the original data fragments (emulating many-to-one communication), a congested or slow server would always be the bottleneck for the whole session.}
 
 \black{With RaptorQ codes, each sender contributes as much as it can, given the current conditions, in terms of network congestion and local load. The rateless nature of RaptorQ codes, enable receivers to successfully decode a source block regardless of which server sent the symbols. The only requirement is to receive the required number of symbols (See Section \ref{raptorQ}). This is a unique characteristic of SCDP (see Section \ref{multi-source}), which `bypasses' network hotspots by having non-congested servers contributing more symbols to the receiver. Crucially, this is done without any central coordination.}
 
 \noindent\black{\textbf{Flow completion time and goodput. }SCDP's benefits discussed above do not come at a cost for traditional unicast flows. This is due to the combination of the systematic nature of RaptorQ codes, MLFQ, and packet trimming. More specifically, FCT for short flows is very small, unaffected by the introduction of coding because senders first send the original data fragments (systematic coding) with the highest priority, minimising loss for them. As a result, decoding is very rarely required for short flows. SCDP performs exceptionally well also for long flows despite the fact that (the otherwise efficient) decoding is needed more often. This is done by employing pipelining of source blocks, which alleviates the decoding overhead for large data blocks and maximises application goodput (see Section \ref{pipelining}). In combination with receiver-driven flow control and packet trimming, SCDP eliminates Incast and Outcast, playing well with switches' shallow buffers}.
 
 \noindent\black{\textbf{Network utilisation. }SCDP ensures high network utilisation for all communication modes; with RaptorQ coding there is no notion of ordering, as all symbols contribute to the decoding (if needed) of source data. As a result, symbols can be sprayed in the network through all available paths maximising utilisation and minimising the formation of hotspots. At the same time, receivers can receive symbols from different interfaces naturally enabling multi-homed topologies (e.g. \cite{BCube,Jellyfish}).}
 
 \section{SCDP Design}
 \label{design}
 
 \black{In this section, we present SCDP's design; we define SCDP's packet types and adopted switch model. We then describe all SCDP's supported communication modes, and how we maximise goodput and minimise flow completion time (FCT) for long and short flows, respectively.}
 
 \vspace{-3mm}
 \subsection{Packet Types}
 \label{packet-types}
 
\begin{figure}[t]
 	
 	\begin{center}
 		\includegraphics[width=1\linewidth,height=0.15\textheight]{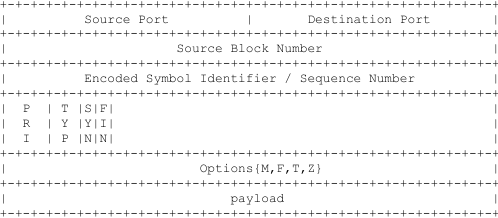} 
 		\caption{\textcolor{black}{SCDP packet format}}
 		\label{pkt-types}
 	\end{center}
 	\vspace{-5mm}
 \end{figure}

 \black{SCDP's packet format is shown in Figure \ref{pkt-types}. Port numbers are used to identify a transport session. The type field (\textsc{typ} in Figure \ref{pkt-types}) is used to denote one of the three SCDP packet types; \textit{symbol}, \textit{header} and \textit{pull} (denoted as \textsc{smbl}, \textsc{hdr} and \textsc{pull}, respectively, in Algorithms~\ref{sender-alg} and \ref{receiver-alg}). The priority field (\textsc{pri} in Figure \ref{pkt-types}) is set by the sender and is used by MLFQ (see Section \ref{service-model}).} 
 
 A \emph{symbol} packet carries in its payload one MTU-sized source or repair symbol. The source block number (SBN) identifies the source block the carried symbol belongs to. The encoding symbol identifier (ESI) identifies the symbol within the stream of source and repair symbols for the specific source block \cite{RFC-6330-RQ}. A sender initiates a transport session by pushing an initial window of symbols with the \emph{syn} flag set, for the first source block. These symbol packets also carry a number of \emph{options}: the \emph{transfer mode} (\textsc{m} in Figure \ref{pkt-types}) can be unicast, many-to-one or one-to-many. The rest of the options are used to define the total length of the session (\textsc{f} in Figure \ref{pkt-types}), number of source blocks (\textsc{z} in Figure \ref{pkt-types}) and the symbol size (\textsc{t} in Figure \ref{pkt-types}). The source block size $K$ is derived from these options as described in RaptorQ RFC~\cite{RFC-6330-RQ}. We adopt the notation used in this RFC\cite{RFC-6330-RQ}.
 
 \emph{Header} packets are trimmed versions of symbol packets. Upon receiving a symbol packet that cannot be buffered, a network switch trims its payload and forwards the header, with the highest priority. Header packets are used to ensure that a window (\textit{w}) of symbol packets is always in-flight.
 
 A \emph{pull} packet is sent by a receiver to request a symbol. The sequence number is only used to indicate how many symbols of the specified source block identifier to send, in case pull requests get reordered. Multiple symbol packets may be sent in response to a single pull request, as described in Section \ref{unicast}. The \emph{fin} flag is used to identify the last pull request; upon receiving such a pull request, a sender sends the last symbol packet for this SCDP session.
 
 \vspace{-2mm}
 
 \vspace{-3.5mm}
 \subsection{Switch Service Model} 
 \label{service-model}
 
 SCDP relies on network switching functionality that is either readily available in today's data centre networks \cite{HOMA} or is expected to be \cite{NDP} when P4 switches are widely deployed. SCDP does not require any more switch functionality than NDP~\cite{NDP}\footnote{\black{As reported in~\cite{NDPRethinking}, there is ongoing work by switch vendors to implement the NDP switch. Moreover, a smartNIC implementation of the NDP end-host stack is also ongoing. This is very promising for the deployability of next-generation protocols, including SCDP, in the real-world.}}, Homa \cite{HOMA}, QJUMP~\cite{qjump}, or PIAS~\cite{PIAS} do.
 
 \noindent\textit{Priority scheduling and packet trimming.} In order to support latency-sensitive flows, we employ MLFQ \cite{PIAS}, and packet trimming \cite{cuttingPayload}. We assume that network switches support a small number of queues with respective priority levels. The top priority queue is only used for header and pull packets. This is crucial for swiftly providing feedback to receivers about loss. Given that both types of packets are very small, it is extremely unlikely that the respective queue gets full and that they are dropped\footnote{SCDP receivers employ a simple timeout mechanism, as in \cite{NDP}, to recover from the unlikely losses of pull and header packets.}. The rest of the queues are small and buffer symbol packets. Switches perform weighted round-robin scheduling between the top-priority (header/pull) queue and the symbol packet queues. This guards against a congestion collapse situation, where a switch only forwards trimmed headers and all symbol packets are trimmed to headers. When a data packet is to be transmitted, the switch selects the head packet from the highest priority, non-empty queue.
 
 \noindent\textit{Multipath routing.} SCDP packets are sprayed to all available equal-cost paths to the destination\footnote{In SCDP's one-to-many transfer mode there are multiple destinations.} in the network. SCDP relies on ECMP and spraying could be done either by using randomised source ports \cite{infocom-morteza}, or the ESI of symbol and header packets and the sequence number of pull packets.

 \begin{algorithm}[t]
 	\setlength{\textfloatsep}{0pt}
 	\scriptsize
 	\DontPrintSemicolon
 	\SetNoFillComment
 	\SetKwFunction{FInit}{initSession()}
 	\SetKwFunction{FCreateHeader}{createHeader(ss, syn, fin, type)}
 	\SetKwFunction{FRecv}{onReceivePullRequest(pullReq)}
 	\SetKwFunction{FCreate}{createPacket(ss)}
 	\SetKwFunction{FSendNet}{sendPacket(header,symbol)}
 	\SetKwProg{Fn}{Function}{}{}
 	
 	SessionState $ss$\;
 	\Fn{\FInit}{
 		{// initialise $ss$: $SBN$, $srcPort$, $dstPort$, $type$, $options$}\\
 		$ss.ESI \leftarrow 0 $\;
 		$ss.expectedPullSeqNum \leftarrow 0 $\;
 		$ss.numSentSymbols \leftarrow 0 $\; 
 		\While{$ss.numSentSymbols < w$} {
 			$hdr \leftarrow createHeader(ss, true, false, SMBL)$\;
 			$symbol \leftarrow getNextSymbol(ss)$ \textit{// source or repair symbol}\;
 			$sendPacket(hdr,symbol)$ \textit{// send to network}\;
 			$ss.ESI \leftarrow ss.ESI + 1 $\;
 			$ss.numSentSymbols \leftarrow s.numSentSymbols + 1$\;
 		}
 	}
 	
 	\Fn{\FRecv}{
 		$gap \leftarrow ss.expectedPullSeqNum $ - $ pullReq.seqNum $\;
 		\While{$gap > 0$} {
 			$hdr \leftarrow createHeader(ss, false, false, SMBL)$\;
 			$symbol \leftarrow getNextSymbol(ss)$ \textit{// source or repair symbol}\;
 			$sendPacket(header,symbol)$ \textit{// send to network}\;
 			$gap \leftarrow gap - 1 $\;
 		}
 		\If{$\textit{pullReq.fin} == true$}
 		{ 
 			\textit{// session will be completed and \;}
 			\textit{// garbage collected soon (in a timeout) \;}
 			$ss.toBeGarbageCollected \leftarrow true$ \;
 		}
 	}
 	
 	\Fn{\FCreateHeader}{
 		$ hdr  \leftarrow createHeader(ss)$ \textit{// sets port numbers}\;
 		$ hdr.\{SBN, syn, fin\} \leftarrow (ss.SBN, syn, fin)$\;
 		$ hdr.\{typ,pri, opts\} \leftarrow (type, getMLFQPriority(), ss.opts)$\;
 		\If{$\textit{type} == SMBL$}{$ hdr.\{ESI\} \leftarrow ss.ESI$\;}
 		\If{$\textit{type} == PULL$}{$ hdr.\{seqNum\} \leftarrow ss.seqNum$\;}
 	}
 	
 	\caption{\textcolor{black}{SCDP Sender}}
 	\label{sender-alg}
 	\vspace{-1mm}
 \end{algorithm}

 \begin{algorithm}[t]
 	\scriptsize
 	\DontPrintSemicolon
 	\SetNoFillComment
 	\SetKwFunction{FInit}{initSession(packet)}
 	\SetKwFunction{FSend}{sendSymbol()}
 	\SetKwFunction{FGetHeader}{getHeaderInfo(pkt.hdr)}
 	\SetKwFunction{FGetHeaderTemp}{getHeaderInfo(hdr)}
 	\SetKwFunction{FRecvPacket}{onReceivePacket(pkt)}
 	\SetKwFunction{FRecvSymbol}{processSymbol(symbol)}
 	\SetKwFunction{FRecvSymbolTemp}{processSymbol(pkt.payload)}
 	\SetKwFunction{FRecvTrim}{processHeader(header)}
 	\SetKwFunction{FDecodeSB}{decodeSrcBlock()}
 	\SetKwFunction{FAddPull}{addPullRequest()}
 	\SetKwProg{Fn}{Function}{}{}
 	\SetKwComment{tcp}{\small // }{}%
 	\SetCommentSty{small}
 	
 	SessionState $ss$\;
 	$ss.established\leftarrow false$\;
 	
 	\Fn{\FRecvPacket}{
 		$syn, type \leftarrow  $ \FGetHeader \;
 		\If{$syn == true$ \&\& $ss.established == false$}{
 			$ ss.\{established, requestMoreSymbols\} \leftarrow (true, true)$\;
 			$ ss.\{seqNum, numRcvdSymbols\} \leftarrow (0, 0)$\;
 			$ss.K  \leftarrow  calcKFromOpts(ss.opts)$ \;\textit{// $K$ is derived from the header options as in RaptorQ RFC~\cite{RFC-6330-RQ}}  
 		}
 		
 		\If{$type == SMBL $}{
 			\FRecvSymbolTemp
 		}
 		
 		\If{$type == HDR $}{
 			\FRecvTrim
 		}
 	}
 	
 	\Fn{\FRecvSymbol}{
 		$ss.storeSymbol($$\textit{symbol}$$)$\;
 		$ss.numRcvdSymbols  \leftarrow  ss.numRcvdSymbols + 1 $\;
 		\eIf {$ss.numRcvdSymbols == ss.K$ \&\& $ss.overhead==0$}{
 			$ss.skipDecoding   \leftarrow true$\;
 			$ss.requestMoreSymbols   \leftarrow false$\;
 			$ss.deliverSBN() $ \textit{ // deliver to application layer}
 		}
 		{\If{$ss.numRcvdSymbols == ss.K+ss.overhead$}{\FDecodeSB}}
 		\If{$ss.numRcvdSymbols == ss.K - 1$}{$ss.Fin \leftarrow  true$}
 		\If{$ss.requestMoreSymbols == true$}{\FAddPull}
 	}
 	
 	\Fn{\FRecvTrim}{
 		$ss.overhead \leftarrow 2$\;
 		\FAddPull
 	}
 	
 	\Fn{\FGetHeaderTemp}{
 		$ (ss.SBN, ss.ESI, ss.opts)  \leftarrow hdr.\{SBN,ESI, opts\} $\;
 		$ (type, syn) \leftarrow hdr.\{typ, syn\} $\;
 		$return$ $(type, syn)$\;
 	}
 	
 	\Fn{\FAddPull}{
 		$ss.seqNum \leftarrow  ss.seqNum+1 $\;
 		$pullReq \leftarrow createHeader(ss, false, ss.Fin, PULL)$\;
 		{// $createHeader$ is defined in Algorithm~\ref{sender-alg} }\\
 		$enqueuePullRequest(pullReq)$\;
 	}
 	
 	\Fn{\FDecodeSB}{
 		$success \leftarrow ss.decode()$\;
 		\eIf{$success == true$}{
 			$ss.requestMoreSymbols  \leftarrow false$\;
 			$ss.deliverSBN()$ \textit{ // deliver to application layer}\;
 		}
 		{
 			$ss.overhead\leftarrow ss.overhead + 1$ \textit{// very rare}\;
 		}
 	}
 	
 	\caption{\textcolor{black}{SCDP Receiver}}
 	\label{receiver-alg}
 	\vspace{-1mm}
 \end{algorithm}
 
 \vspace{-4mm}
 \subsection{Unicast Transport Sessions}
 \label{unicast}
 A sender implicitly opens a unicast SCDP transport session by pushing an initial window \black{of $w$ (\textit{syn}-enabled) symbol packets tagged with the highest priority (Lines $2 - 12$ in Algorithm~\ref{sender-alg}\footnote{\black{For clarity, Algorithms \ref{sender-alg} and \ref{receiver-alg} illustrate a slightly simplified version of SCDP for unicast data transport for a single source block without pipelining.}})}. Senders tag outgoing symbol packets with a priority value, which is used by the switches when scheduling their transmission (\textsection\ref{service-model}). The priority of outgoing symbol packets is gradually degraded when specific thresholds are reached. Calculating these thresholds can be done as in PIAS \cite{PIAS} or AuTO \cite{AUTO} (\black{Line 30 in Algorithm ~\ref{sender-alg}}). The receiver establishes a new session upon receiving the first symbol that carries the $syn$ flag (\black{Lines $5 - 10$ in Algorithm~\ref{receiver-alg}}). After receiving the initial window of packets, the receiver takes control of the flow of incoming packets by pacing pull requests to the sender (\black{Lines 33 and 37 in Algorithm~\ref{receiver-alg}}). A pull request carries a sequence number which is auto-incremented for each incoming symbol packet (\black{Line $43$ in Algorithm~\ref{receiver-alg}}). The sender keeps track of the sequence number of the last pull request and, upon receiving a new pull request, it sends one or more packets to fill the gap between the sequence numbers of the last and current request (\black{Lines $14 - 26$ in Algorithm~\ref{sender-alg}}). Such gaps may appear when pull requests are reordered due to packet spraying. Senders ignore pull requests with sequence numbers that have already been `served'; i.e. when they had previously responded to the respective pull requests.
 
 Receivers maintain a single queue of pull requests for all active transport sessions. Flow control's objective is to keep the receiver's incoming link as fully utilised as possible at all times. This dictates the pace at which receivers send pull requests to all different senders. Receivers buffer encoding symbols along with their ESI and SBN and start decoding a source block upon receiving either $K$ source symbols (\black{Lines $20 - 24$ in Algorithm~\ref{receiver-alg}}), where $K$ is the total number of source symbols, or $K+o$ source and repair symbols, when loss occurs \black{($o$ is the induced network overhead in number of symbols) (\black{Lines $25 - 27$ in Algorithm~\ref{receiver-alg}}). As discussed in Section \ref{raptorQ}, RaptorQ codes perform exceptionally well in terms of decoding failure probability; with $o = 2$, which is the value we have chosen for SCDP, the decoding failure is very rare (in the order of $10^{-6}$) and when it happens the penalty is one RTT for requesting one more symbol and the extra latency for attempting decoding twice. It would be extremely unlikely for decoding to fail with $o = 3$\label{key}.}\\
 The receiver sets the \textit{fin} flag in the pull request for the last symbol (a source or repair symbol at that point) that sends to the sender. Note that this may not actually be the last request that the receiver sends, because the symbol packet that is sent in response to that request may get trimmed. All pull requests for the last required symbol (not a specific one) are sent with the \textit{fin} flag on (\black{Lines $29 - 31$ in Algorithm~\ref{receiver-alg}}). The sender responds to fin-enabled pull requests by sending the next symbol in the potentially very large stream of source and repair symbols, with the highest priority. It finally releases the transport session only after a time period that ensures that the last prioritised symbol packet was not trimmed (\black{Lines $22 - 26$ in Algorithm~\ref{sender-alg}}). This time period is very short; in the very unlikely case that the prioritised symbol packet was trimmed, the respective header would be prioritised along with the pull packet subsequently sent by the receiver.
 
 \vspace{-5mm}
 \subsection{One-to-many Transport Sessions}
 \label{multicast}
 One-to-many transport sessions exploit support for network-layer multicast (e.g. with\cite{ElmoMulticast,rfcMulticast2,reliablemulticast, infocom-multicast, DualStructure, ScalingIP}) and coordination at the application layer; for example, in a distributed storage scenario, multicast groups could be pre-established for different replica server groups or setup on demand by a metadata storage server. This would eliminate the associated latency overhead for establishing multicast groups on the fly and is practical for other data centre multicast workloads, such as streaming telemetry \cite{sflow-streaming,gangliaDistributed} and distributed messaging \cite{Akka,JGroups}, where destination servers are known at deployment time. With recent advances in scalable data centre multicasting, a very large number of multicast groups can be deployed with manageable overhead in terms of switch state and packet size. For example, Elmo \cite{ElmoMulticast} encodes multicast group information inside packets, therefore minimising the need to store state at the network switches. With small group sizes, as in the common data centre use cases mentioned above, Elmo can support an extremely large number of groups, which can be encoded directly in packets, eliminating any maintenance overhead associated with churn in the multicast state.  ``In a three-tier data centre topology with 27K hosts, Elmo supports a million multicast groups using a 325-byte packet header, requiring as few as 1.1K multicast group-table entries on average in leaf switches, with a traffic overhead as low as 5\% over ideal multicast'' \cite{ElmoMulticast}.
 
 As with unicast transport sessions, an SCDP sender initially pushes \black{$w$ (\textit{syn}-enabled) symbol packets tagged with the highest priority}. Receivers then request more symbols by sending respective pull packets. The sender sends a new symbol packet only after receiving a request from all receivers within the same multicast group. \black{In Algorithm \ref{sender-alg}, this would only require a simple extension where the sender counts the number of pending pull requests from each receiver (not shown in order to maintain clarity). Receivers queue and pace pull packets as in the unicast transport mode depicted in Algorithm \ref{receiver-alg}. Network hotspots, (e.g. when incoming symbols are frequently trimmed at the ToR switch), can prevent specific receivers from receiving symbols as fast as other receivers of the same one-to-many session do.} The rateless property of RaptorQ codes is ideal for such situation; within a single transport session, receivers may receive a different set of symbols but they will all decode the original source block as long as the required number of symbols is collected, regardless of which symbols they missed (see Section \ref{raptorQ}). \black{Receivers successfully decode the original data as soon as they receive the necessary number of symbols and they are not slowed down by receivers that are behind a hotspot. This is an important property for applications that only require a specific subset of receivers (e.g. some form of quorum) to receive the data before notifying a user or some other service.} 
 
 \black{Some receivers may end up receiving more symbols than what would be required to decode the original source block. This is unnecessary network overhead induced by SCDP but, in Section \ref{unnecessary-overhead}, we show that even under severe congestion, SCDP performs significantly better than NDP, exploiting the support for network-layer multicast. Dealing with situations where receivers are extremely slow or unresponsive is an important problem. We argue that dealing with such a situation is a policy issue and should be handled at the application rather than the data transport layer. For example, the data transport protocol could notify the application of a straggler server (e.g. in a high-performance, user-space stack deployment), which, in turn, could either ignore the notification and leave the data transport session unchanged or update the multicast group used by the data transport layer. Different applications may have different requirements and consistency constraints that are related to dealing with unresponsive servers. Exploring such policies is outside the scope of this work.}
 
 \vspace{-2.3mm}
 \subsection{Many-to-one Transport Sessions}
 \label{multi-source}
 Many-to-one data transport is a generalisation of the unicast transport discussed in Section \ref{unicast}. \black{Each sender $i$ pushes an initial window $w_i$ of (\textit{syn}-enabled) symbol packets to the receiver, as shown in Algorithm \ref{sender-alg} (Lines $2 - 13$)}. These packets are tagged with the highest priority and may contain source or repair symbols. The total number of initially pushed symbol packets $w_{total}=\sum_{i=1}^{n_s}w_i$, where $n_s$ is the total number of senders, is selected to be larger than the initial window $w$ used in unicast transport sessions\footnote{We assume that the value of $w$ is decided at the application layer.}. This is to enable natural load balancing in the data centre in the presence of slow senders or hotspots in the network. In that case, SCDP ensures that a subset of senders (e.g. $2$ out of $3$ in a 3-replica scenario) can still fill the receiver's downstream link. In Section \ref{window-size-eval}, we show that initial window sizes that are greater than $10$ symbol packets result in the same (high) goodput performance. A large initial window would inevitably result in more trimmed symbol packets, which however would not affect short flows that are prioritised over longer multi-source sessions. As discussed in Section \ref{raptorQ}, RaptorQ codes are rateless and all symbols contribute to the decoding process, therefore the receiver is agnostic to their origin. As a result, efficient data transport can be achieved by partitioning the potentially large stream of source and repair symbols amongst all senders, so that each one produces unique symbols. These can be done through coordination at the application layer or randomness. \black{Receivers behave as shown in Algorithm \ref{receiver-alg}.}
 
 
 \vspace{-2mm}
 \subsection{Maximising Goodput for Long Flows - Block Pipelining}
 \label{pipelining}
 With RaptorQ codes, if loss occurs, the receiver must perform decoding on the collected source and repair symbols (\textsection\ref{raptorQ}). This induces latency before the data can become available to the application. For large source blocks, SCDP masks this latency by splitting the large source block to many smaller blocks, instead of encoding and decoding the whole block. The smaller blocks are then pipelined over a single SCDP session. With pipelining, a receiver decodes each smaller block while receiving symbol packets for the next one, effectively masking the latency induced by decoding, except for the last source block. The latency for decoding this last smaller block is considerably smaller compared to decoding the whole block at once. For short, latency-sensitive flows, this could be a serious issue, but SCDP strives to eliminate losses, resulting in fast, decoding-free completion of short flows (see section below). 
 
 \vspace{-1.2mm}
 \subsection{Minimising Completion Time for Short Flows}
 \label{latency-overhead-tricks}
 SCDP ensures that a window of $w$ symbol packets are on the fly throughout the lifetime of a transport session. The window decreases by one symbol packet for each one of the last $w$ symbol packets that the sender sends. As long as no loss occurs (detected by the receiver through receiving a trimmed header), a receiver sends $K - w$ pull requests in total, where $K$ is the number of source symbols (or original fragments) and $w$ is the size of the initial window. For every received trimmed header (i.e. observed loss), the receiver sends a pull request, and, subsequently, the sender sends a new symbol, which equally contributes to the decoding of the source block. This ensures that SCDP does not induce any unnecessary overhead; i.e. symbol packets that are redundant in decoding the source block. The target for the total number of received symbols also changes when loss is detected. Initially, all receivers aim at receiving $K$ source symbols. Upon receiving the first trimmed header, \black{the target changes to $K+o$ (where $o$ is the overhead discussed in Section \ref{unicast}), which ensures that decoding failure is extremely unlikely to occur (see Section \ref{raptorQ}).}
 
 By prioritising earlier packets of a session over later ones through MLFQ, SCDP minimises loss for short flows. This has an extremely important corollary in terms of SCDP's computational cost; no decoding is required for the great majority of short flows, therefore completion times are almost always near-optimal. We evaluate this aspect of SCDP's design in Section \ref{network-overhead}. Note that for all supported types of communication, encoding latency can be masked either (1) by pre-encoding a number of repair symbols or (2) by generating repair symbols while sending source and previously generated repair symbols. The latter is possible due to the systematic nature of RaptorQ coding that enables senders to begin transmission before generating any repair symbols, by sending the original data fragments (i.e. source symbols).

 
 \begin{figure}[t]
 	\setlength{\belowcaptionskip}{-4pt}
 	\centering
 	\includegraphics[scale=0.364]{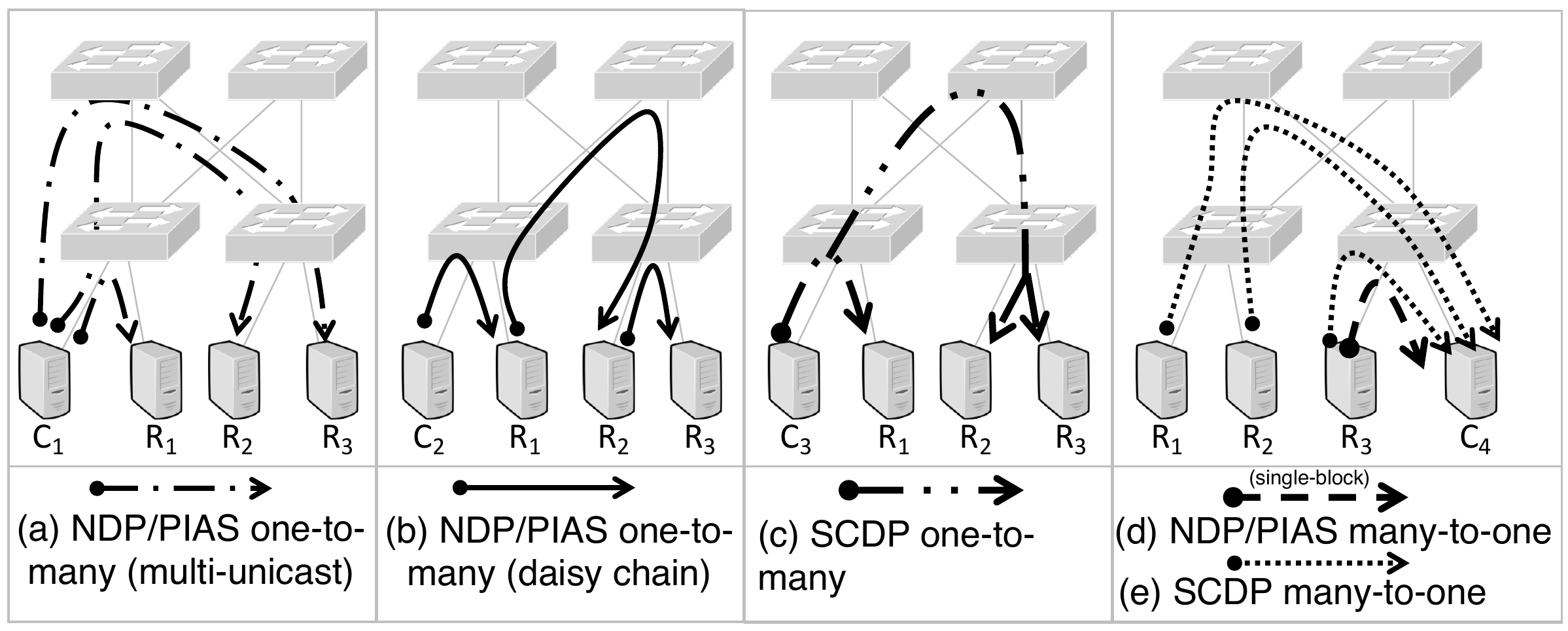}\quad
 	\caption{Read/write workloads and replica placement policy used in evaluation. In our simulations, the selection of remote racks to store data blocks is random and racks in different pods can be selected (i.e. core switches are involved).}
 	\label{modern-workloads}
 	\vspace{-3mm}
 \end{figure}
 
 \begin{figure*}[t]
 	
 	\subcaptionbox{rs = 1MB, $\lambda$ = 2000}[.231\linewidth][c]{%
 		\includegraphics[scale=0.2]{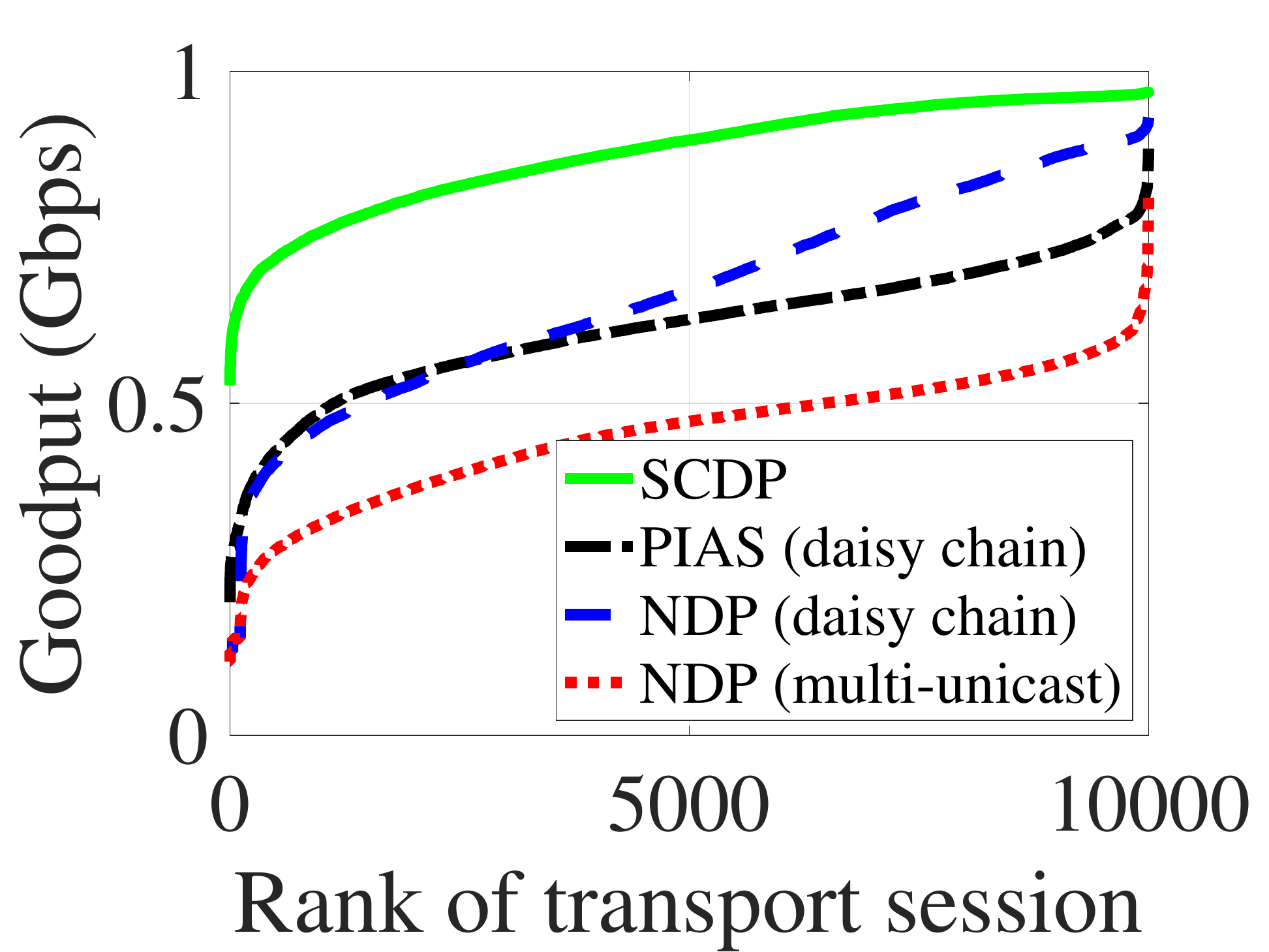}}\quad
 	\subcaptionbox{rs = 1MB, $\lambda$ = 4000}[.231\linewidth][c]{%
 		\includegraphics[scale=0.2]{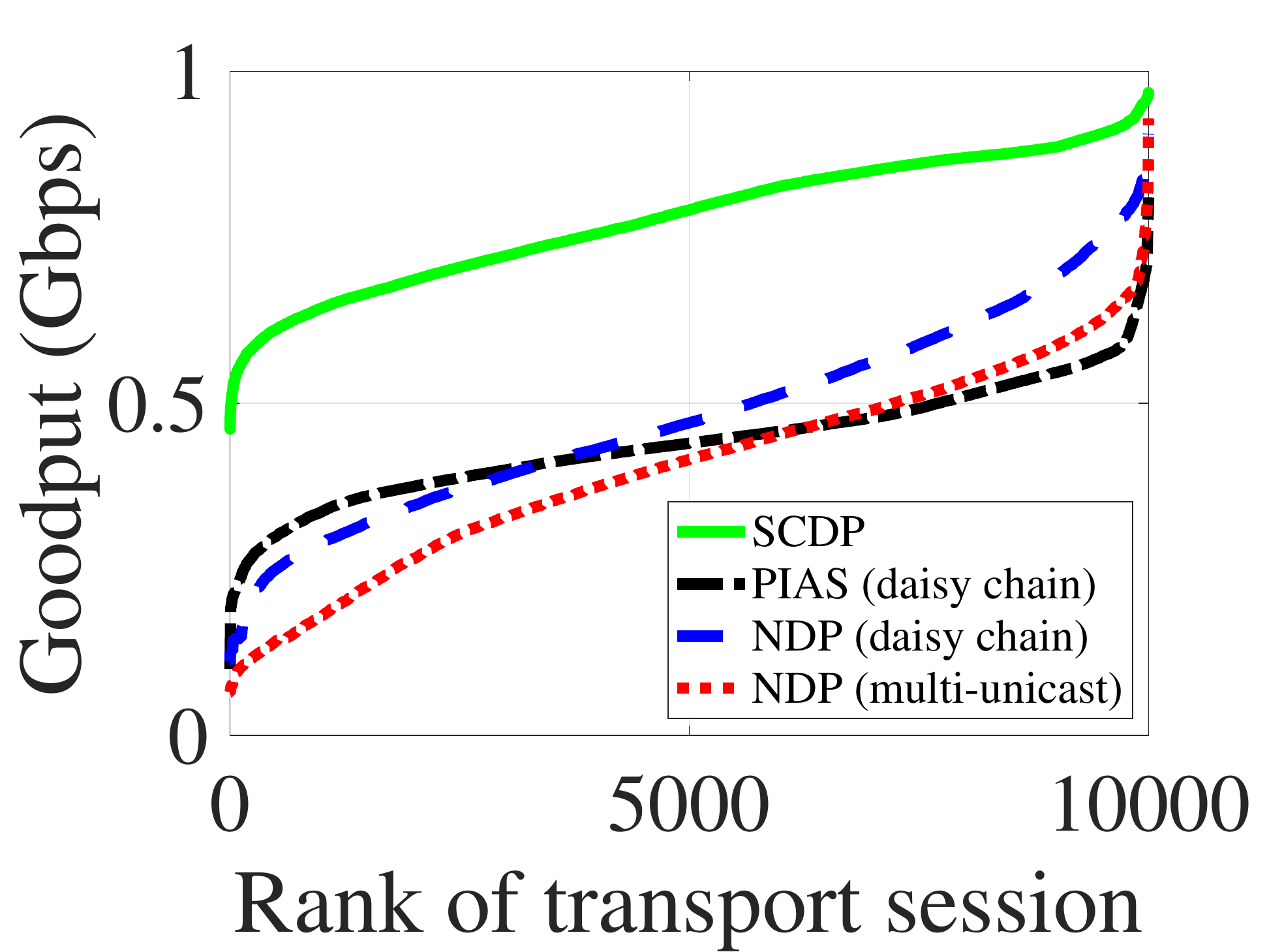}}\quad
 	\subcaptionbox{rs = 4MB, $\lambda$ = 2000}[.231\linewidth][c]{%
 		\includegraphics[scale=0.2]{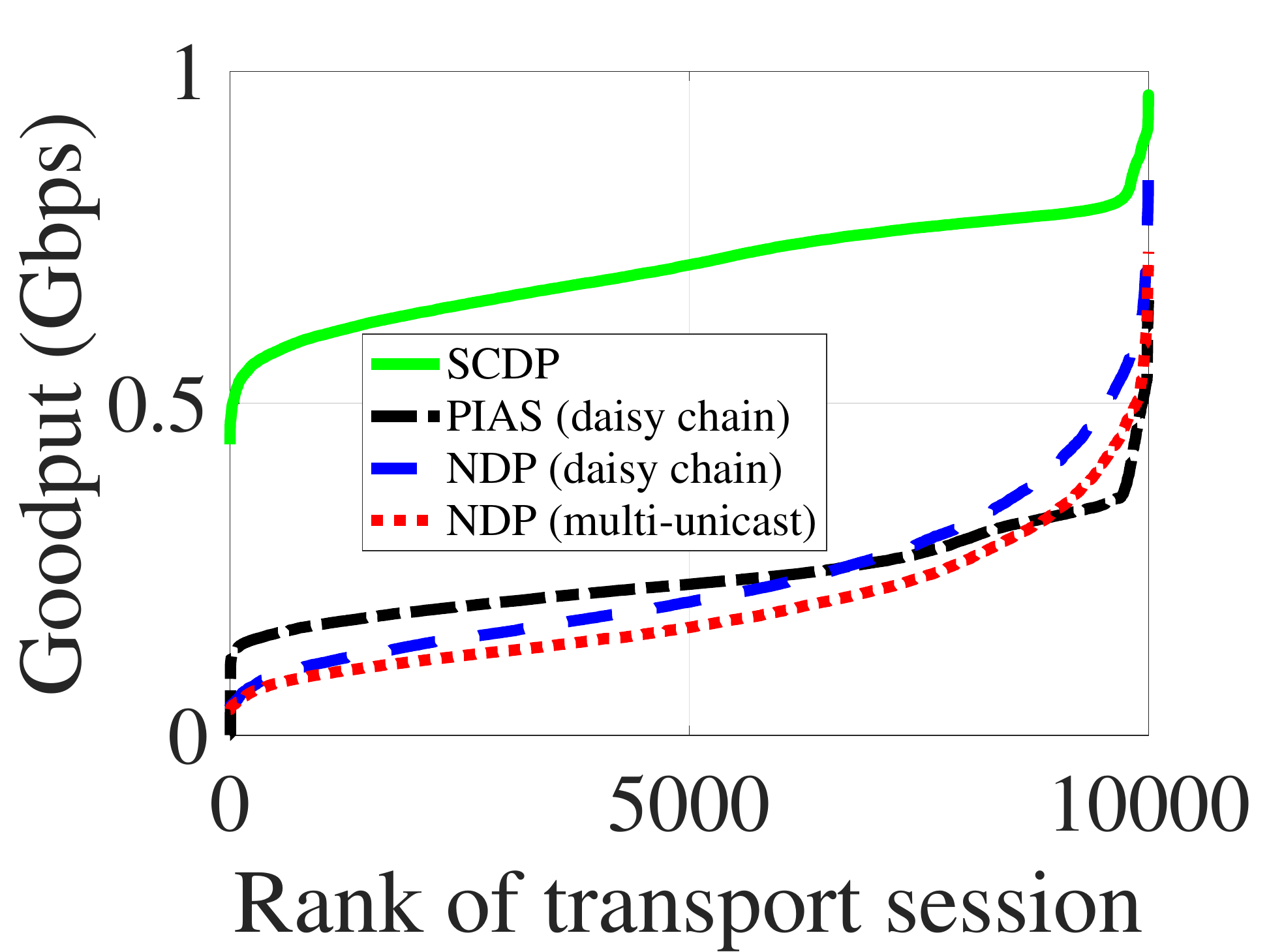}}\quad
 	\subcaptionbox{rs = 4MB, $\lambda$ = 4000}[.231\linewidth][c]{%
 		\includegraphics[scale=0.2]{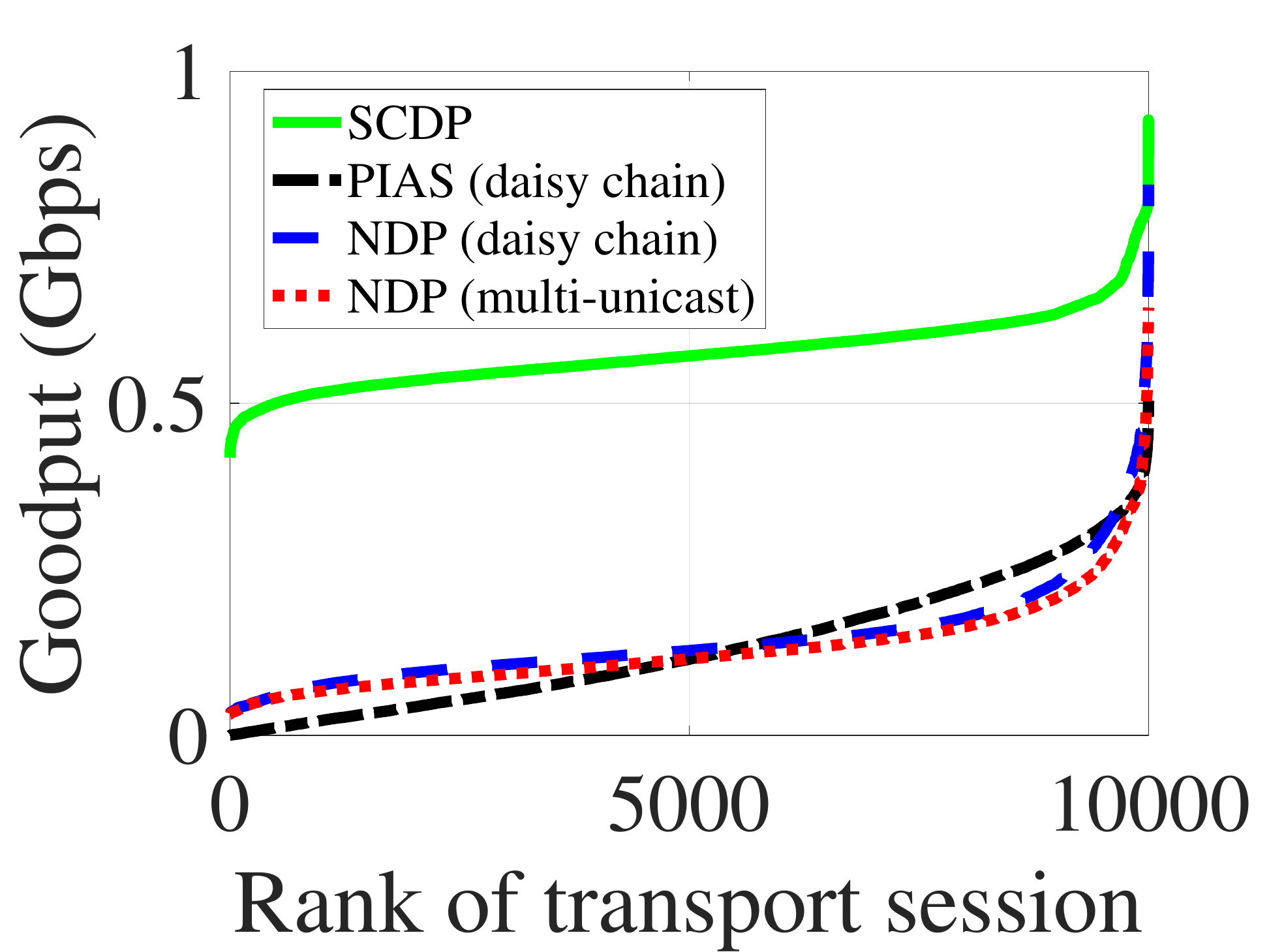}}\quad
 	\caption{Performance comparison for SCDP, NDP and PIAS - write I/O with $3$ replicas (one-to-many)}
 	\label{3replica-multicast}
 	\vspace{-5mm}
 \end{figure*}
 
 \begin{figure*}[t]
 	
 	\subcaptionbox{rs = 1MB, $\lambda$ = 2000}[.231\linewidth][c]{%
 		\includegraphics[scale=0.2]{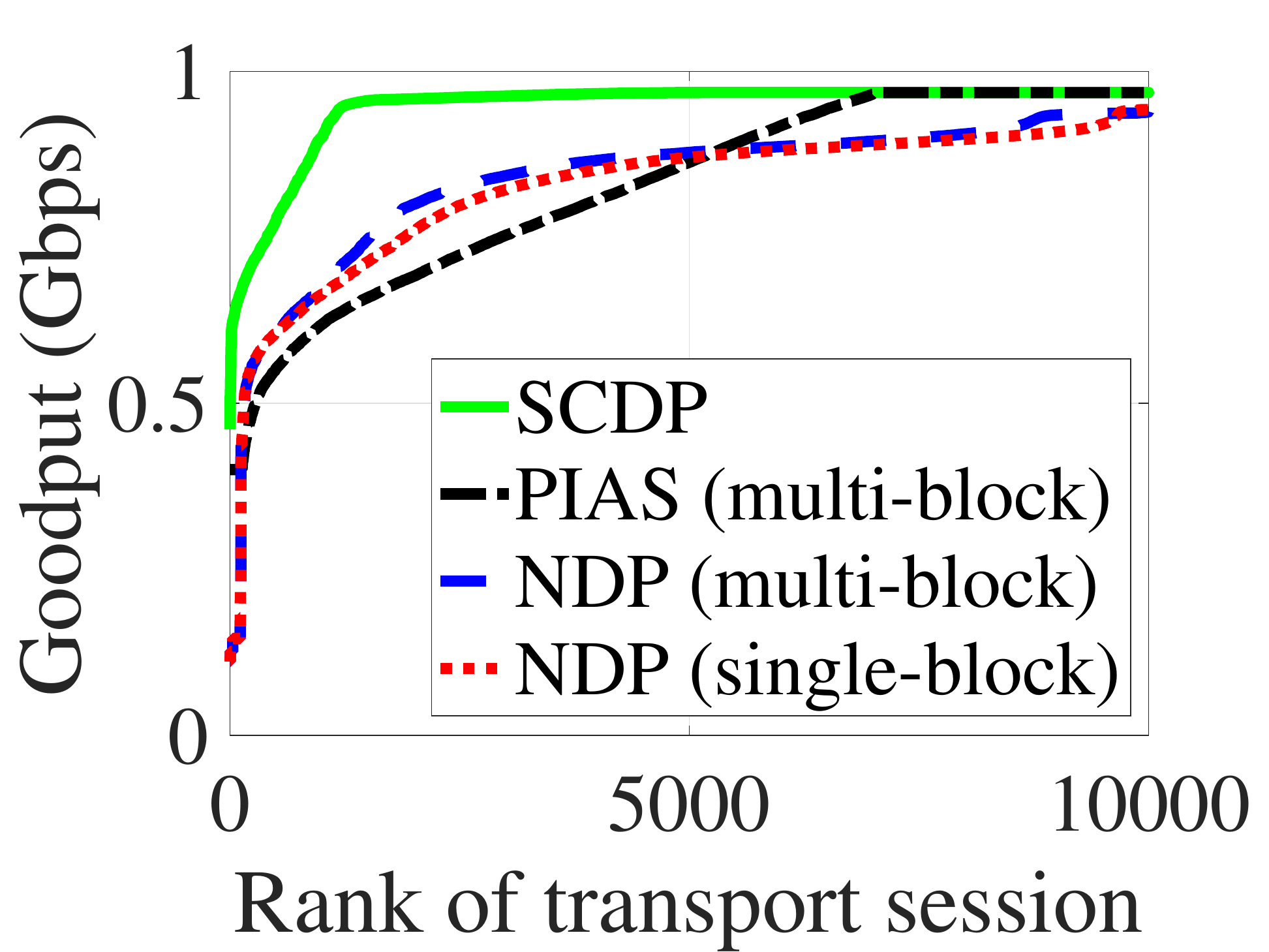}}\quad
 	\subcaptionbox{rs = 1MB, $\lambda$ = 4000}[.231\linewidth][c]{%
 		\includegraphics[scale=0.2]{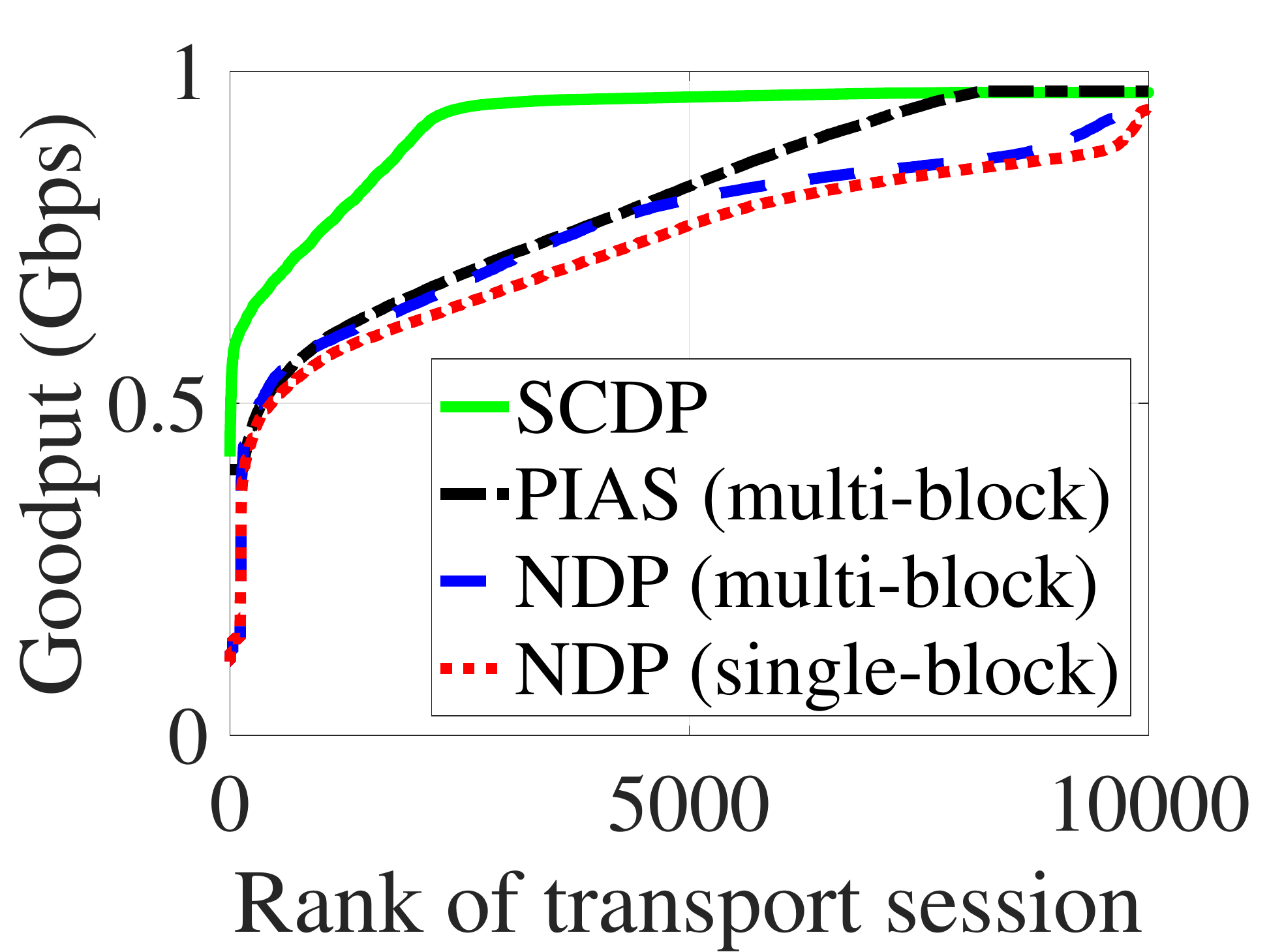}}\quad
 	\subcaptionbox{rs = 4MB, $\lambda$ = 2000}[.231\linewidth][c]{%
 		\includegraphics[scale=0.2]{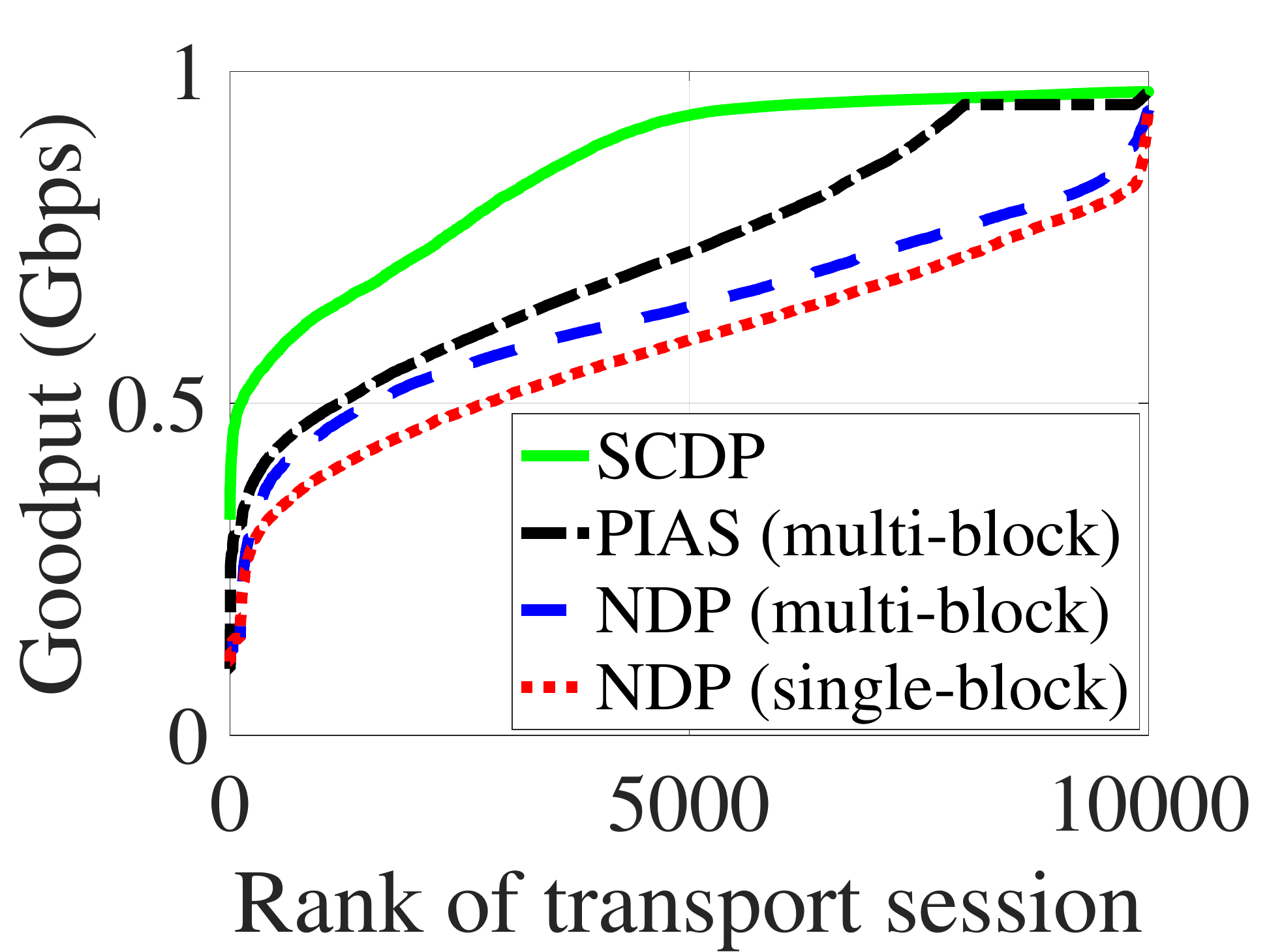}}\quad
 	\subcaptionbox{rs = 4MB, $\lambda$ = 4000}[.231\linewidth][c]{%
 		\includegraphics[scale=0.2]{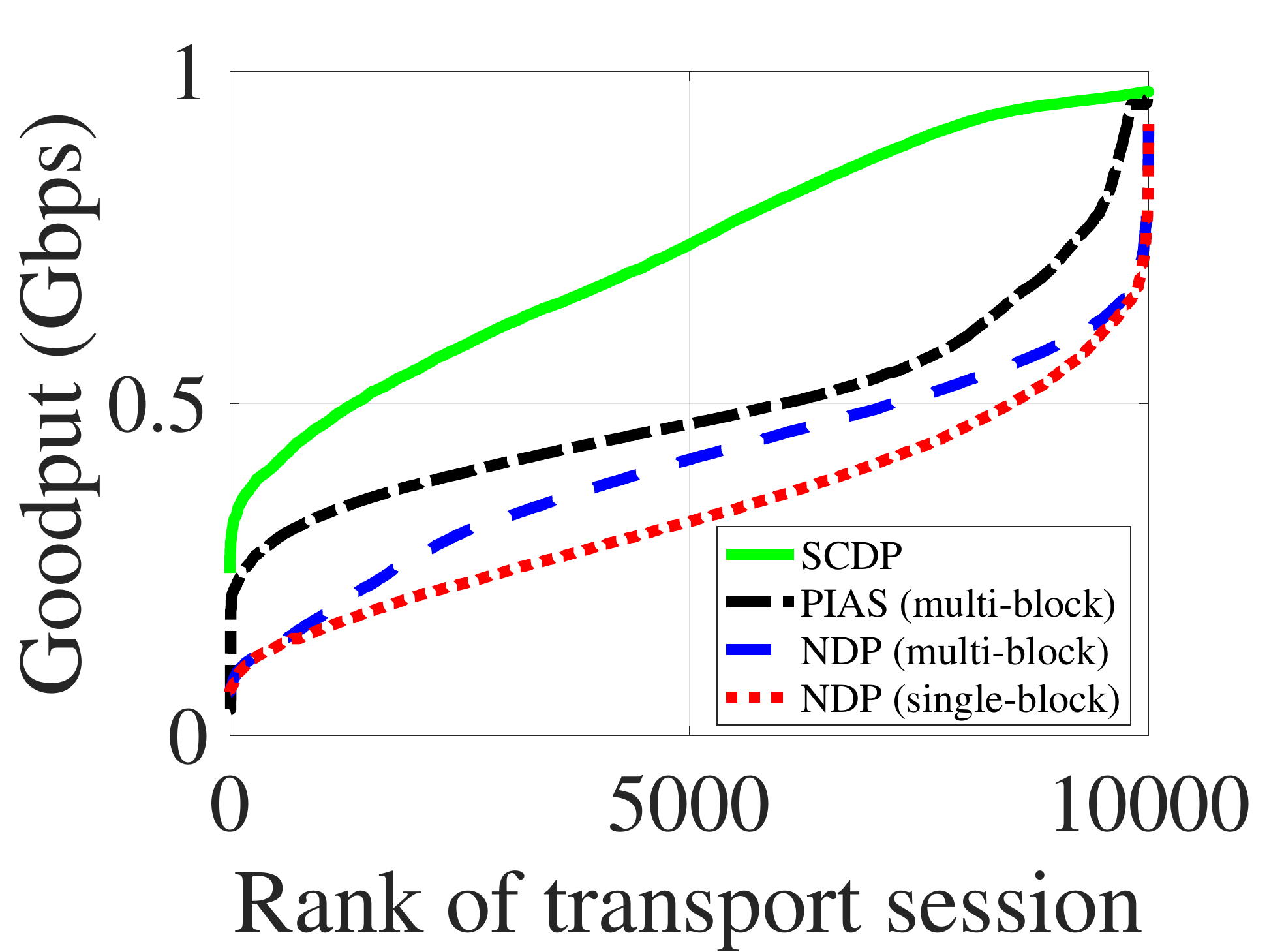}}\quad
 	\caption{Performance comparison for SCDP, NDP and PIAS - read I/O with $3$ replicas (many-to-one)}
 	\label{3replica-multisource} 
 	\vspace{-5mm}
 \end{figure*}
 
 \section{Experimental Evaluation}
 \label{sec:evaluation}
 
 We have extensively evaluated SCDP's performance through large scale, packet-level simulations and compared it to the state-of-the-art. To do so, we have developed OMNeT++ models for SCDP, NDP, PIAS, the respective switch service models, including MLFQ, and network-layer multicast \cite{multicastfattree}\footnote{Some of our models that we use in this paper have been published at the OMNeT++ Community Summit\cite{omnetpp-ndp-model}. More introductory details in~\cite{moThesis}.}.  
 
 \noindent\black{\textbf{Simulation setup.} For our experimentation we have used a $250$-server FatTree topology with $25$ core switches and $5$ aggregation switches in each pod ($50$ aggregation switches in total). This is a typical size for a simulated data centre topology, also used in the evaluation of recently proposed protocols \cite{HOMA,PIAS,pHost-CoNEXT-2015, pfabric}. The default values for the link capacity, link delay and switch buffer size are 1 Gbps, $10\mu$s and $20$ packets, respectively. We have run each simulation $5$ times with different seeds and report average (with $95\%$ confidence intervals) or aggregate values.}
 
 \noindent\black{\textbf{Multi-Level Feedback Queuing.} For protocols that rely on MLFQ, the switch buffer is allocated to $5$ packet queues with different scheduling priorities. The thresholds for demoting the priority for a specific session are statically assigned to 10KB, 100KB, 1MB and 10MB, respectively. In a real-world deployment these would be set dynamically, e.g. as in AuTO \cite{AUTO}.} \black{In the following, we briefly discuss details specific to the developed protocol models.}
 
 \noindent\black{\textbf{SCDP.} We have implemented SCDP in full, as described in Section \ref{design}. For the MLFQ mechanism, the top priority queue is for pull and header packets, which are very small. 
 	We model the decoding latency based on the results reported in~\cite{CodornicesRqNew}, by fitting the worst-case decoding latencies for different number of $K$ source symbols into a polynomial function.When calculating the completion time or goodput for a given SCDP session, we use the fitted model to extrapolate a decoding latency for the last block in the pipeline, and add it to the total time. We do not model the encoding latency as this can be easily masked by either (i) pre-computing repair symbols or (ii) encoding repair symbols while sending source symbols given that RaptorQ codes are systematic. The size of an encoding symbol (source and repair) is $1500$ bytes long (i.e. one MTU). Unless otherwise stated, the initial window $w$ for one-to-one and one-to-many sessions is set to $12$ symbol packets. For many-to-one sessions $w$ is set to $6$ symbol packets per sender. For all experiments we set the block size for pipelining to $100$ MTU-sized symbols.}
 
 \noindent\black{\textbf{NDP}~\cite{NDP} is a receiver-driven, unicast data transport protocol. A sender initiates a flow by sending an initial window of data at line rate, as in SCDP. The receiver then pulls packets from the sender by sending pull requests. If a switch queue overflows, the packet data is trimmed and the header is priority-forwarded. The receiver adds a pull packet for each received data or header packet, which are then paced from a single pull queue shared by all applications, based on the receiver's downlink link rate. In our NDP model the initial window value is set to 12 packets and all packets are $1500$ bytes long (i.e. one MTU). It has been shown (e.g. in \cite{Aeolus20}, \cite{AMRT20}, \cite{polo20} and \cite{opera20}) that NDP outperforms other modern data transport protocols (e.g. Homa~\cite{HOMA} and pHost~\cite{pHost-CoNEXT-2015}), therefore it constitutes a good baseline for our experimental evaluation.}
 
 \noindent\black{\textbf{NDP+} is a simple extension of NDP that uses MLFQ and is included here to understand how MLFQ affects the performance of SCDP in relation to NDP. Note that results for NDP+ are not included in the plots to maintain clarity, but are reported when appropriate. We use the same priority demoting thresholds for NDP+ as in SCDP. Packets are set to be MTU-sized.}
 
 \noindent\black{\textbf{PIAS.}~\cite{PIAS} is a flow scheduling mechanism that leverages MLFQ and employs DCTCP \cite{DCTCP} for end-to-end data transport, which relies on Explicit Congestion Notification (ECN) in the network to provide multi-bit feedback to end hosts. For uniformity, we use the same priority demoting thresholds for PIAS as in SCDP and NDP+. Packets are set to be MTU-sized.}
 
 \vspace{-3mm}
 \subsection{Goodput for One-to-Many and Many-To-One Sessions}
 \label{goodput-performance}
 
 In this section we measure the application goodput for SCDP, NDP, NDP+ and PIAS in a distributed storage setup with $3$ replicas (as depicted in Figure \ref{modern-workloads}). The setup involves many-to-one and one-to-many communication. In each run, we simulate $2000$ transport sessions (or I/O requests at the storage layer) with sizes 1MB and 4MB each (denoted as \emph{rs} in the figures). Transport session arrival times follow a Poisson process \black{with inter-arrival rate $\lambda$}; we have used different $\lambda$ values ($2000$ and $4000$) to assess the performance of the studied protocols under different loads. Each I/O request is `assigned' to a host in the network (denoted as $C_i$ in Figure \ref{modern-workloads}), which is selected uniformly at random and acts as the client. Replica  selection and placement is based on HDFS' default policy. More specifically, we assume that clients are not data nodes themselves, therefore a data block is placed on a randomly selected node (denoted as $R_i$ in Figure \ref{modern-workloads}). One replica is stored on a node in a different remote rack, and the last replica is stored on a different node in the same remote rack. A client will read a block from a server located in the same rack, or a randomly selected one, if no replica is stored in the same rack. In order to simulate congestion in the core of the network, $30\%$ of the nodes run background long flows, the scheduling of which is based on a permutation traffic matrix.

 \noindent\textbf{One-to-many transport sessions. } We evaluate SCDP's performance in one-to-many traffic workloads and assess how it benefits from the underlying support for network-layer multicast, compared to NDP, NDP+ and PIAS. One-to-many communication with these protocols is implemented through (1) multi-unicasting data to multiple recipients (Figure \ref{modern-workloads}a) or (2) daisy-chaining the transmission of replicas through the respective servers (Figure \ref{modern-workloads}b). In daisy-chaining, each replica starts transmitting the data to the next replica server (according to HDFS's placement policy), as soon as it starts receiving data from another replica server. Daisy-chaining eliminates the bottleneck at the client's uplink. We measure the overall goodput from the time the client initiates the transmission until the last server receives the whole data. The results for various loads and I/O request sizes are shown in Figure \ref{3replica-multicast}. In all figures, flows are ranked according to the measured goodput (shown on the y axis). SCDP, with its natural load balancing and the support of multicast (Figure \ref{modern-workloads}c), significantly outperforms NDP and PIAS even when daisy-chaining is used for replicating data. Daisy-chaining is effective compared to multi-unicasting when the network is not heavily loaded. With SCDP, around $50\%$ of the sessions experience goodput that is over $90\%$ of the available bandwidth for 1MB sessions and $\lambda = 2000$. The remaining $50\%$ sessions still get a goodput performance over $60\%$ of the available bandwidth. When the network load is heavier, daisy-chaining does not provide any significant benefits over multi-unicasting because data needs to be moved in the data centre multiple times and congestion gets severe. For $\lambda = 4000$ and 4MB sessions, NDP's and PIAS' performance is significantly worse for most sessions, whereas SCDP still offers an acceptable transport service to all sessions. SCDP fully exploits the support for network-layer multicasting providing superior performance to all storage clients because the required network bandwidth is minimised. Minimising the bandwidth requirements for one-to-many flows that are extremely common in the data centre, makes space for regular short and long flows.  For the experimental setup with the heaviest network load ($\lambda = 4000$ and 4MB sessions), we have measured the average goodput for SCDP background traffic to be 0.408 Gbps, compared to 0.252 Gbps and 0.182 Gbps for NDP and PIAS experiments, respectively\footnote{This improvement for background flows is despite these running at the lowest possible priority, and they span the whole duration of the simulation.}. This is 15.6\% of the available bandwidth freed up for regular unicast flows. We evaluate the positive effect that SCDP has with respect to network hotspots in Section \ref{minimising-hotspots}. 
 \textcolor{black}{NDP+ is on average $14\%$ better than NDP and $21\%$ worse than SCDP, in terms of measured goodput.  This reinforces our argument that the performance gains in one-to-many communication is mostly due to exploiting the supported network-level multicast. PIAS performs worse than SCDP, NDP and NDP+  because it relies on DCTCP for data transport and as a result it suffers from the limitations of a single-path protocol (i.e. lack of support for multi-path transport and packet spraying).}

 \noindent\textbf{Many-to-one transport sessions. }In the many-to-one scenario, clients read previously stored data from the network. SCDP naturally balances this load according to servers' capacity and network congestion, as discussed in Section \ref{multi-source} (see Figure \ref{modern-workloads}e). With NDP and PIAS, clients read data either from a replica server located in the same rack or a randomly selected server, if there is no replica stored in the same rack and we simulate both a single-block (see Figure \ref{modern-workloads}d) and multi-block request workload. The latter enables parallelisation at the application layer (e.g. the read-ahead optimisation where a client reads multiple consecutive blocks under the assumption that they will soon be requested). Here, we simulate a $3$-block read-ahead policy and measure the overall goodput from the time the I/O request is issued until all $3$ blocks are fetched. To make the results as comparable to each other as possible, for the $3$-block setup we use blocks the size of which is one third of the size of the single-block scenario (as reported in Figure \ref{3replica-multisource}). We do not include multi-block results for SCDP as they are almost identical to the single-block case, confirming the argument that it naturally distributes the load without any application-layer parallelisation. In Figure \ref{3replica-multisource} we observe that SCDP significantly outperforms NDP and PIAS for all different request sizes and $\lambda$ values. Even under heavy load, SCDP provides acceptable performance to all transport sessions. This is the result of (1) the natural and dynamic load balancing provided to SCDP's many-to-one sessions and (2) MLFQ; long background flows run at the lowest priority to boost the performance of shorter flows. Around $82\%$ of the sessions experience goodput that is above $90\%$ of the available bandwidth for 1MB sessions and $\lambda = 2000$. In contrast, NDP and PIAS offer this good performance to only $10\%$ and $23\%$ of the sessions, respectively. For $\lambda = 4000$ and 4MB sessions, NDP's and PIAS' performance is significantly worse for most sessions, whereas SCDP still offers good performance to all sessions. Notably, the performance difference between SCDP and both NDP and PIAS increases with the congestion in the network, with SCDP being able to provide acceptable levels of performance where NDP and PIAS would not (e.g. in the presence of hotspots or in over-subscribed networks). \black{NDP+ outperforms both NDP and PIAS, providing on average $7\%$ improvement in goodput over the goodput that can be provided by NDP and PIAS. This shows that only a small part of SCDP's performance gains over NDP come from MLFQ. The key differentiator is the natural load balancing that is enabled by RaptorQ codes; a congested server will not slow down the session because the rest of the senders will contribute most of the needed source and repair symbols. It is worth noting that PIAS shows better performance for some sessions compared to NDP (for $\lambda = 2000$ and 1MB sessions). The benefit becomes clearer for larger flows, at the highest inter-arrival rate. However, this does not come for free; instead, background traffic paid for this. For the experimental setup with $\lambda = 4000$ and 4MB sessions, we have measured the average goodput for NDP background traffic (not shown in the figures) to be 0.342 Gbps, compared to 0.152 Gbps for the respective PIAS experiment.}

 \begin{figure*}[t]
 	
 	\centering
 	\subcaptionbox{ (0, 100KB]}[.231\linewidth][c]{%
 		\includegraphics[scale=0.2]{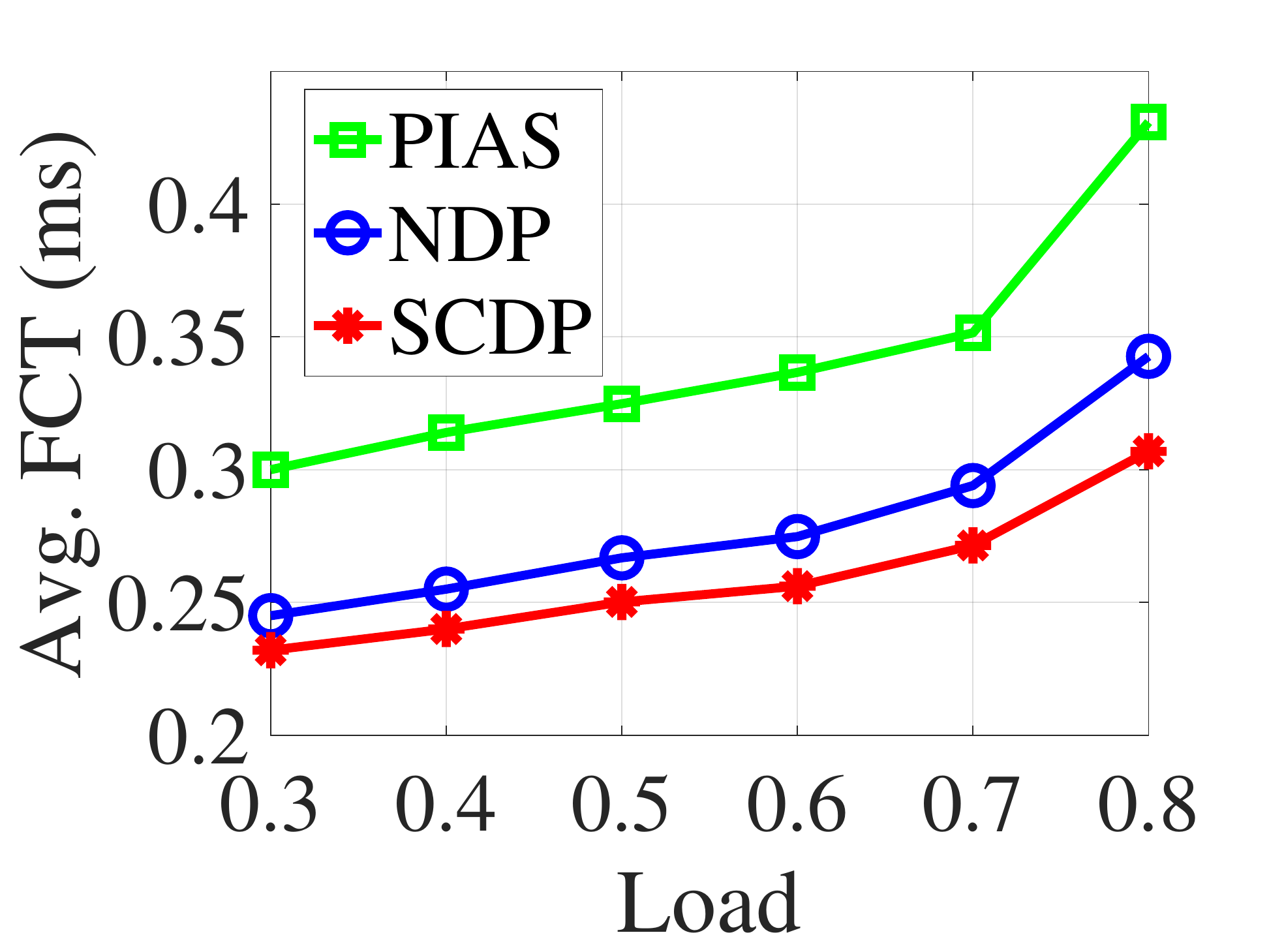}}\quad
 	\subcaptionbox{ (0, 100KB] 99th percentile}[.231\linewidth][c]{%
 		\includegraphics[scale=0.2]{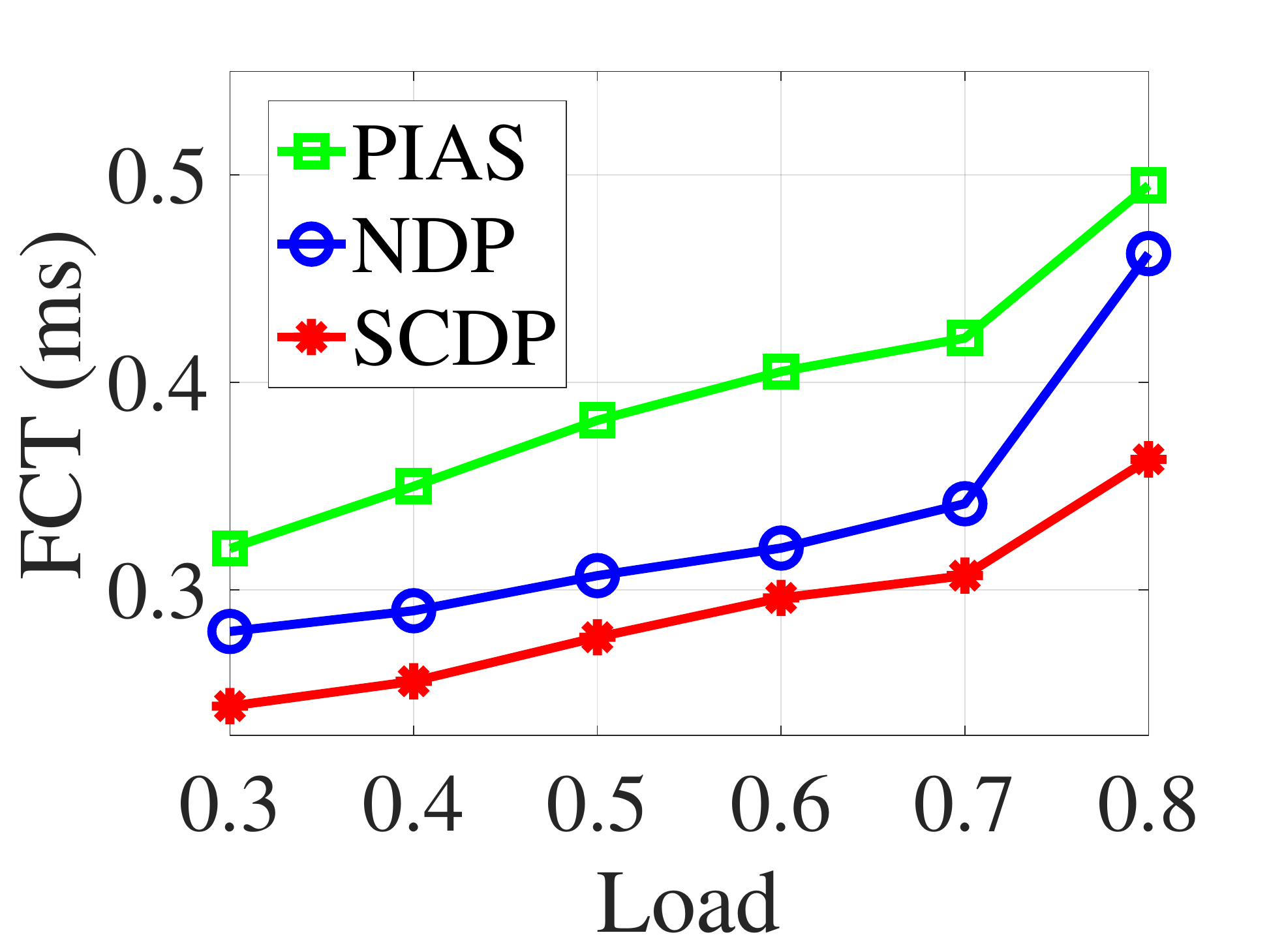}}\quad
 	\subcaptionbox{ (100KB, 1MB]}[.231\linewidth][c]{%
 		\includegraphics[scale=0.2]{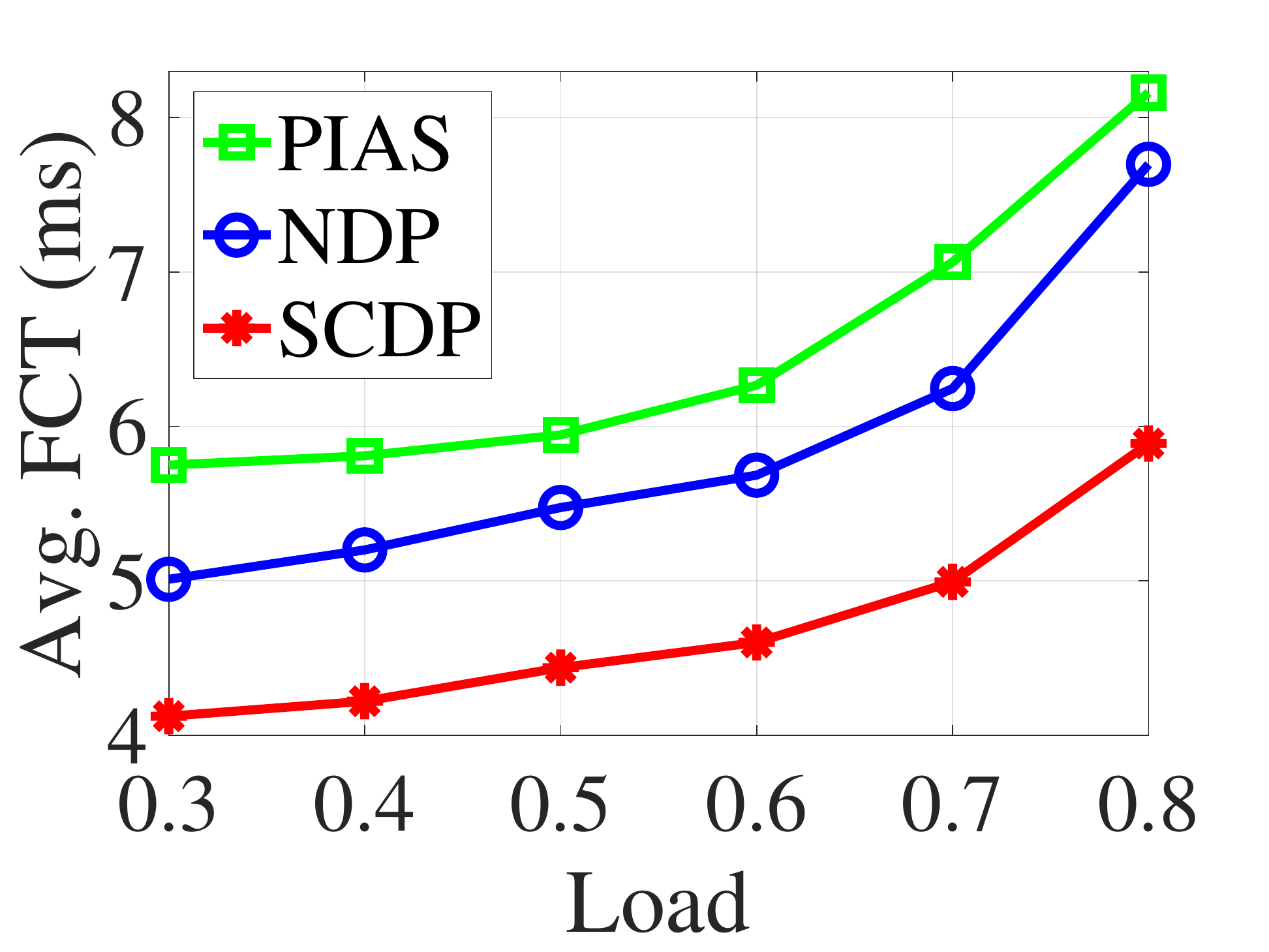}}\quad
 	\subcaptionbox{ (1MB, 10MB]}[.231\linewidth][c]{%
 		\includegraphics[scale=0.2]{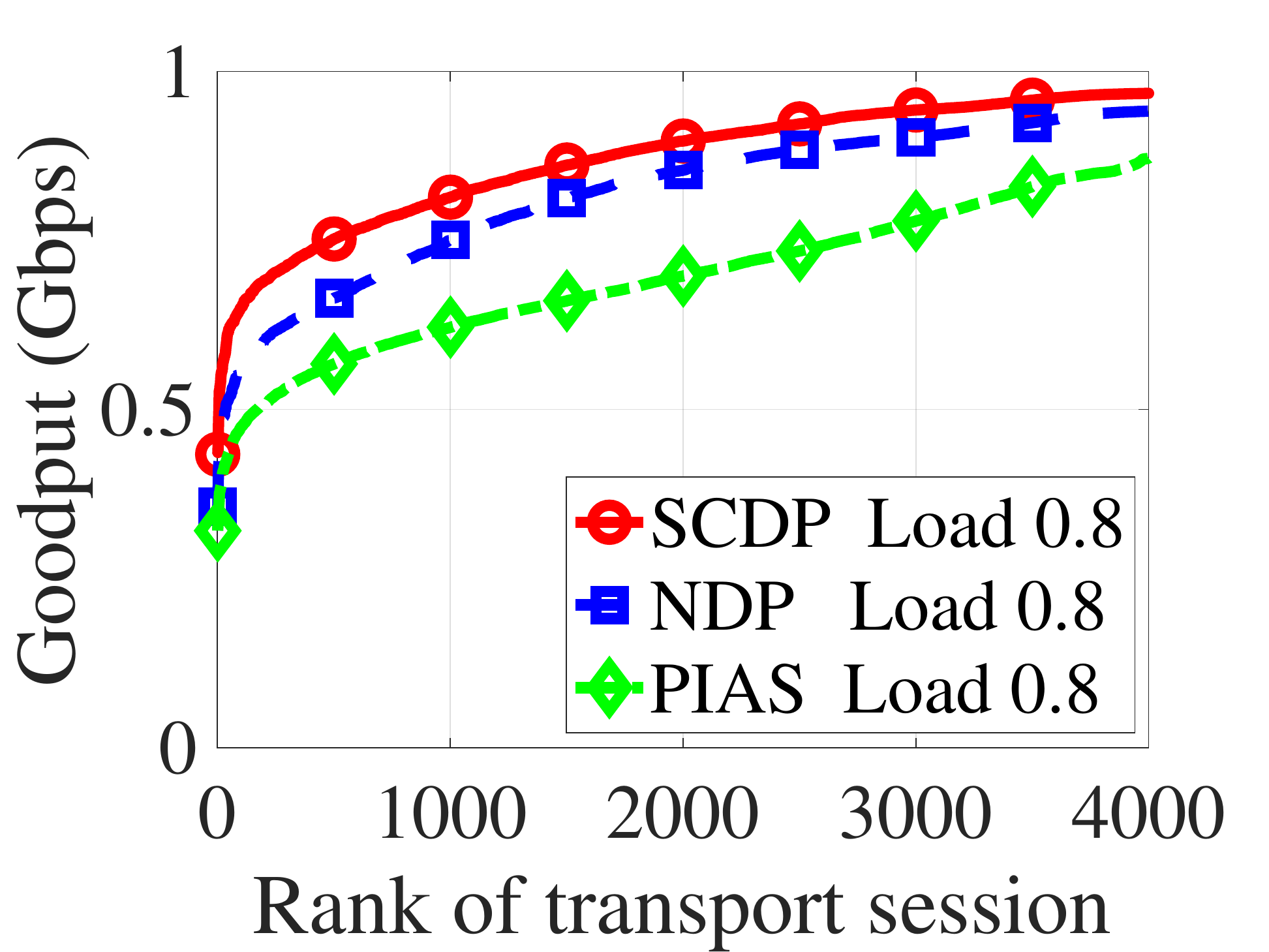}}\quad
 	\caption{Web search workload with unicast flows as background traffic}
 	\label{fct-ws-workloads-type-1} 	
 	\vspace{-3mm}
 \end{figure*}
 
 \begin{figure*}[t]
 	\setlength{\belowcaptionskip}{-3pt}
 	\subcaptionbox{ (0, 100KB]}[.231\linewidth][c]{%
 		\includegraphics[scale=0.2]{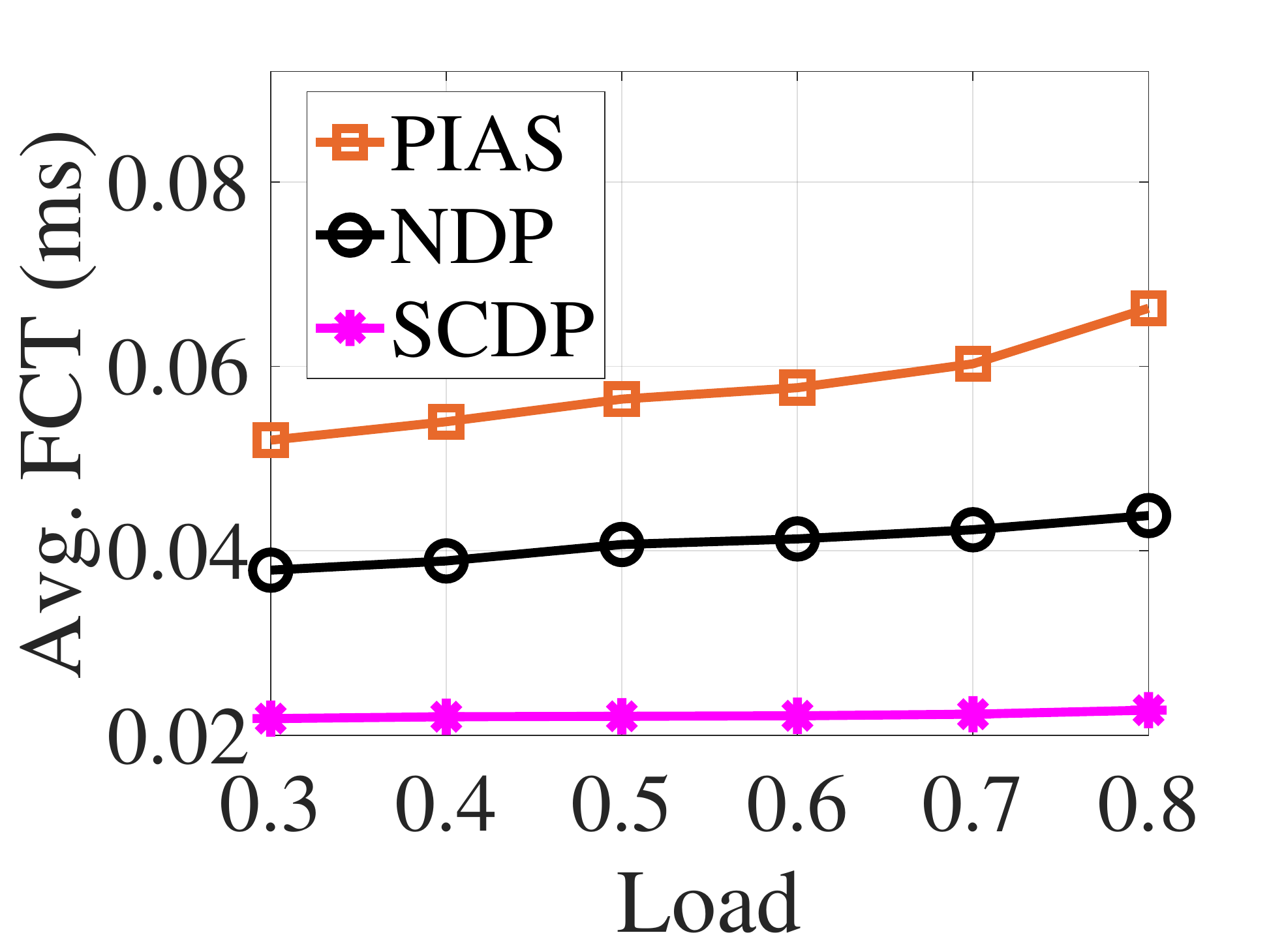}}\quad
 	\subcaptionbox{(0, 100KB] 99th percentile}[.231\linewidth][c]{%
 		\includegraphics[scale=0.2]{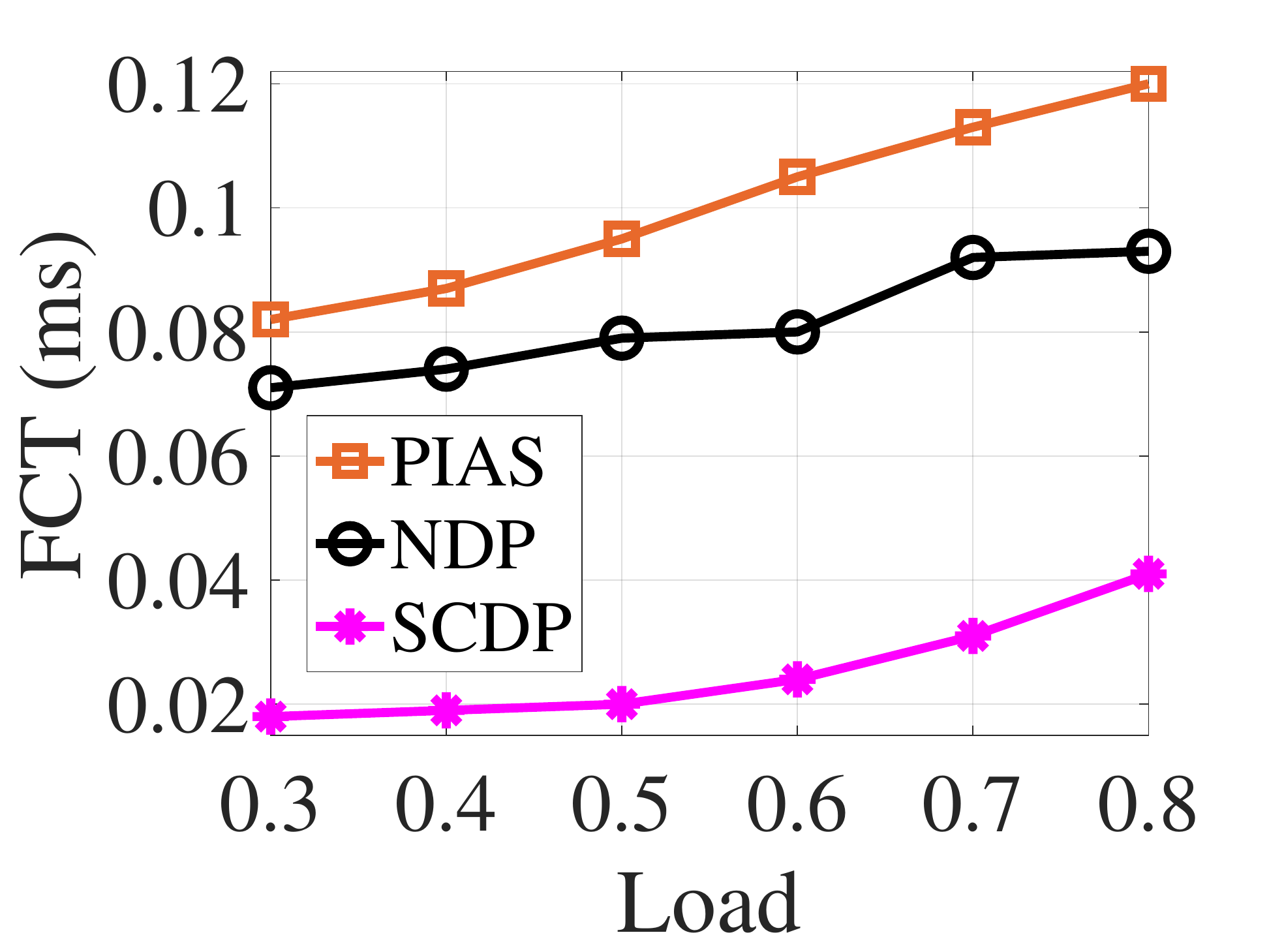}}\quad
 	\subcaptionbox{ (100KB, 1MB]}[.231\linewidth][c]{%
 		\includegraphics[scale=0.2]{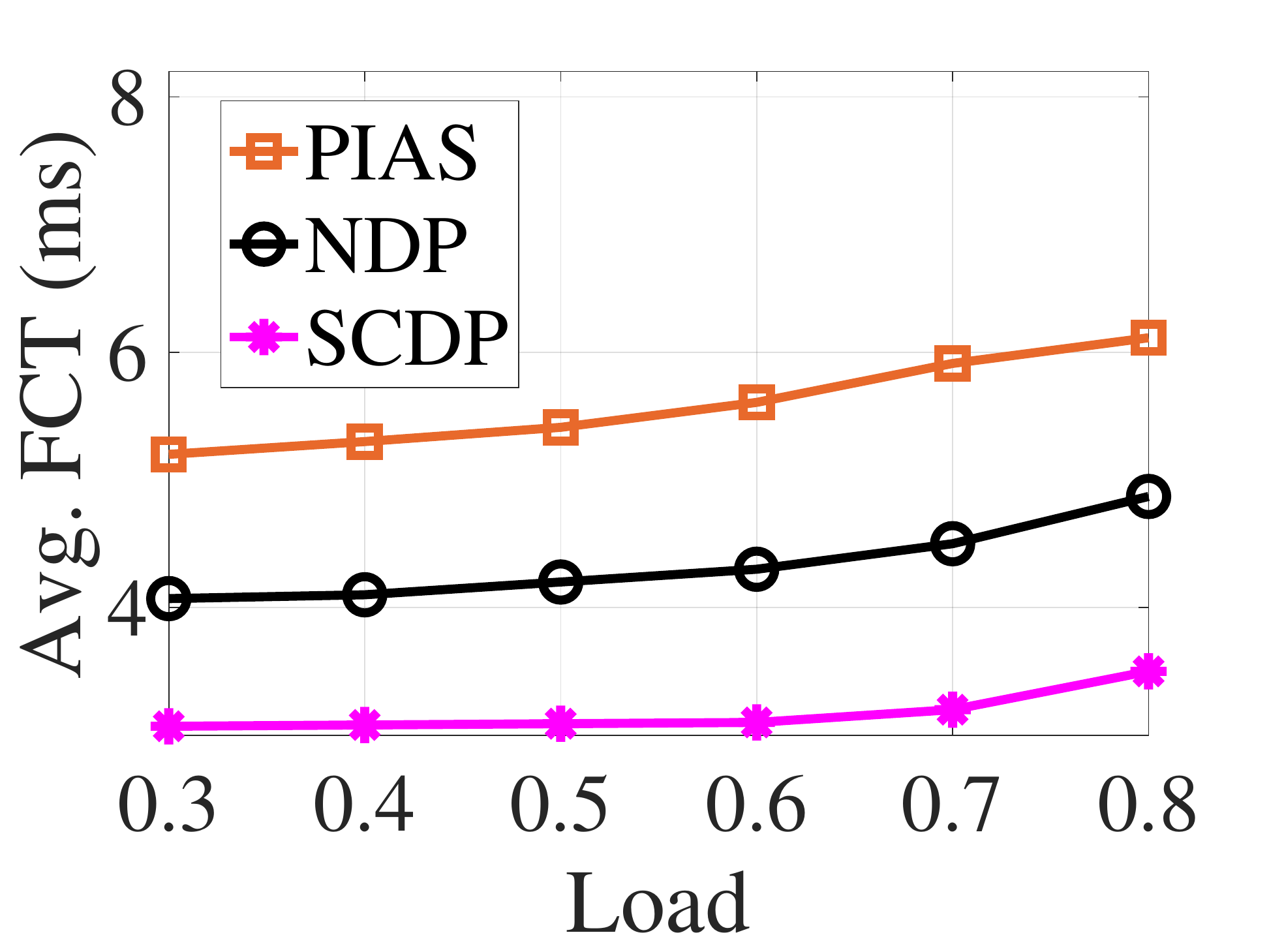}}\quad
 	\subcaptionbox{ (1MB, 10MB]}[.231\linewidth][c]{%
 		\includegraphics[scale=0.2]{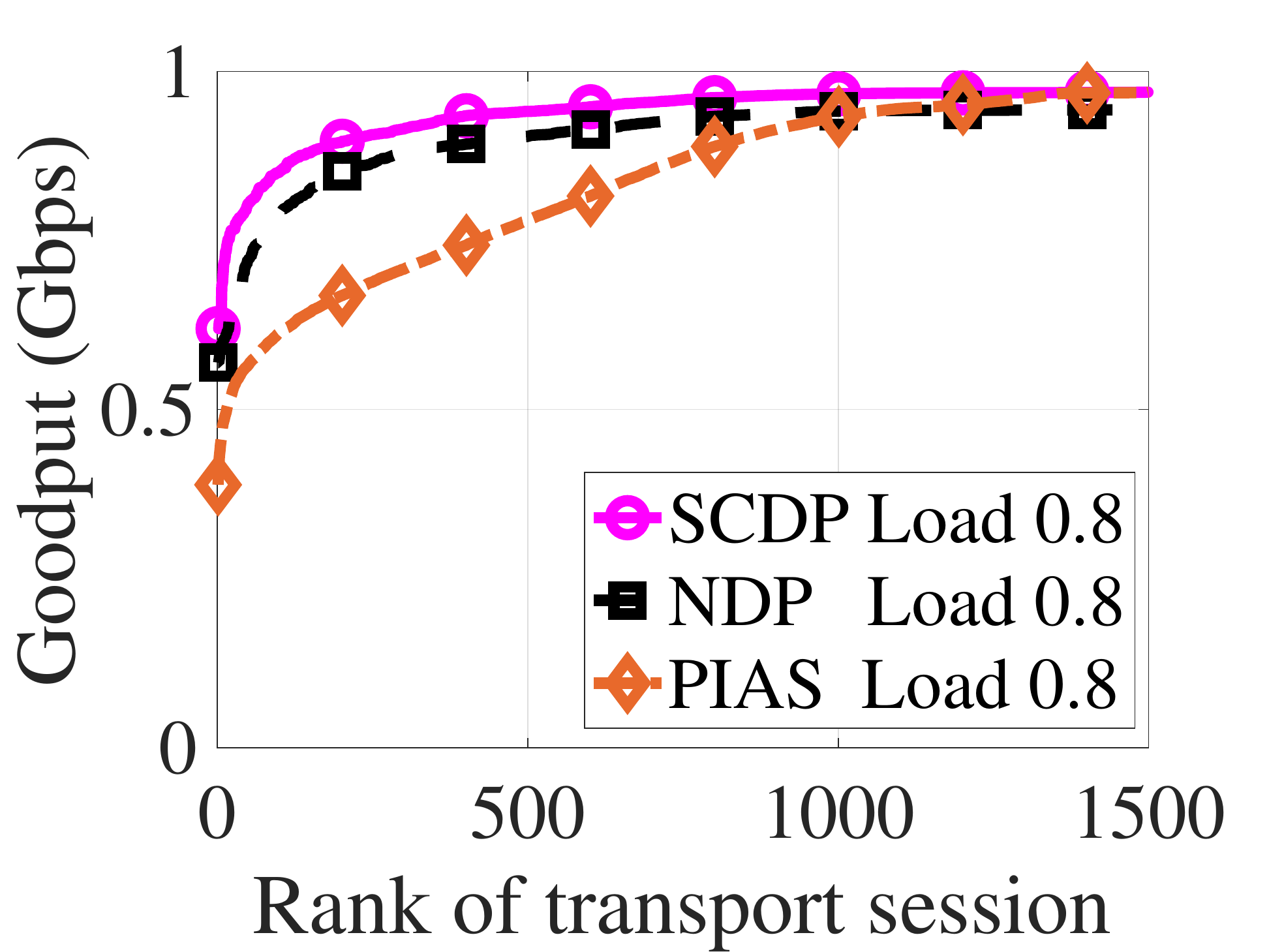}}\quad
 	\caption{Data mining workload with unicast flows as background traffic}
 	\label{fct-dm-workloads-type-1} 	
 	\vspace{-3mm}
 \end{figure*}

 \begin{figure}[htp]
 	\setlength{\belowcaptionskip}{-3pt}
 	\centering
 	\subcaptionbox{(0, 100KB]}[.48\linewidth][c]{%
 		\includegraphics[width=0.9\linewidth,height=0.12\textheight]{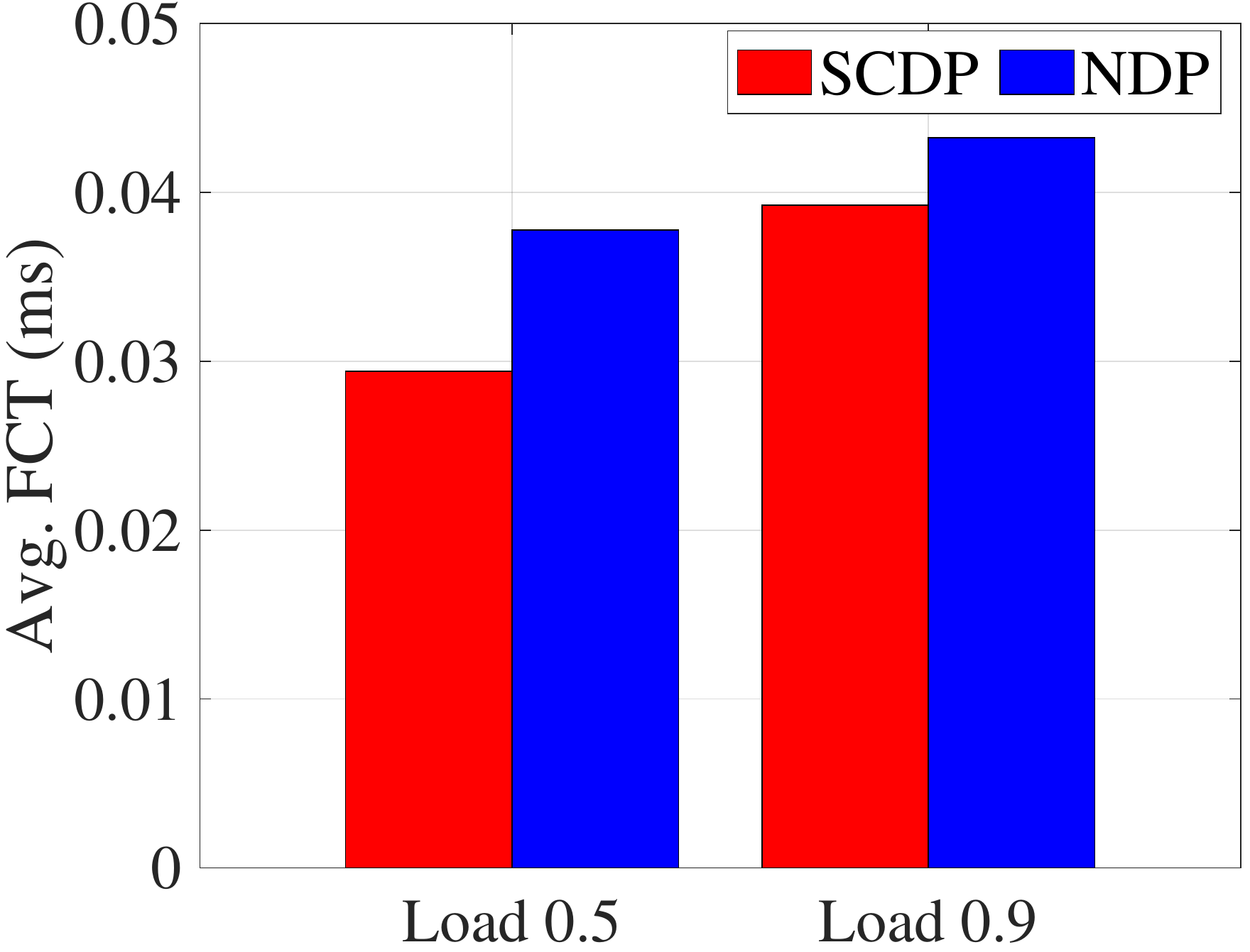}}\quad
 	\subcaptionbox{(100KB, 1MB]}[.48\linewidth][c]{%
 		\includegraphics[width=0.9\linewidth,height=0.12\textheight]{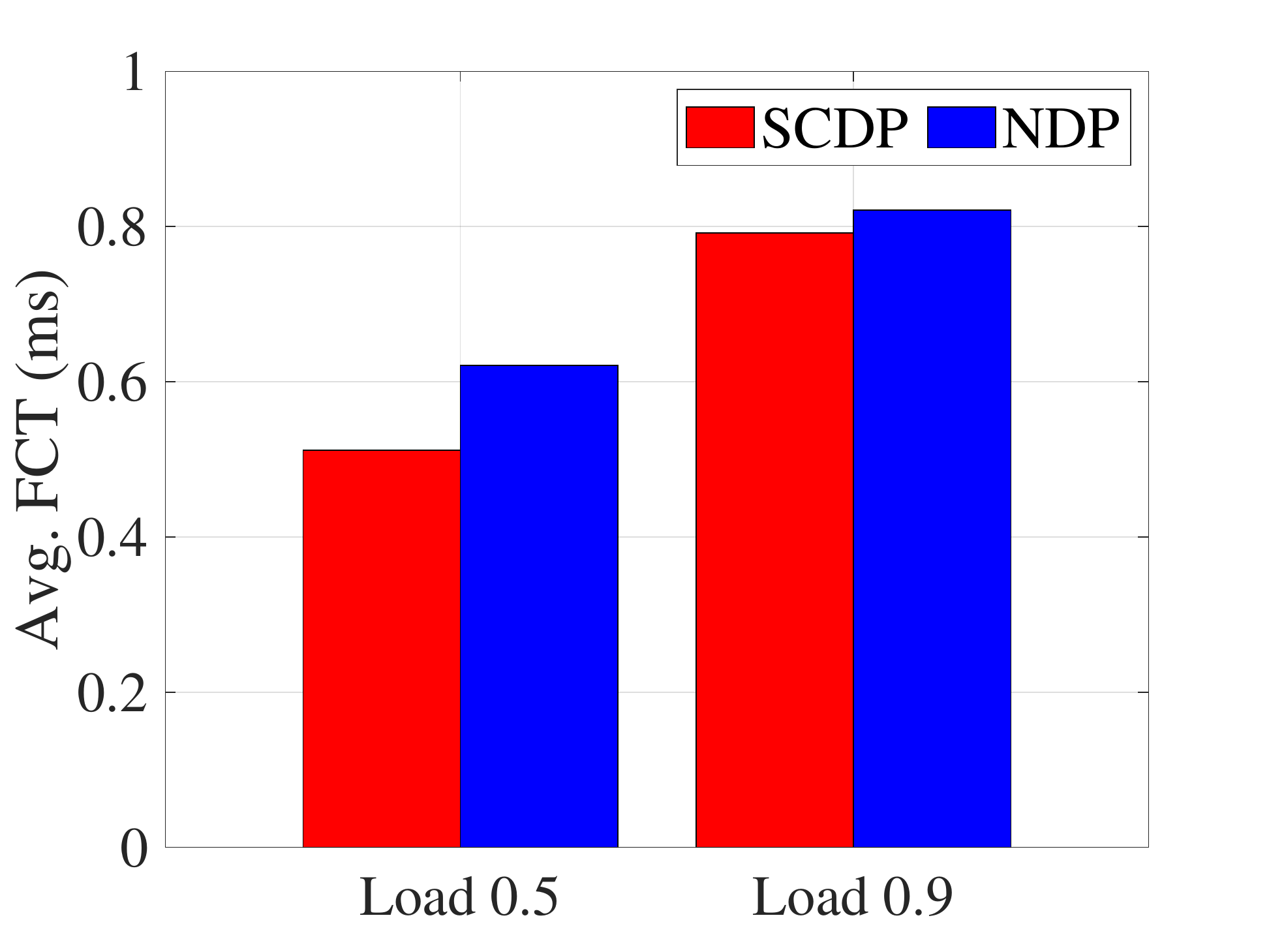}}\quad
 	\caption{Web search workload with 10Gbps links}
 	\label{highspeedlinks} 
 	\vspace{-5.5mm}
 \end{figure}
 
 \vspace{-4mm}
 \subsection{Performance Benchmarking with Realistic Workloads}
 \label{realistic-workloads}
 SCDP is designed to be a general-purpose transport protocol for data centres therefore it is crucial that it provides high performance for all supported transport modes and traffic workloads. In this section, we use realistic workloads reported by data centre operators to evaluate SCDP's applicability and effectiveness beyond one-to-many and many-to-one sessions. Here, we consider two typical services; \textit{web search} and \textit{data mining} \cite{VL2,DCTCP}. The respective flow size distributions are shown in Table \ref{table-fct-workloads}. They are both heavy-tailed; i.e. a small fraction of long flows contribute most of the traffic. We have chosen the workloads to cover a wide range of average flow sizes ranging from $64$KB to $7.4$MB. \black{We simulate six target loads of background traffic ($0.3$, $0.4$, $0.5$, $0.6$, $0.7$ and $0.8$).} We generate $20000$ transport sessions, the inter-arrival time of which follows a Poisson process with $\lambda = 2500$. In Figures \ref{fct-ws-workloads-type-1}a and \ref{fct-ws-workloads-type-1}c and \ref{fct-dm-workloads-type-1}a and \ref{fct-dm-workloads-type-1}c, we report the average flow completion time (FCT) of flows with sizes in ($0-1$MB). For the shortest flows ($0-100$KB) we also report the 99th percentile of the measured FCTs (Figures \ref{fct-ws-workloads-type-1}b and \ref{fct-dm-workloads-type-1}b). Finally, Figures \ref{fct-ws-workloads-type-1}d and \ref{fct-dm-workloads-type-1}d illustrate the measured goodput for flows with sizes in (1MB, 10MB] ($4000$ and $1500$ flows in web search and data mining workloads, respectively) (for load value of 0.8).

 \begin{table}[ht!]
 	\scriptsize
 	\centering
 	\begin{tabular}{c|c|c|c|c|c|}
 		\cline{2-6}
 		& \begin{tabular}[c]{@{}c@{}}0 - \\ 10KB\end{tabular} & \begin{tabular}[c]{@{}c@{}}10KB - \\ 100KB\end{tabular} & \begin{tabular}[c]{@{}c@{}}100KB -\\  1MB\end{tabular} & \begin{tabular}[c]{@{}c@{}}1M-\\ ..\end{tabular} & \begin{tabular}[c]{@{}c@{}}Average  \\ flow size\end{tabular} \\ \hline
 		\multicolumn{1}{|c|}{\begin{tabular}[c]{@{}c@{}}Web \\ Search \cite{DCTCP}\end{tabular}}     & 19\%                                                & 43\%                                                     & 18\%                                                   & 20\%                                             & 1.6MB                                                         \\ \hline
 		\multicolumn{1}{|c|}{\begin{tabular}[c]{@{}c@{}}Data \\ Mining \cite{VL2}\end{tabular}} & 78\%                                                & 5\%                                                     & 8\%                                                    & 9\%                                              & 7.4MB                                                        \\ \hline
 	\end{tabular}
 	\caption{Flow size distribution of realistic workloads}
 	\label{table-fct-workloads}
 	
 \end{table}

 \begin{figure*}[t]
 	\setlength{\belowcaptionskip}{-3pt}
 	\centering
 	\subcaptionbox{ (0, 100KB]}[.231\linewidth][c]{%
 		\includegraphics[scale=0.2]{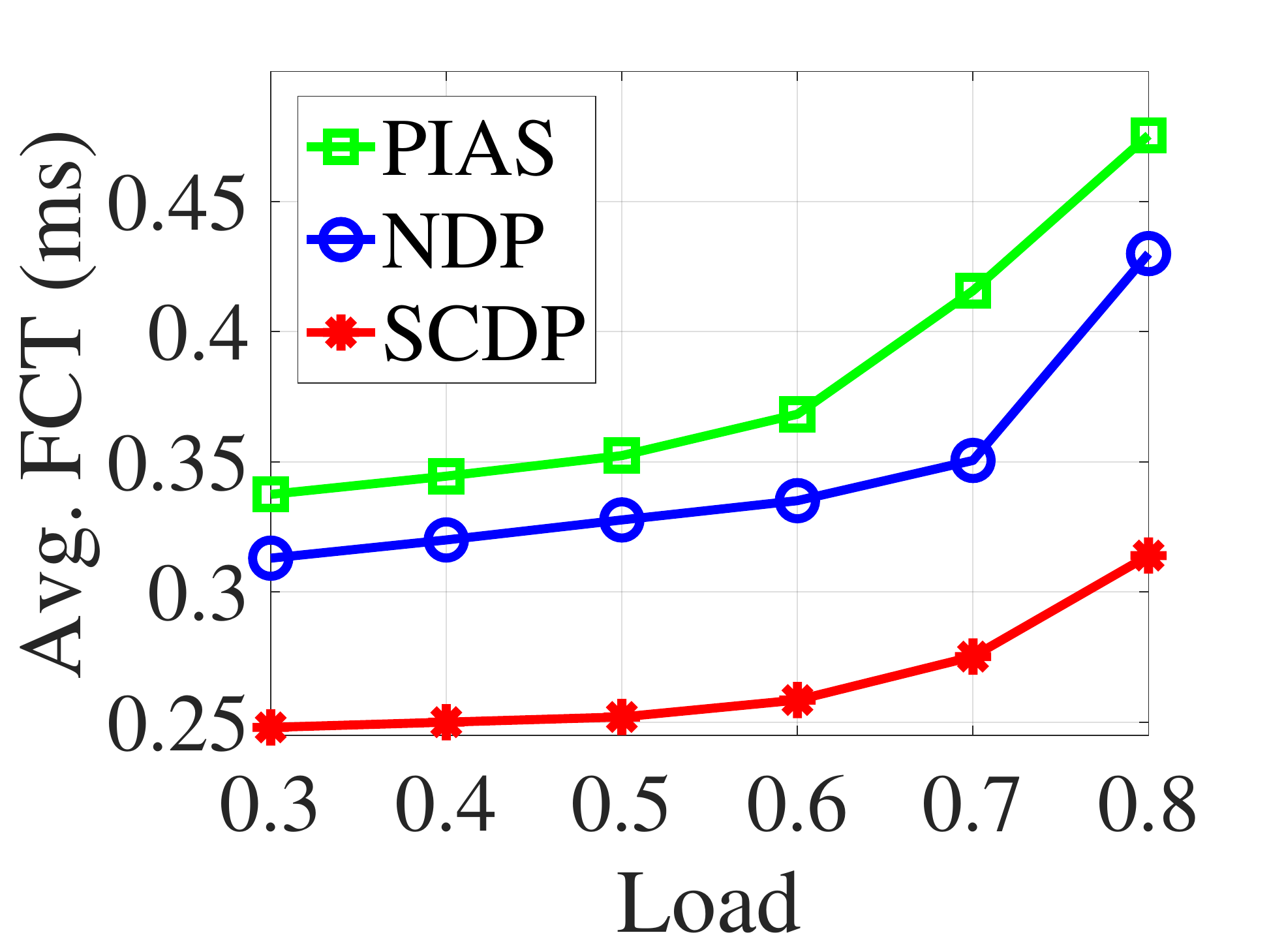}}\quad
 	\subcaptionbox{ (0, 100KB] 99th percentile}[.231\linewidth][c]{%
 		\includegraphics[scale=0.2]{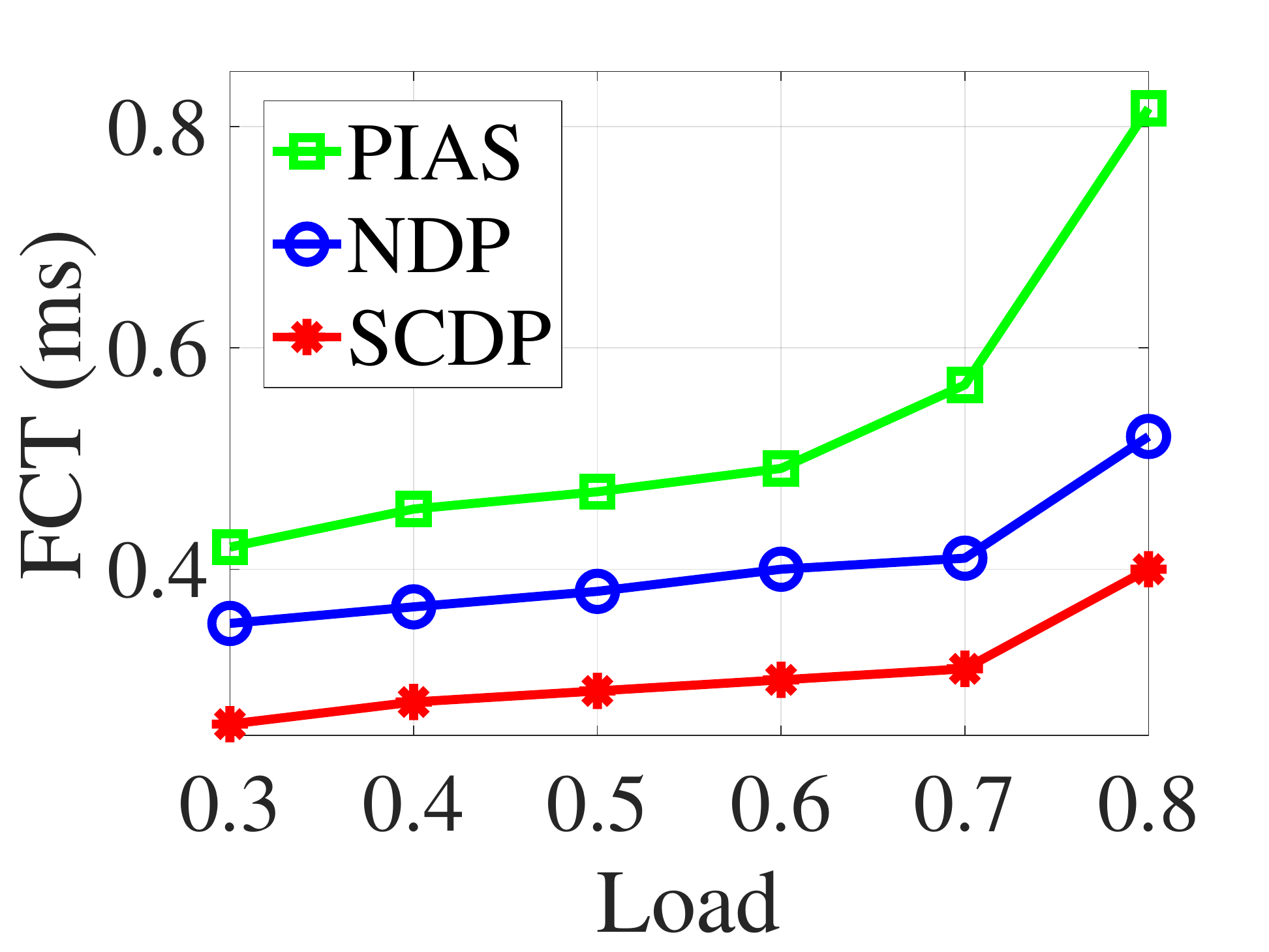}}\quad
 	\subcaptionbox{ (100KB, 1MB]}[.231\linewidth][c]{%
 		\includegraphics[scale=0.2]{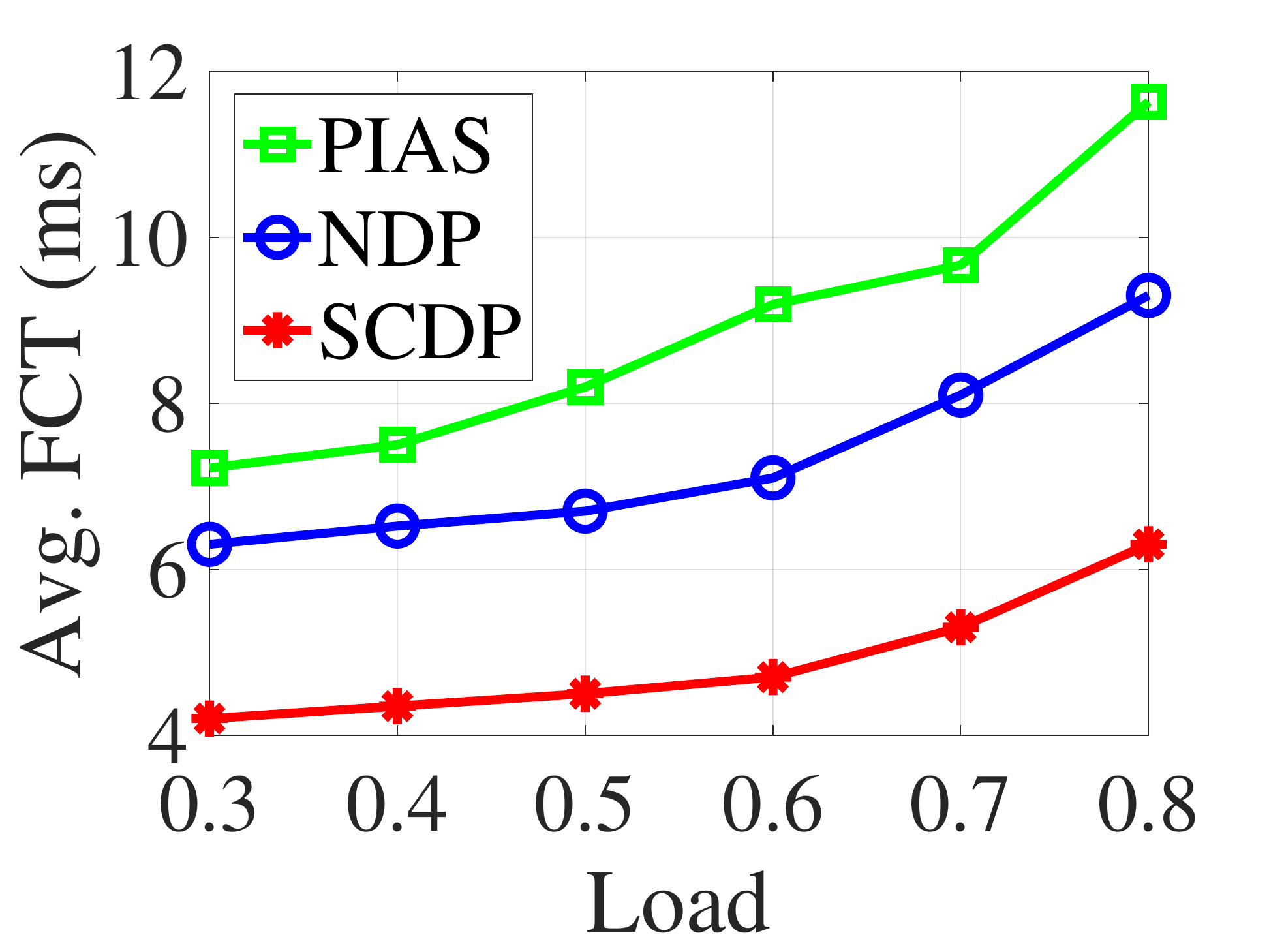}}\quad
 	\subcaptionbox{ (1MB, 10MB]}[.231\linewidth][c]{%
 		\includegraphics[scale=0.2]{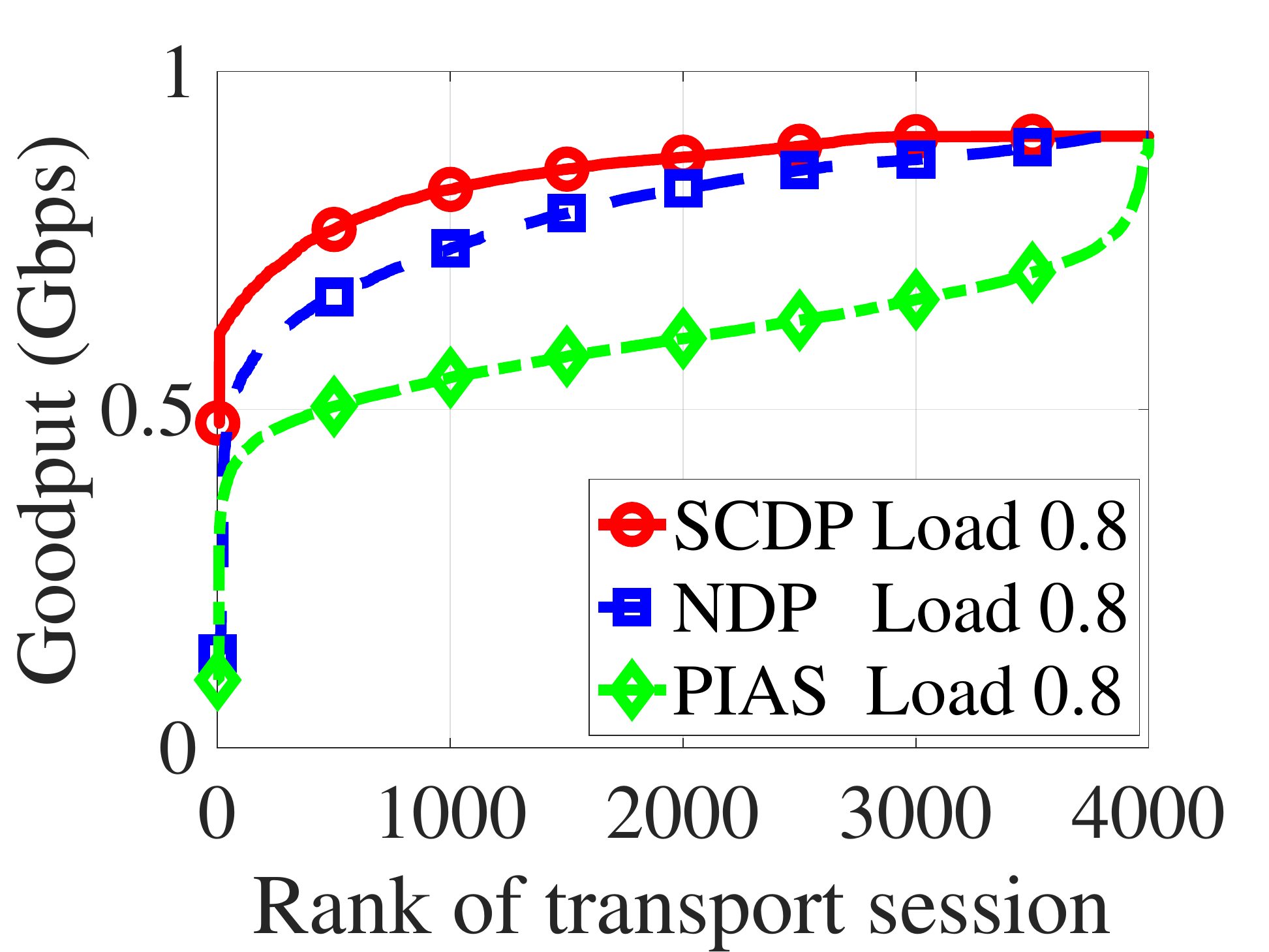}}\quad
 	\caption{Web search workload with a mixture of one-to-many and many-to-one sessions as background traffic}
 	\label{fct-ws-workloads-type-2} 	
 	\vspace{-3mm}
 \end{figure*}
 
 \begin{figure*}[t]
 	\setlength{\belowcaptionskip}{-2pt}
 	\subcaptionbox{ (0, 100KB]}[.231\linewidth][c]{%
 		\includegraphics[scale=0.2]{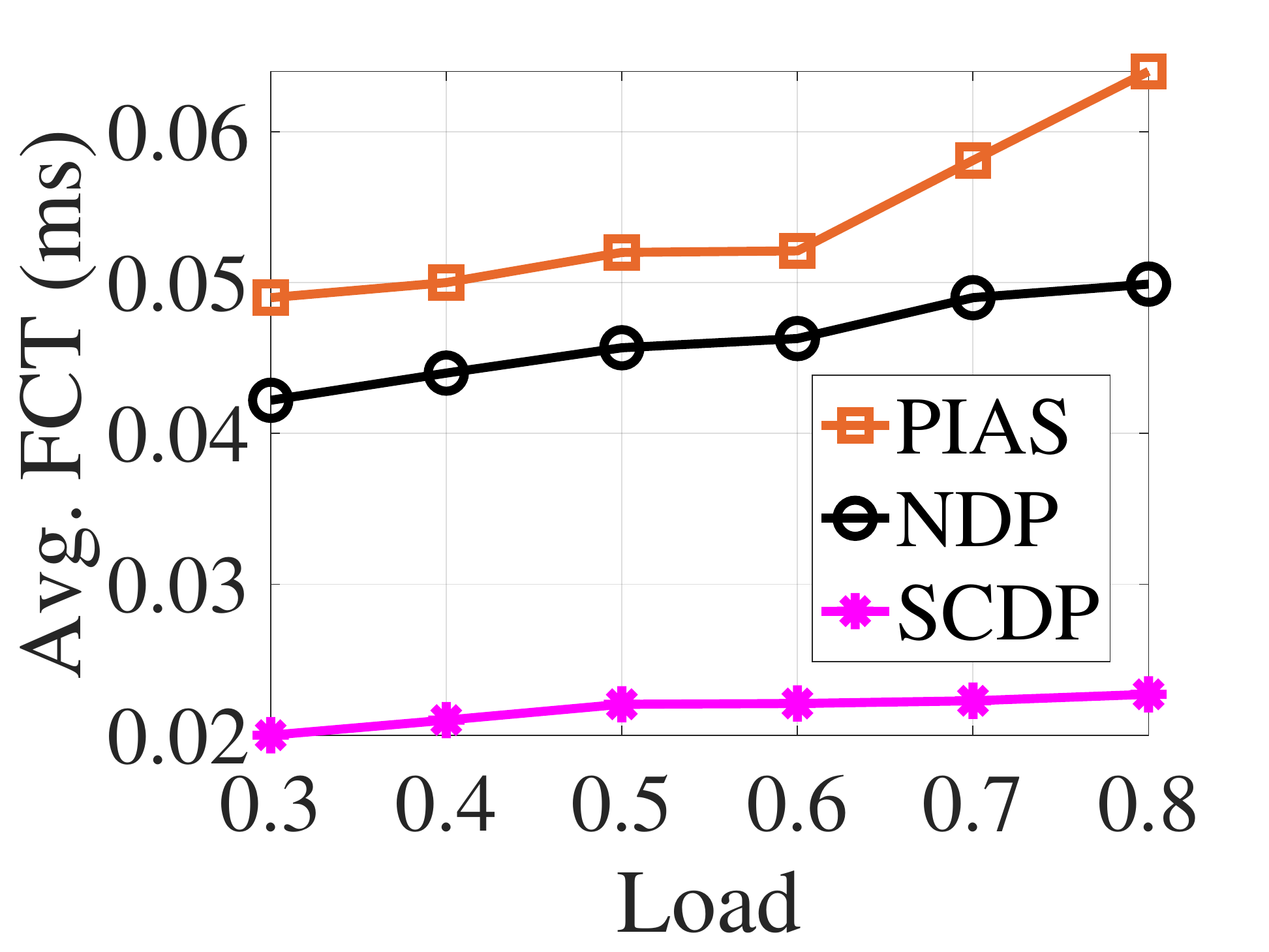}}\quad
 	\subcaptionbox{(0, 100KB] 99th percentile}[.231\linewidth][c]{%
 		\includegraphics[scale=0.2]{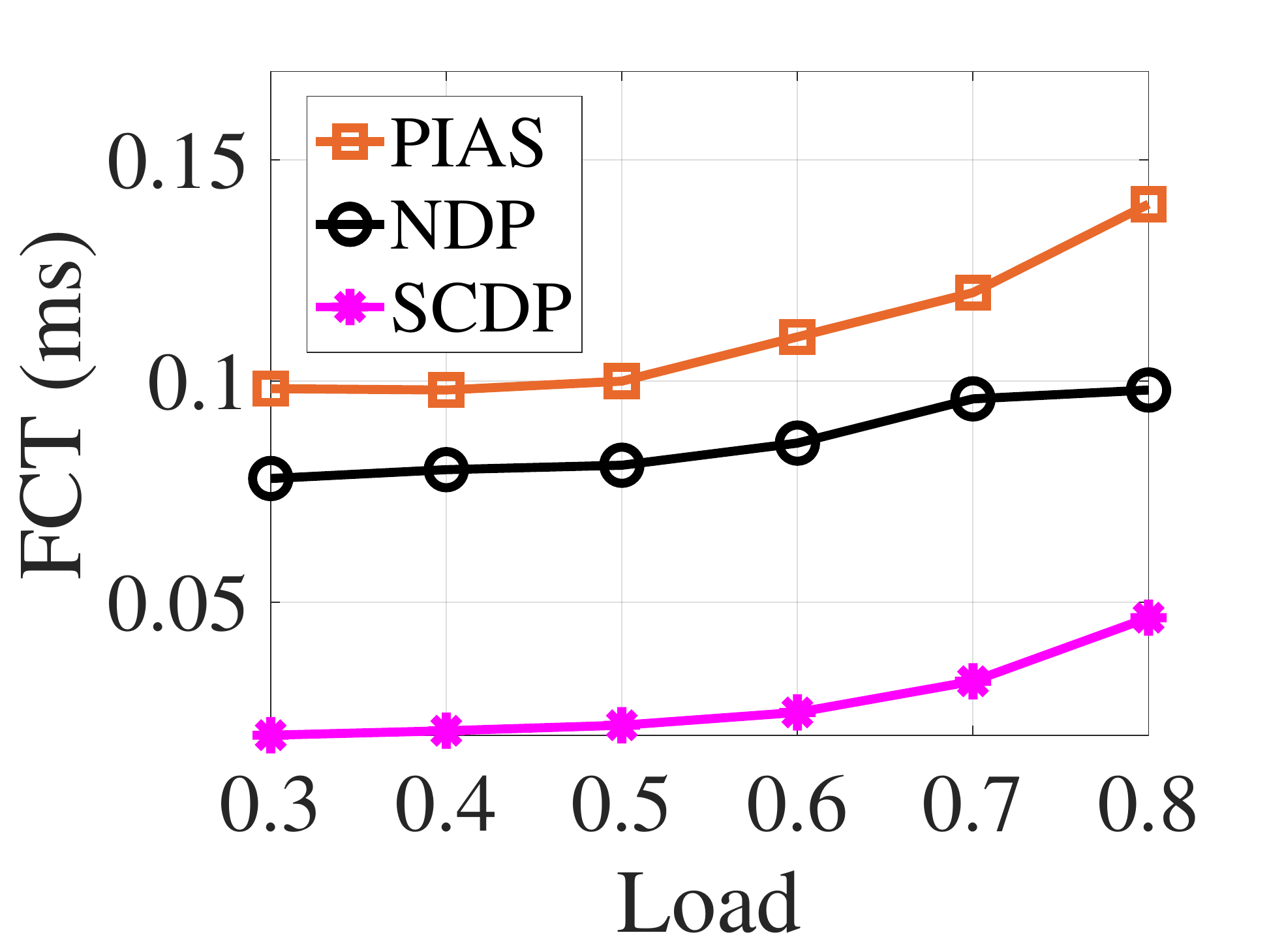}}\quad
 	\subcaptionbox{ (100KB, 1MB]}[.231\linewidth][c]{%
 		\includegraphics[scale=0.2]{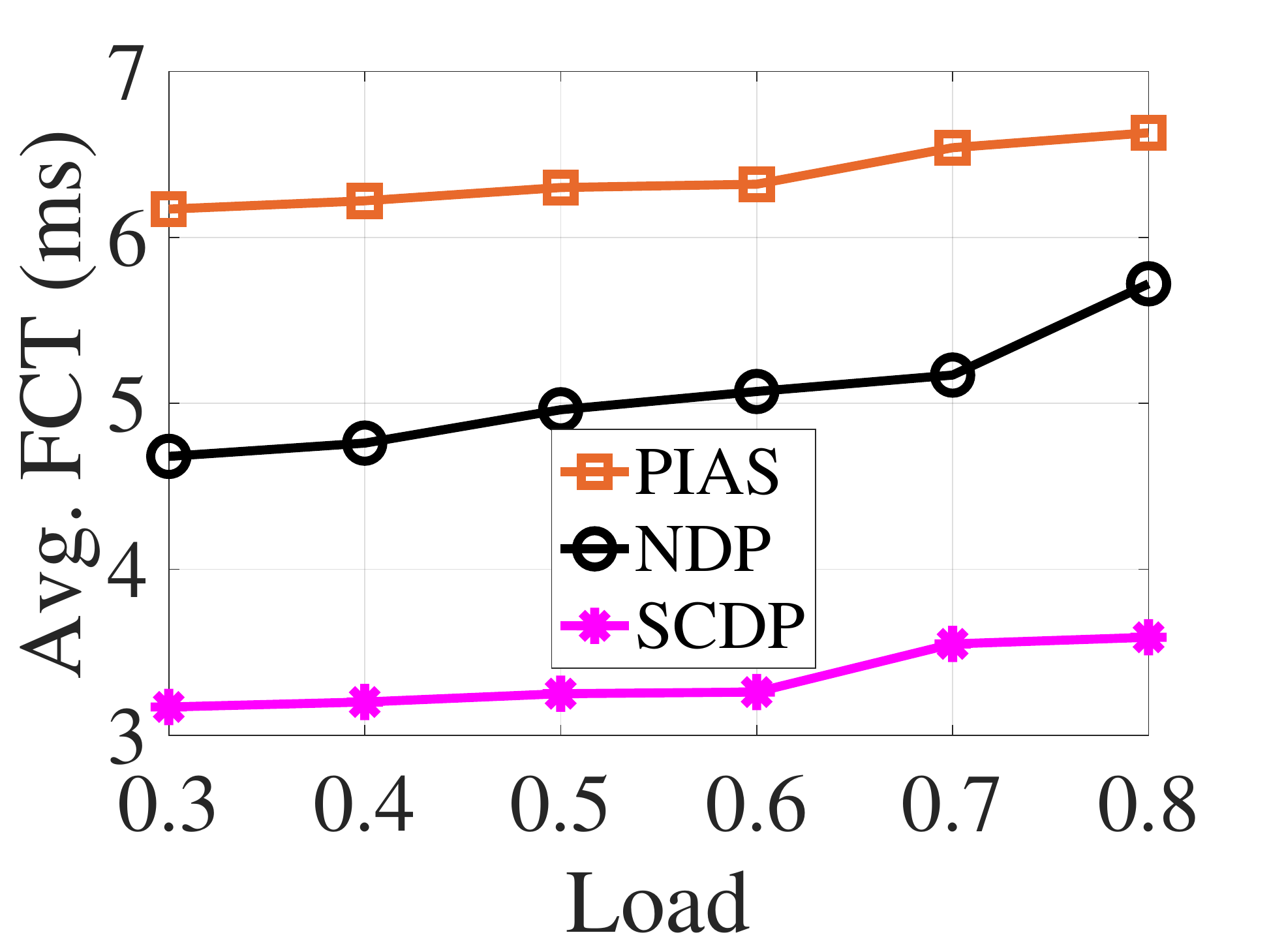}}\quad
 	\subcaptionbox{ (1MB, 10MB]}[.231\linewidth][c]{%
 		\includegraphics[scale=0.2]{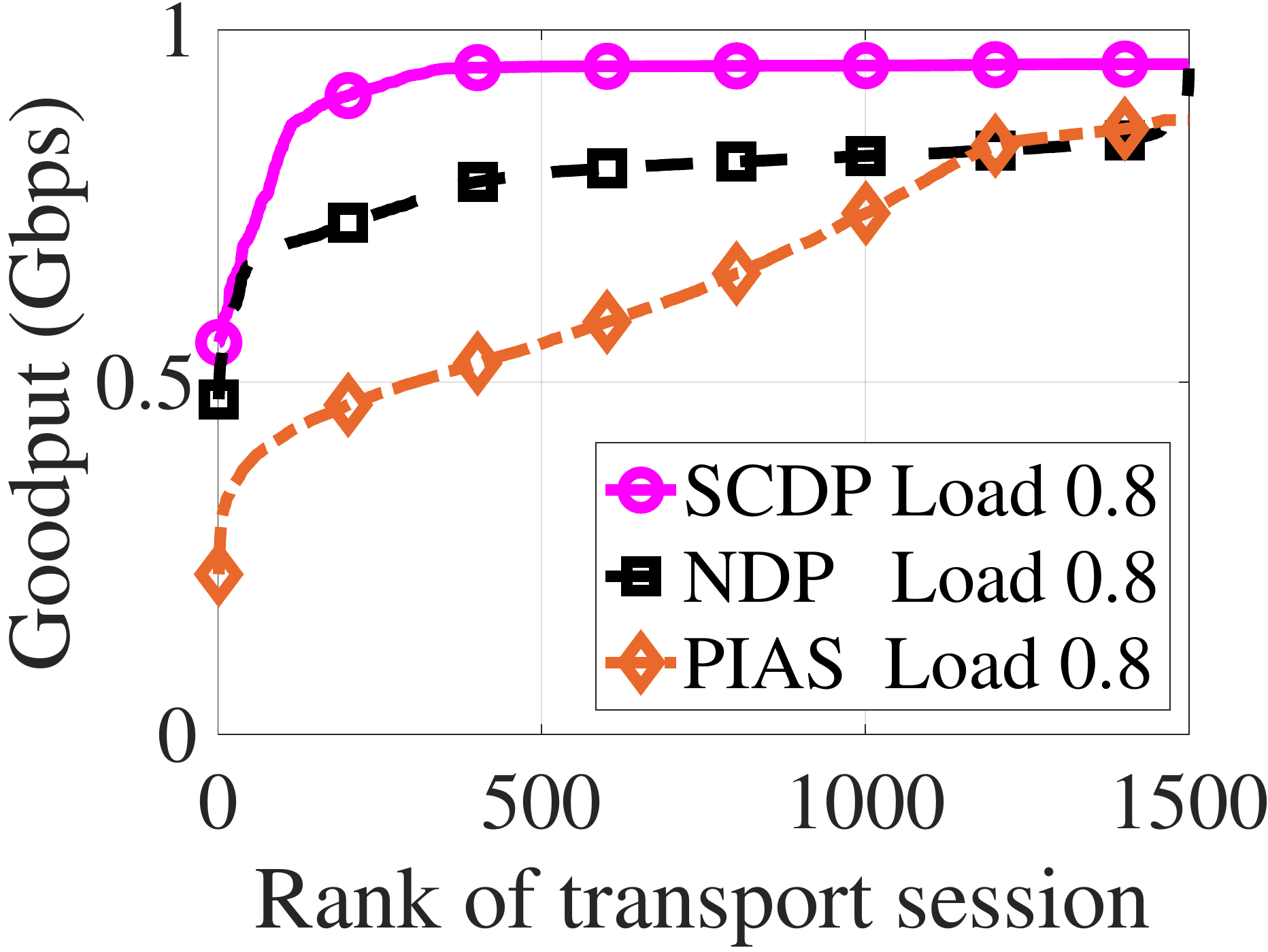}}\quad
 	\caption{Data mining workload with a mixture of one-to-many and many-to-one sessions as background traffic}
 	\label{fct-dm-workloads-type-2} 	
 	\vspace{-3mm}
 \end{figure*}

 \begin{figure*}[t]
 	\setlength{\belowcaptionskip}{-3pt}
 	\centering
 	\subcaptionbox{Incast: goodput comparison}[.23\linewidth][c]{%
 		\includegraphics[scale=0.2]{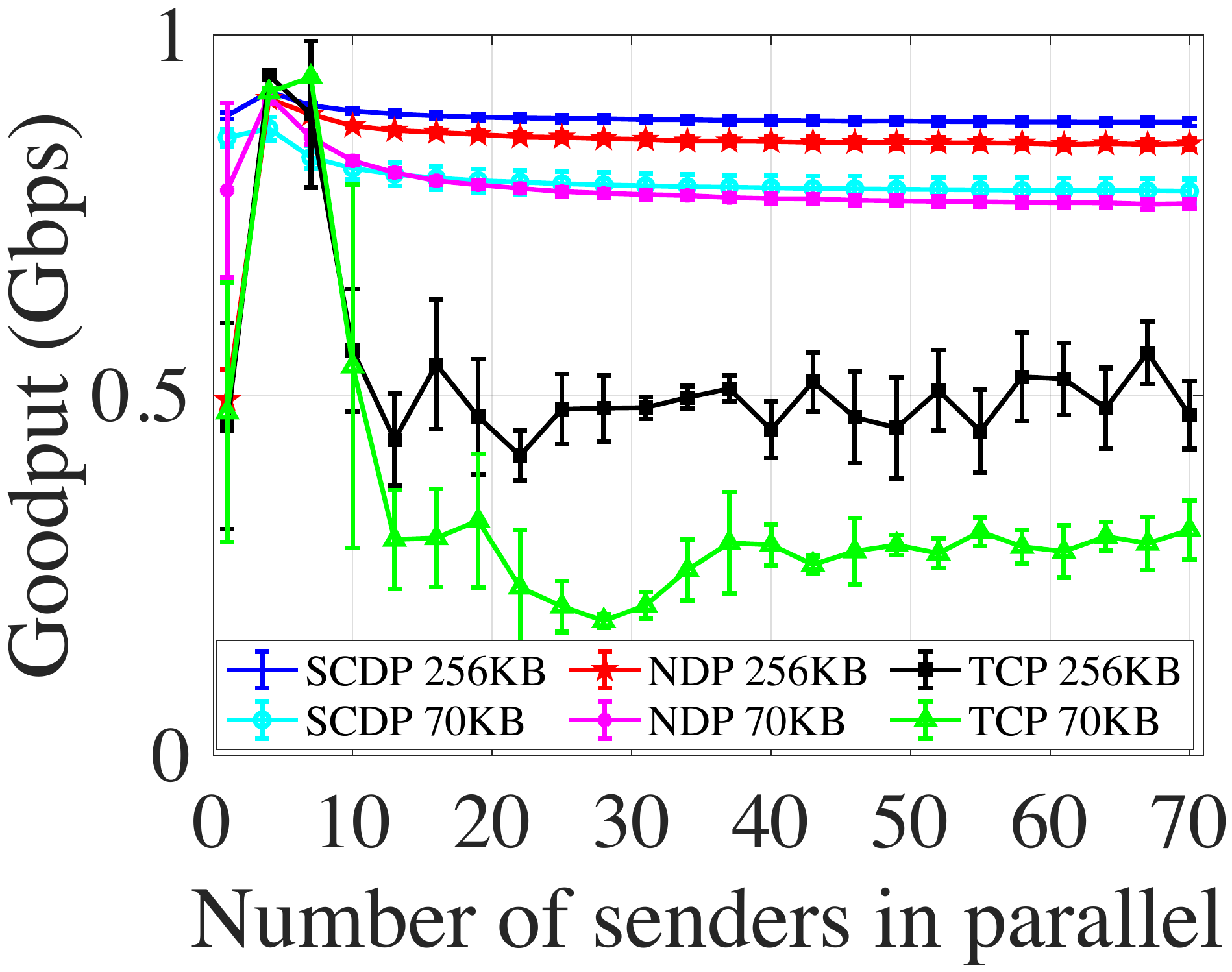}}\quad
 	\subcaptionbox{Incast: FCT with 70 senders}[.23\linewidth][c]{%
 		\includegraphics[scale=0.2]{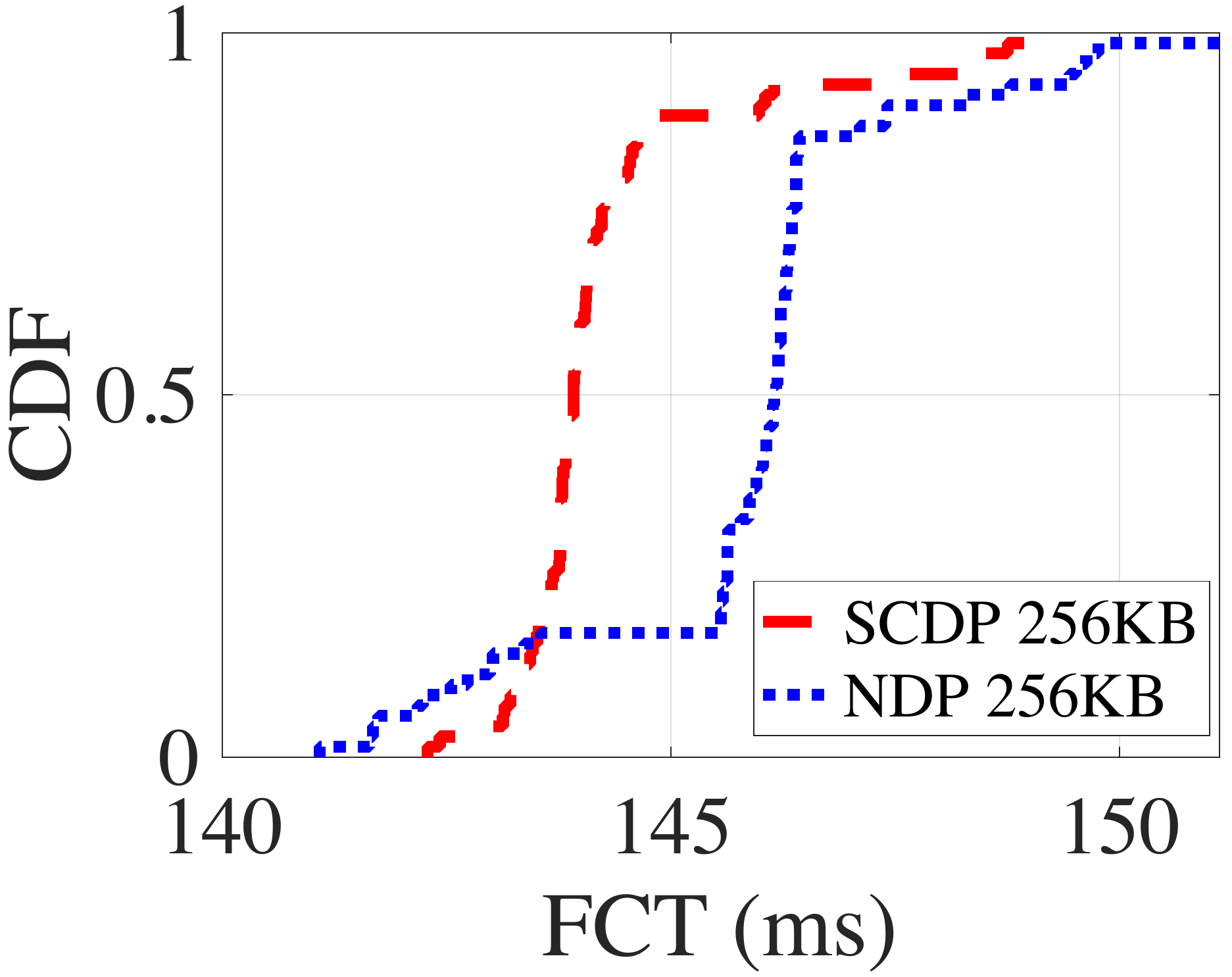}}\quad
 	\subcaptionbox{Outcast: setup}[.19\linewidth][c]{%
 		\includegraphics[scale=0.17]{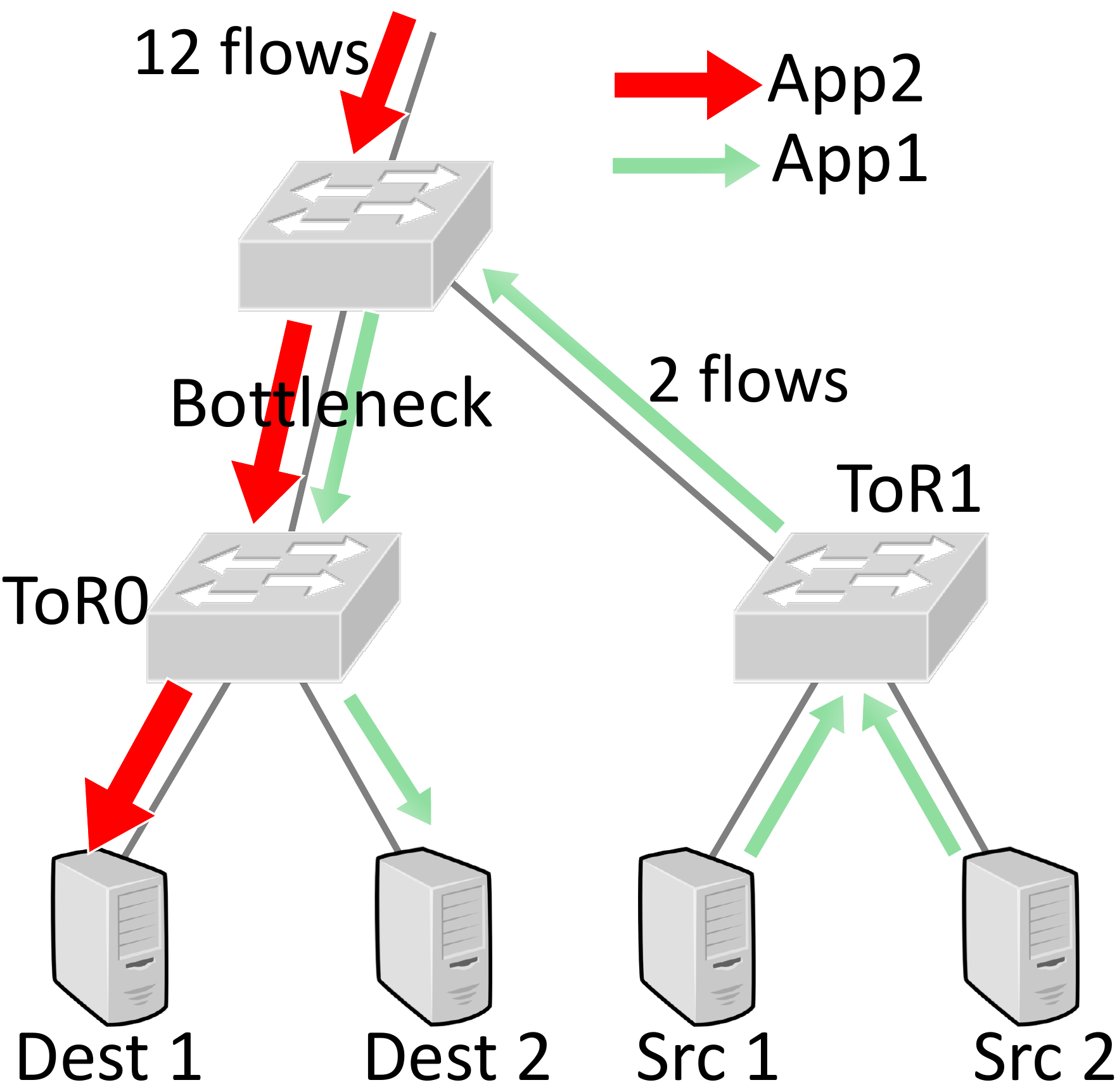}}\quad
 	\subcaptionbox{Outcast: goodput}[.23\linewidth][c]{%
 		\includegraphics[scale=0.2]{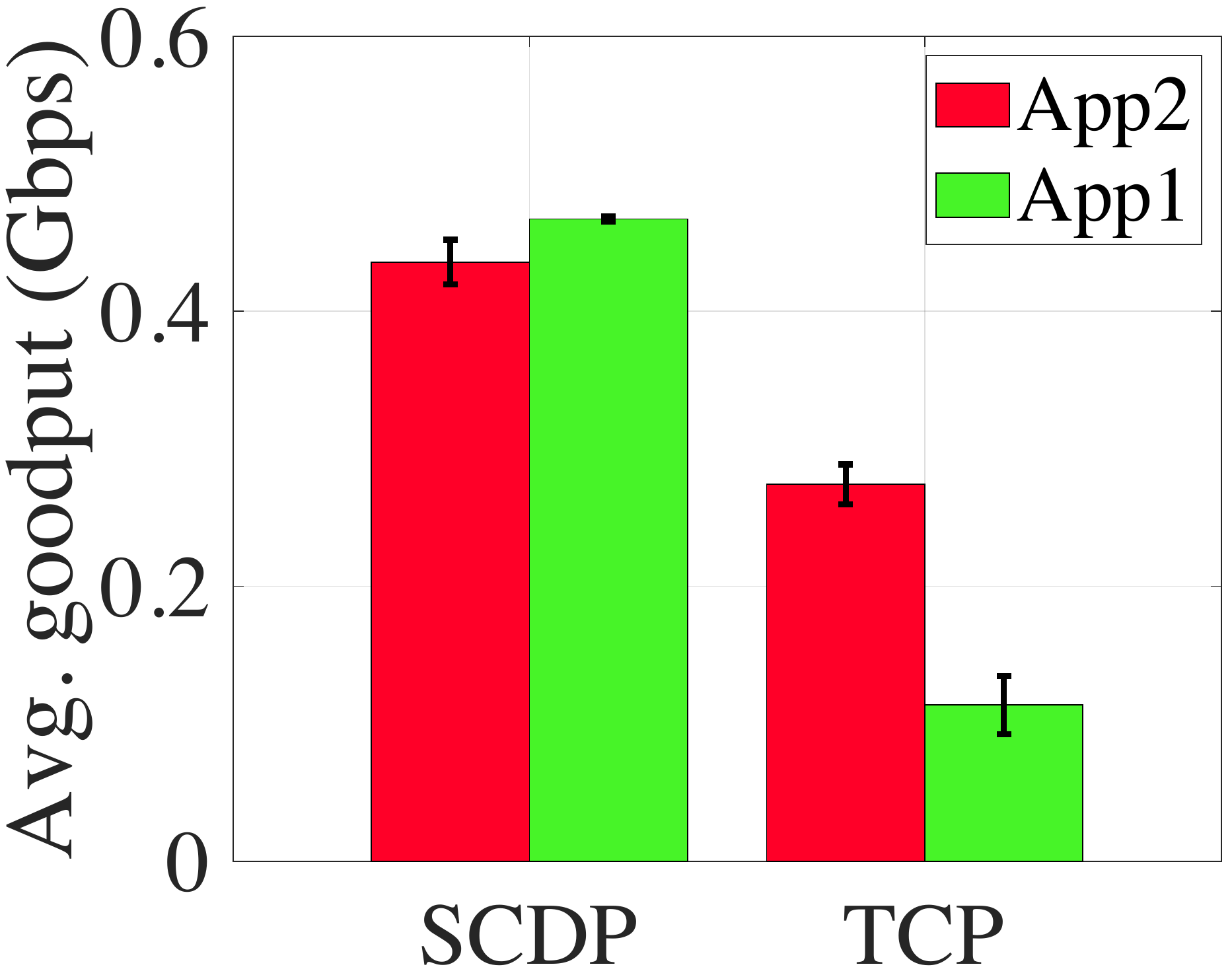}}\quad
 	\label{key}	\caption{Incast and Outcast evaluation}
 	\label{incast-outcast-fig} 	
 	\vspace{-3mm}
 \end{figure*}

 SCDP performs better in all scenarios due to the decoding-free completion of (almost all) short flows and the supported MLFQ. Note that when loss occurs, SCDP sessions must exchange $2$ additional symbols; they also pay the `decoding latency' price. For very short flows, the 99th percentile FCT is close to the average one for all loads, which indicates that this is rarely happening. We study the extent that this overhead and the associated decoding latency is required in Section~\ref{network-overhead}. For higher loads, NDP performs even worse than SCDP because of the lack of support for MLFQ, which results in the trimming of more packets belonging to short flows.
 
 Note that the FCT of short flows in web search is larger than in data mining. This is mainly because the percentage of long flows in the former workload is larger than in the latter, resulting in a higher overall load (for all fixed loads of background traffic). A key message here is that SCDP provides significantly better tail performance for short flows compared to NDP and PIAS, especially as the network load increases, despite the (very unlikely) potential for decoding and network overhead. For flows with sizes in (1MB, 10MB], we observe that goodput with SCDP is better compared to NDP and PIAS; tail performance is also better.\\
 \textcolor{black}{For all realistic workloads, NDP+ performance is better than NDP and on par with SCDP when the load is not very high. NDP+ performs slightly better than SCDP when the load is very high, due to SCDP’s induced decoding latency when loss occurs. For example, for $0$-$100$KB flows and 0.8 load, the average FCT for SCDP is 0.31ms and 0.292ms for NDP+. This reinforces that the latency penalty due to decoding is almost negligible because MLFQ prevents losses for very short flows.}
 
 It is worth noting here that in our experiments, NDP outperforms PIAS. This is contradicting the results presented in \cite{HOMA} for the same workload. We believe that this happens because the experimental setup in \cite{HOMA} is such that packet losses for short messages are very rare (if not non-existent) by having large buffers (in contrast to the, admittedly more realistic, experimental setup in this paper), therefore short flows use the highest priority queue to complete quickly without any losses. This is also mentioned in \cite{Aeolus20} where it is stated that ``one possible reason is that Homa assumes infinite switch buffers in their simulations. In contrast, in our simulations, we allocate 500KB buffer for each switch port''. We have reproduced the experiment in \cite{HOMA} using our OMNeT++ models by allocating a very large buffer to all queues, eliminating losses for short flows. In this experimental setup we observed that the average FCT for PIAS, when the network load is 0.8, drops from 0.43ms when using 100KB switch buffer (Figure~\ref{fct-ws-workloads-type-1}a) to 0.32ms, when using a very large buffer. This is indeed better than the average FCT observed for NDP. It is however worth noting, that this improvement does not come for free; instead, the average goodput for longer flows drops from ~0.7Gbps (Figure~\ref{fct-ws-workloads-type-1}d) to ~0.5Gbps. In other words, by eliminating loss for short flows, loss becomes more frequent in the lower priority queues occupied by packets belonging to longer flows. In general, we argue that PIAS performs worse than NDP because (1) it relies on DCTCP for data transport and as a result it suffers from the limitations of a single-path protocol (i.e. lack of support for multi-path transport and packet spraying); (2) connection establishment requires a three-way handshake and senders start with a small window, both of which can severely hurt FCTs for short flows; and (3) buffer occupancy in NDP is significantly lower than in PIAS \cite{NDP} which also affects performance for short slows.
 
 \vspace{-3mm}
 \subsection{Experimentation with 10Gbps Links}
 \label{10-links}
 
 Our decision to use 1Gbps links was solely driven by the very expensive nature of simulations, in terms of computational and memory resources. OMNeT++ is a packet-level simulator which means that by increasing the supported link rates by one order of magnitude (or more), the number of ‘live’ packets in the simulated network would dramatically increase, requiring extremely large amounts of memory and processing power to store and process all simulated packets. We are confident that our results are representative of SCDP's general behaviour and performance, compared to the state of the art. There are two aspects of SCDP that would need to be considered when deployed in faster networks; (1) the decoding latency would be more prominent in FCTs of short flows, because the actual data transmission would be faster; (2) the value of the initial window would need to be larger, in order for receivers to be able to run their links at capacity. We have performed experimentation to explore these two aspects; (1) regarding decoding latency, we have experimented with short flows in the context of the ‘web search’ workload in a simulated network with 10Gbps links and two different network loads (0.5 and 0.9). The results are shown in Figure \ref{highspeedlinks}. We observe that the flow completion times for both SCDP and NDP are (as expected) roughly an order of magnitude smaller compared to the respective results in Figures \ref{fct-ws-workloads-type-1}a and \ref{fct-ws-workloads-type-1}c. SCDP still performs better compared to NDP despite the fact that the decoding latency is now more prominent in the flow completion time. When the network load is very high, the gap between SCDP and NDP is at its smallest, because losses (trimmed packets) and therefore decoding are more frequent. It is worth pointing out that, in this experiment, we have not changed our underlying model for decoding latency, which is based on the results presented in~\cite{CodornicesRqNew}. In~\cite{CodornicesRqNew}, receivers were able to decode (roughly) at 1.3Gbps. However, in \cite{LiquidCloudStorage}, the authors report substantially higher decoding throughputs (up to 10 Gbps), which provides confidence that, in combination with SCDP’s pipelining mechanism, decoding will not be a bottleneck. Future hardware offloading approaches could potentially render decoding of small blocks negligible. The issue of selecting the value of the initial window in a 10Gbps setup is discussed in Section \ref{window-size-eval}.
 
 \vspace{-3mm}
 \subsection{Minimising Hotspots in the Network}
 \label{minimising-hotspots}
 SCDP increases network utilisation by exploiting support for network-layer multicasting and enabling load balancing when data is fetched simultaneously from multiple servers, as demonstrated in Section \ref{goodput-performance}. This, in turn, makes space in the network for regular short and long flows. In this section, we evaluate this performance benefit. We use as background traffic a 50\%/50\% mixture of write and read I/O requests (4MB each) that produce one-to-many and many-to-one traffic, respectively. We repeat the experiment of the previous section and evaluate the performance benefits of SCDP over NDP and PIAS with respect to minimising hotspots and maximising network utilisation for regular short and long flows. 
 
 In Figures \ref{fct-ws-workloads-type-2}a and \ref{fct-ws-workloads-type-2}c, we observe that SCDP's performance is almost identical to the one reported in Figures \ref{fct-ws-workloads-type-1}a and \ref{fct-ws-workloads-type-1}c (similarly between Figure \ref{fct-dm-workloads-type-1} and Figure \ref{fct-dm-workloads-type-2}). In contrast, NDP's and PIAS' performance deteriorates significantly because the background traffic requires more bandwidth (one-to-many) and results in hotspots at servers' uplinks (many-to-one). Tail performance for SCDP gets only marginally worse (the 99th percentile increases from 0.277ms to 0.287ms for the web search workload in load 0.5), whereas NDP's and PIAS' performance get significantly worse (the 99th percentile increases from 0.306ms to 0.381ms in NDP and from 0.386ms to 0.48ms in PIAS in load 0.5). The observed behaviour is more pronounced in the web search workload which, as described in the previous section, results in higher overall network utilisation compared to the data mining workload.

 \begin{figure*}[tb]
 	\setlength{\belowcaptionskip}{-6pt}
 	\minipage{0.5\textwidth}
 	\centering
 	\subcaptionbox{Goodput performance}[.42\linewidth][c]{%
 		\includegraphics[scale=0.22]{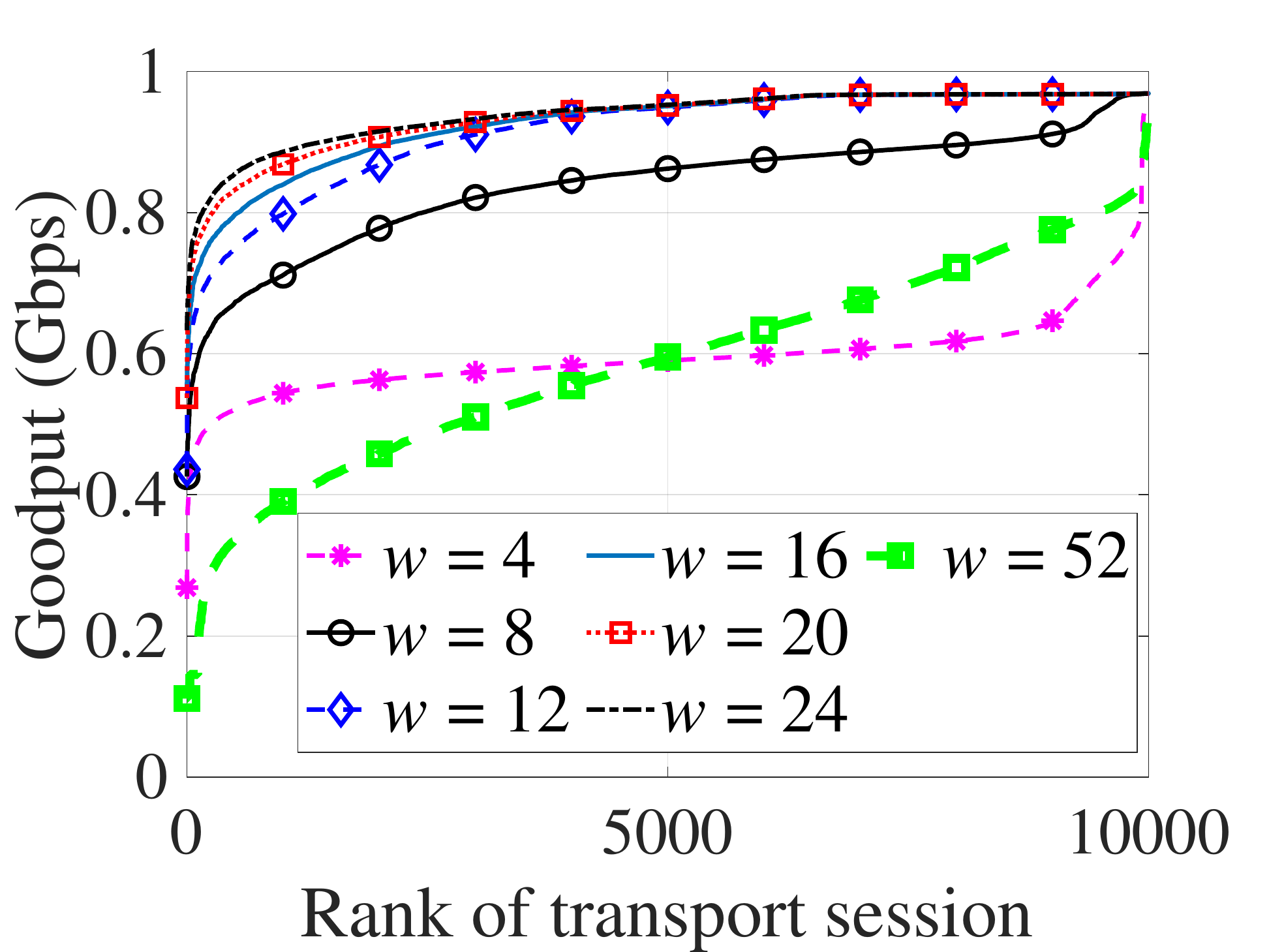}}\quad\quad
 	\subcaptionbox{\# trimmed pkts}[.42\linewidth][c]{%
 		\includegraphics[scale=0.22]{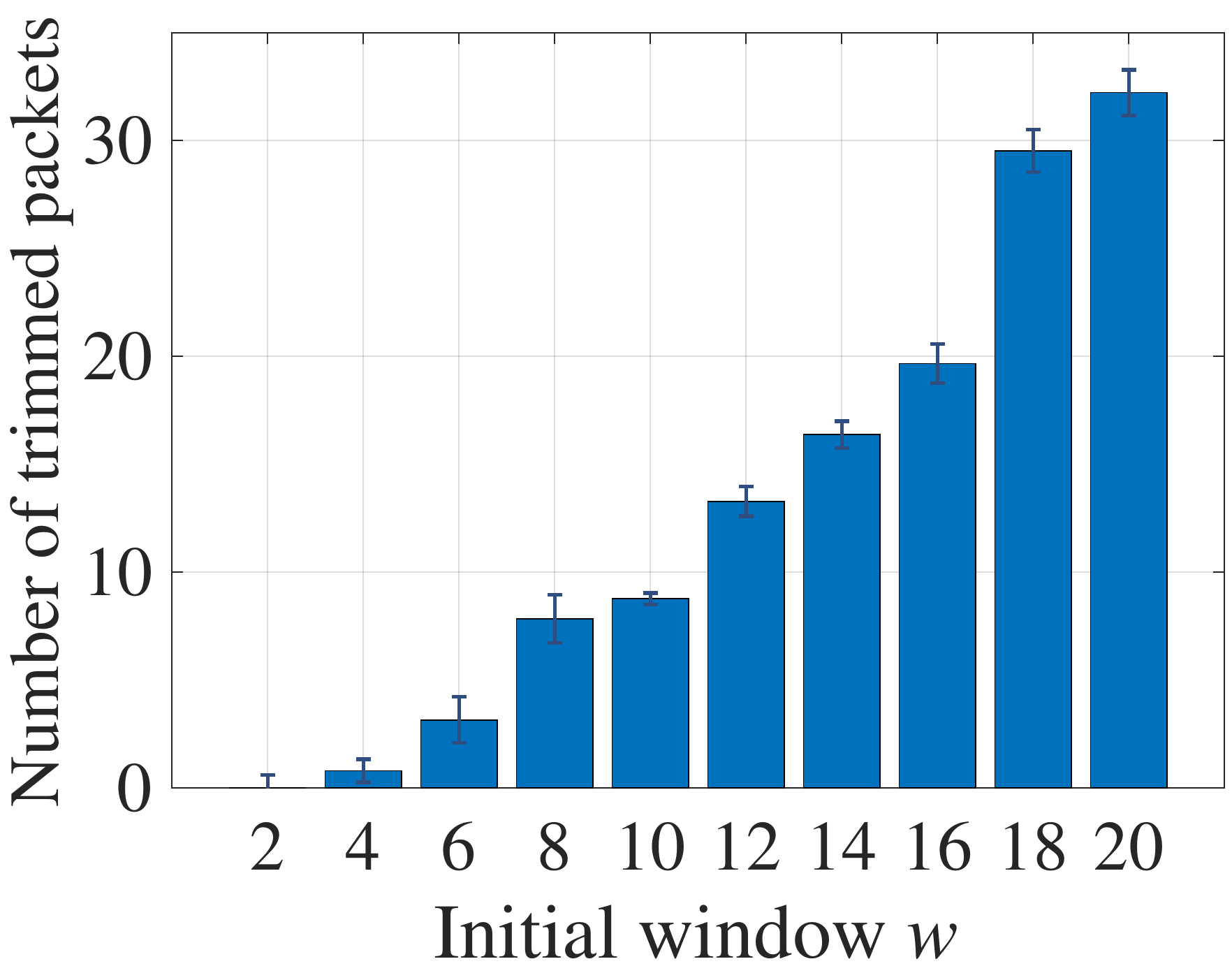}}\quad
 	\caption{The effect of the initial window size}
 	\label{IWeffect}
 	\endminipage\hfill
 	\minipage{0.25\textwidth}
 	\centering
 	\includegraphics[scale=0.22]{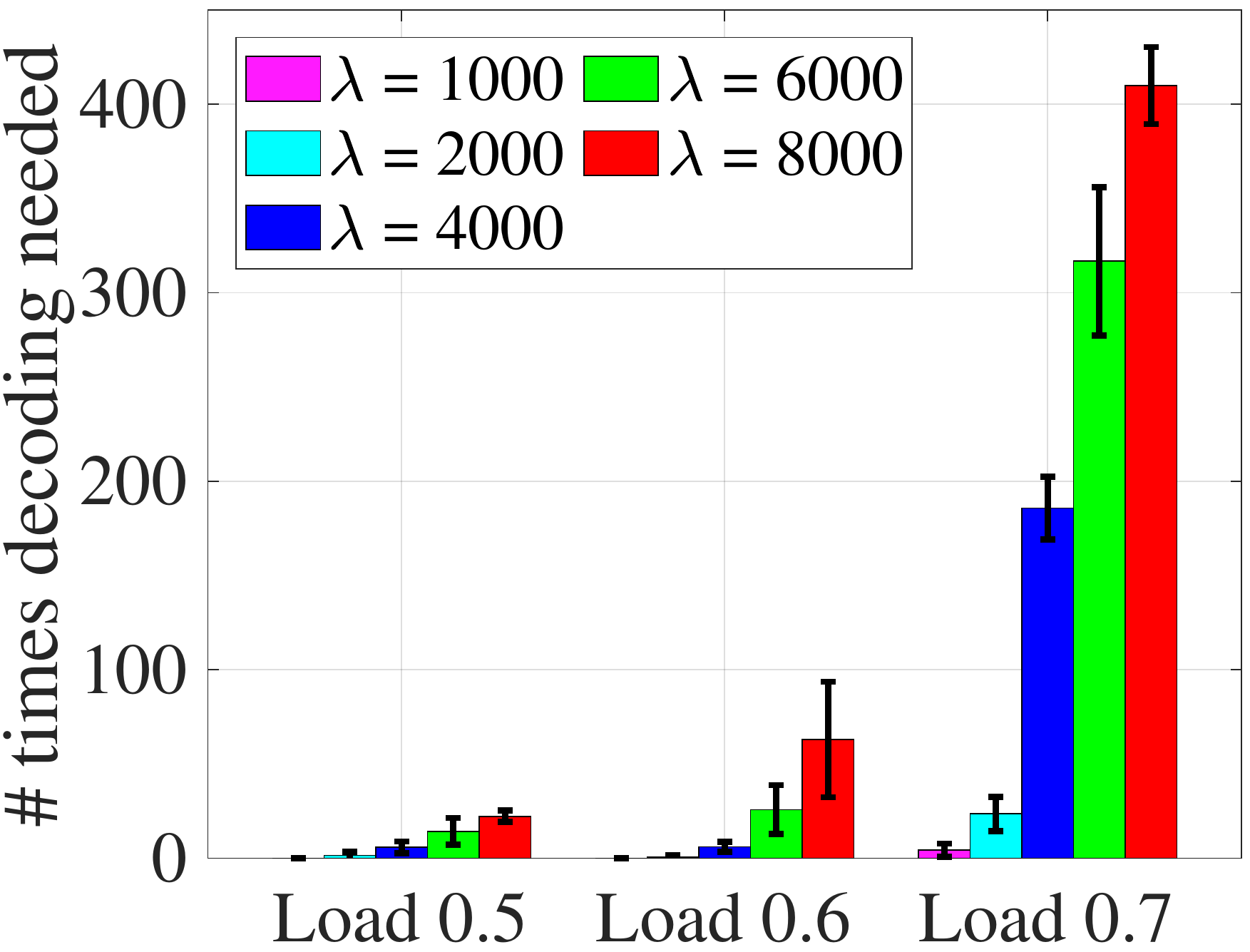}\quad
 	\quad
 	\caption{\# decoded sessions}
 	\label{eval-overhead-sessions}
 	\endminipage\hfill
 	\minipage{0.25\textwidth}
 	\centering
 	\includegraphics[scale=0.22]{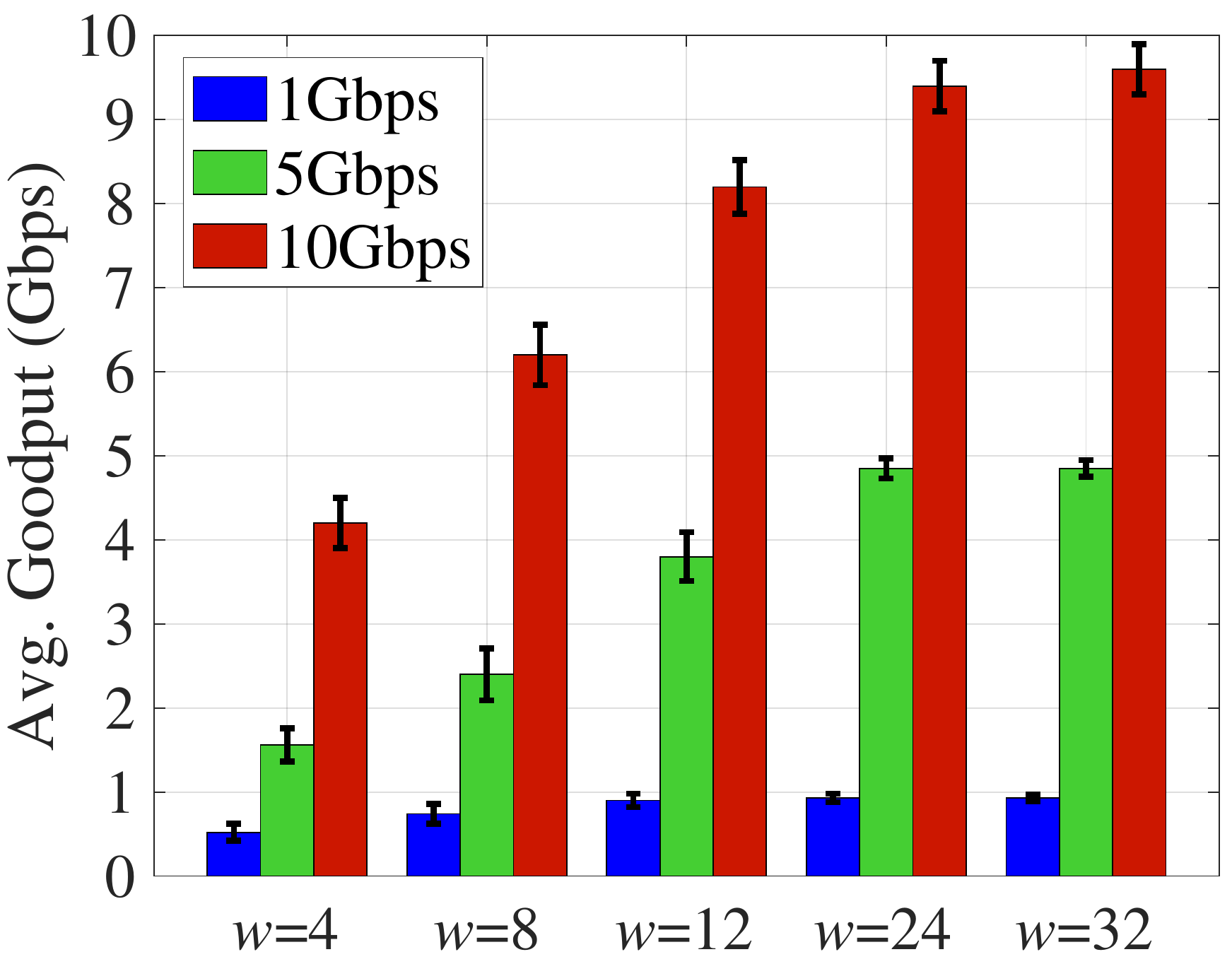} 
 	\caption{varying \textit{w} \& link rates}
 	\label{iw-speeds}
 	\endminipage\hfill
 \end{figure*}

 \begin{figure*}[t]
 	\setlength{\belowcaptionskip}{-4pt}
 	\minipage{0.75\textwidth}
 	\centering
 	\subcaptionbox{One-to-many - $1$MB}[.3\linewidth][c]{%
 		\includegraphics[width=1\linewidth]{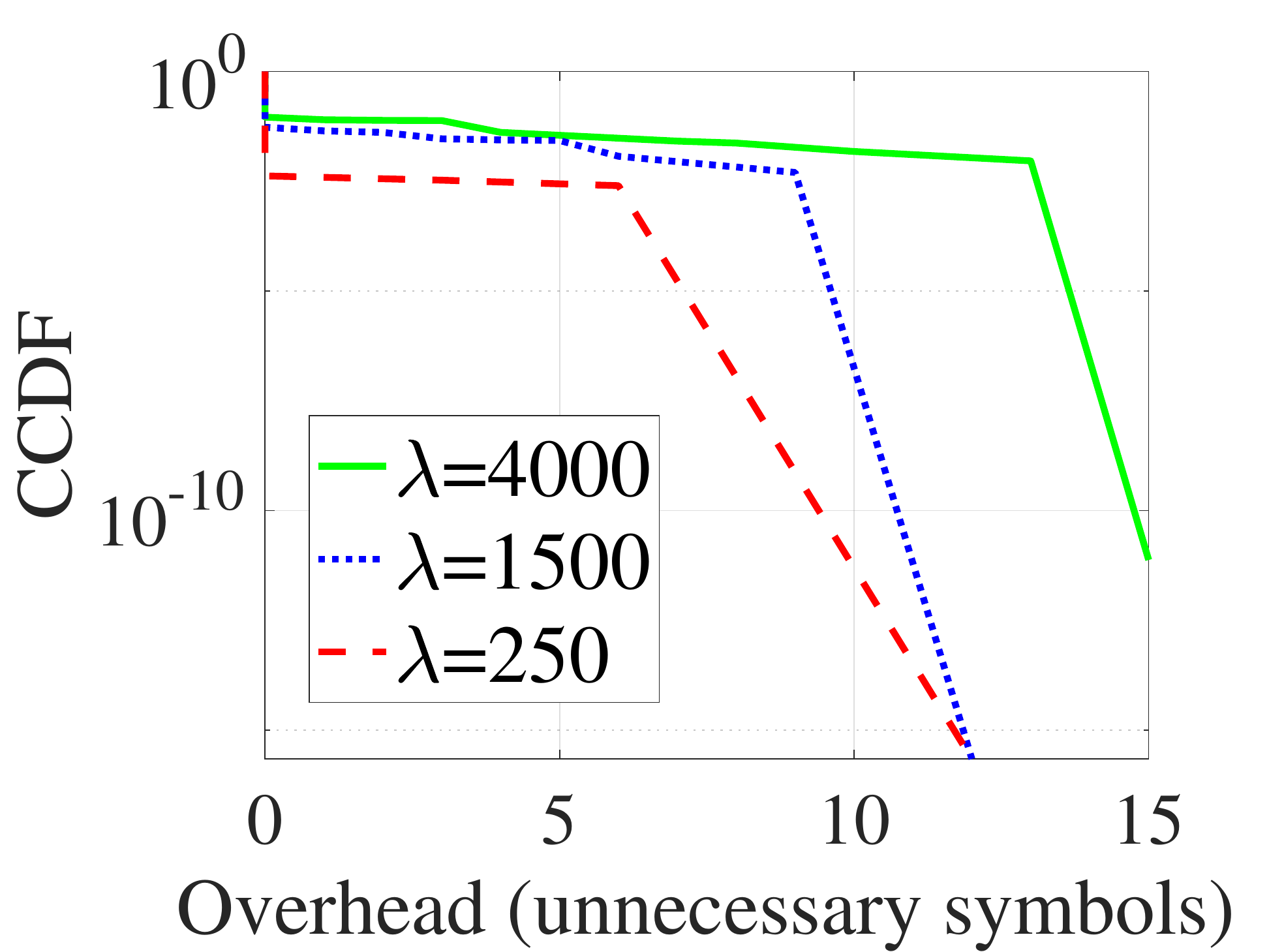}}\quad
 	\subcaptionbox{One-to-many - $3$MB}[.3\linewidth][c]{%
 		\includegraphics[width=1\linewidth]{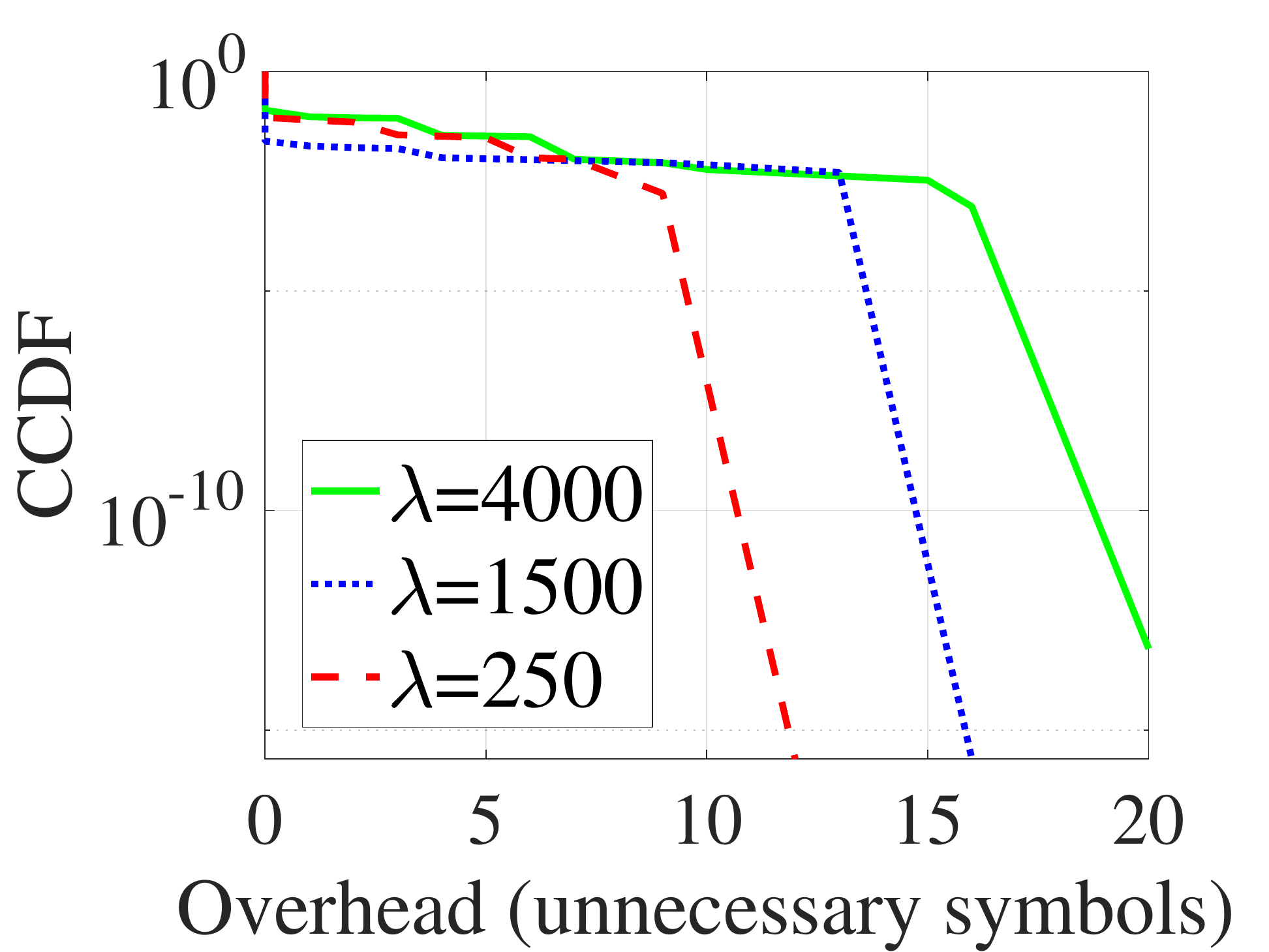}}\quad
 	\subcaptionbox{Goodput - $\lambda = 4000$}[.3\linewidth][c]{%
 		\includegraphics[width=1\linewidth]{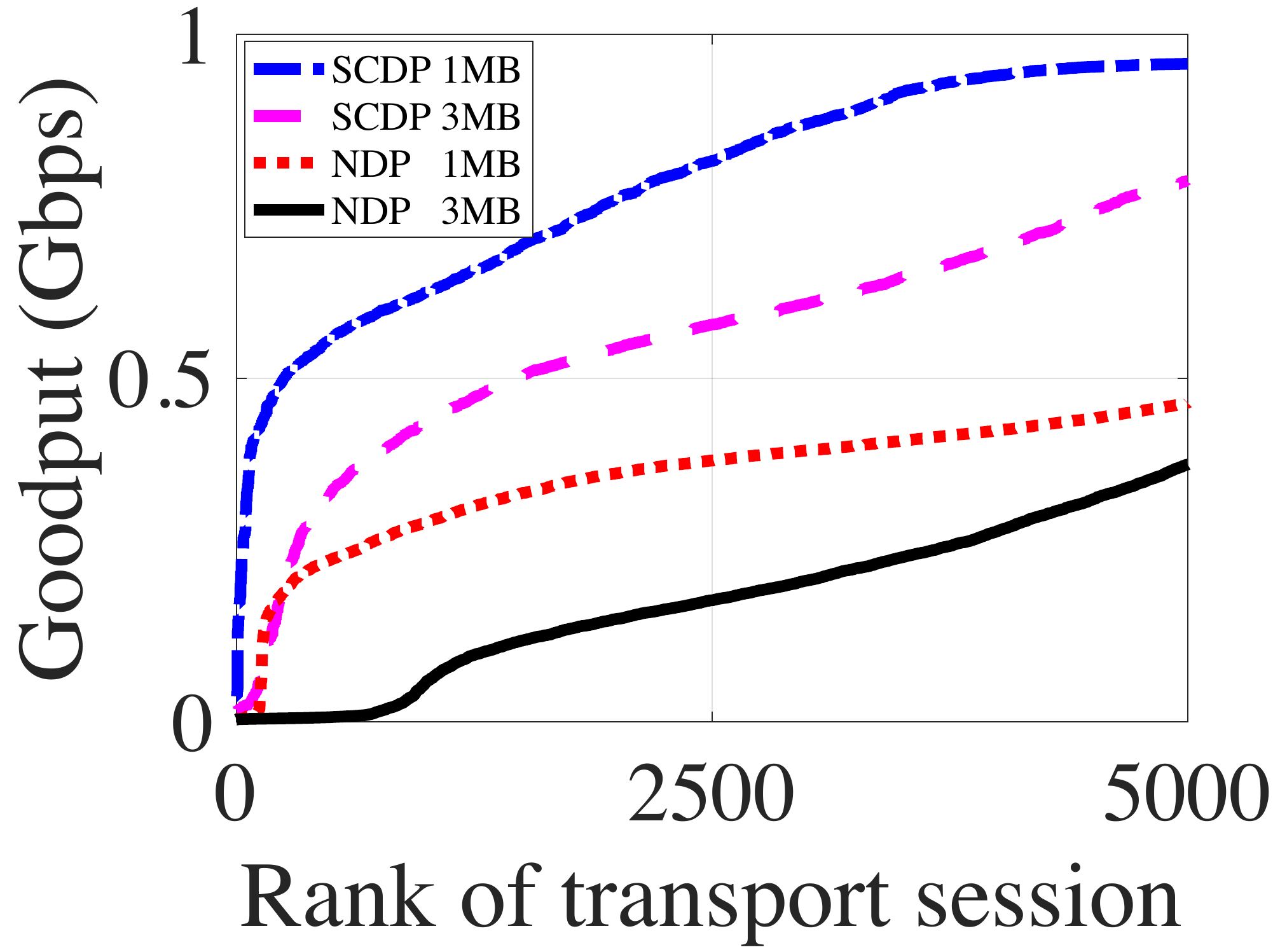}}\quad
 	\caption{Unnecessary network overhead in one-to-many sessions}
 	\vspace{-2mm}
 	\label{eval-unnecessary-overhead}
 	\endminipage\hfill
 	\minipage{0.23\textwidth}
 	\centering
 	\includegraphics[width=1.0\linewidth]{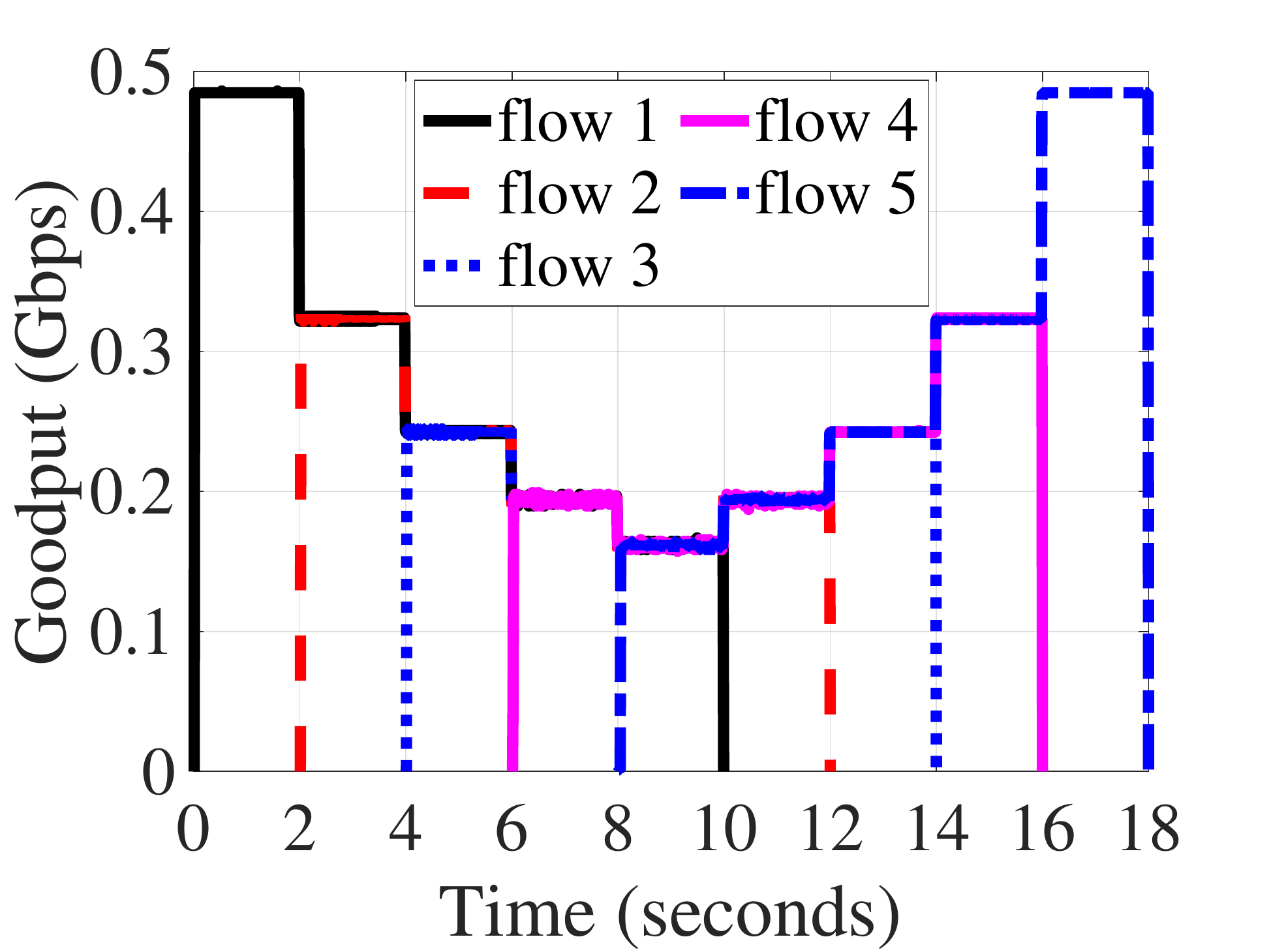}	
 	\caption{Convergence test}
 	\label{fairness}
 	\endminipage\hfill
 	\vspace{-2mm}
 \end{figure*}
 
 \vspace{-3mm}
 \subsection{Eliminating Incast and Outcast}
 \label{incast-section}
 SCDP eliminates Incast by integrating packet trimming and not relying on retransmissions of lost packets due to the rateless nature of RaptorQ codes. We simulated Incast by having multiple senders (ranging from $1$ to $70$) sending blocks of data ($70$KB and $256$KB, each, in two separate experiments) to a single receiver. All sessions were synchronised and background traffic was present to simulate congestion. Figure \ref{incast-outcast-fig}a illustrates the measured aggregated goodput for all SCDP, NDP and TCP flows. As expected, TCP's performance collapses when the number of senders increases. SCDP performs slightly better compared to NDP even when a large number of servers send data to the receiver at the same time. This is attributed to the decoding-free completion of these flows, in combination with the packet trimming and the lack of retransmissions for SCDP. Figure \ref{incast-outcast-fig}b shows the CDF of the FCTs in the presence of Incast with 70 senders. We observe that for the vast majority of transport sessions, SCDP provides superior performance compared to NDP.\\
 SCDP eliminates outcast by employing receiver-driven flow control and packet trimming, which prevent port blackout. We have simulated a classic outcast scenario, where two receivers that are connected to the same ToR switch receive traffic from senders located in the same pod (2 flows crossing 4 hops) and different pods (12 flows crossing 6 hops), respectively. Flow size is 200KB and all flows start at the same time. This is illustrated in Figure \ref{incast-outcast-fig}c. Here, the bottleneck link lies between the aggregate switch and the ToR switch, which is different from the Incast setup. Figure \ref{incast-outcast-fig}d shows the aggregate goodput for the two groups of flows, for SCDP and TCP. TCP Outcast manifests itself through (1) unfair sharing of the bottleneck bandwidth (around  113 and 274 Mbps for the groups of flows, respectively) and (2) suboptimal overall performance (around  0.387 Gbps). SCDP eliminates Outcast as the bottleneck is shared fairly between the two groups of flows (around  460  and 435 Mbps for the groups of flows, respectively, and the overall goodput is around 0.9 Gbps).

 \subsection{The effect of the initial window size}
 \label{window-size-eval}
 A key parameter of SCDP is the \black{initial window $w$ of symbol packets that a sender pushes to the network. The window is maintained throughout the lifetime of a session and is only decreased for the last $w$ pull packets}. Here, we evaluate the effect that the initial window has in the performance of SCDP. The experimental setup is as described in Section \ref{goodput-performance}, with $1.5$MB unicast sessions (we evaluated one-to-many and many-to-one sessions as well, which showed similar results as the unicast sessions). In Figure \ref{IWeffect}a, we observe that for very small values of the initial window, goodput is very low and the receiver's downlink underutilised. As the window increases, utilisation approaches the maximum available link capacity (for $12$ symbol packets). 
 
 For larger values of the initial window (up to 24 symbol packets), the measured goodput is consistently high (i.e. downlink runs at full capacity). Increasing the window inevitably leads to more trimmed packets due to the added network load when pushing symbol packets. This is illustrated in Figure \ref{IWeffect}b, where the average number of trimmed packets for session sizes of $1.5$MB grows from $13$ for an initial window of $12$ to $32$ trimmed packets for an initial window of $20$. We can therefore assume that there is relatively wide range of window values for which performance can be consistently high. In order to further explore this point, we have repeated the same experiment by setting the initial window to $52$ packets; in Figure \ref{IWeffect}a, we observe that goodput deteriorates significantly. This is because the initial `push' phase results in severe congestion and loss, which, in turn, results in (1) significant network overhead induced by the large number of trimmed packets (39 packets on average for each SCDP session) that are forwarded with priority over all other symbol packets; (2) latency decoding being induced to a larger number of SCDP sessions; (3) large batches of pull requests potentially that block pull requests belonging to other sessions.
 
 We also explore the effect of the initial window value with different link rates (otherwise keeping the experimental setup unchanged). In Figure~\ref{iw-speeds}, we clearly observe that, as the supported link rate increases, the value of the initial window must also be increased in order to fully utilise the receivers’ downlink (12 symbols for 1Gbps link, 24 symbols for  5Gbps link and 32 symbols for 10Gbps).

 

 \vspace{-3.5mm}
 \subsection{Network Overhead and Induced Decoding Latency}
 \label{network-overhead}
 SCDP provides zero-overhead data transport when no loss occurs. In the opposite case, \black{there is an overhead $o$ of $2$ extra symbols (compared to the number of original fragments $K$)} are required by the decoder to decode the source block (with extremely high probability). Additionally, the required decoding induces latency in receiving the original source block. Short flows in data centres are commonly latency sensitive so SCDP must be able to provide decoding-free completion of such flows. To asses the efficacy of our MLFQ-based approach, we measure the number of unicast flows that suffer symbol packet loss for different network loads ranging from $0.5$ to $0.7$. For each network load, we examine different $\lambda$ values for the Poisson inter-arrival rate of the studied short flows ($150$KB). In each simulation, we generate $5000$ sessions with the respective $\lambda$ value as their inter-arrival time. In Figure \ref{eval-overhead-sessions}, we observe that for load values of $0.5$ and $0.6$, the times that a short flow would require decoding and extra $2$ symbol packets is very small (0.44\% and 1.2\% of the flows, respectively, when $\lambda$ = 8000), rendering the respective overhead negligible.
 
 \vspace{-3mm}
 \subsection{Overhead in One-to-Many Sessions}
 \label{unnecessary-overhead}
 In Section \ref{multicast}, we identified a limitation of SCDP with respect to unnecessary network overhead which may occur in one-to-many transport sessions in the presence of congestion. This is due to receivers getting behind with the reception of symbols. Consequently, up-to-date receivers will be receiving more symbols than what they actually need. In order to evaluate the extent of this limitation we set up a similar experiment to the one presented in Section \ref{goodput-performance}. Figures \ref{eval-unnecessary-overhead}a and \ref{eval-unnecessary-overhead}b depict the CCDF of the number of symbols that were sent unnecessarily for different values of $\lambda$, and session sizes. We observe that as the network load increases, the number of sessions that induce unnecessary network overhead increases. It is important to note that, even when this happens, the measured goodput for SCDP is significantly better than that of NDP. Figure \ref{eval-unnecessary-overhead}c illustrates the measured goodput for the examined session sizes and highest network load ($\lambda = 4000$). Clearly, SCDP significantly outperforms NDP despite the potential for some unnecessary network overhead. The benefit of exploiting network-layer multicast makes this potential overhead negligible.
 \vspace{-4mm}
 \subsection{Resource Sharing}
 \label{fairness-evaluation}
 SCDP achieves excellent fairness due to the following design principles: (1) receivers pull symbol packets from one or more senders in the data centre at a pace that matches their downlink bandwidth. Given that servers are uniformly connected to the network with respect to link speeds, SCDP enables fair sharing of the network to servers. (2) A receiver pulls symbol packets for each SCDP session on a round robin basis.  As a result, SCDP enables fair sharing of its downlink to all transport sessions running at a specific receiver. It would be straightforward to support priority scheduling at the receiver. (3) SCDP employs MLFQ in the network. Obviously, this prioritisation scheme provides fairness between competing flows only within the same priority level. In Figure \ref{fairness} we report goodput results with respect to the convergence behaviour of 5 SCDP unicast sessions that start sequentially with 2 seconds interval and 18 seconds duration, from 5 sending severs to the same receiving server under the same ToR switch. SCDP performs equally well to DCTCP in that respect \cite{DCTCP}. Clearly, flows acquire a fair share of the available bandwidth very quickly. Each incoming flow is initially prioritised over the ongoing flows (MFLQ) but, given the reported time scales, this cannot be shown in Figure \ref{fairness}. We have repeated this experiment with larger number of flows, and we find that SCDP converges quickly, and all flows achieve their fair share.

\section{Conclusion}
\label{sec:conclusion}

In this paper, we proposed SCDP, a general-purpose transport protocol for data centres that is the first to exploit network-layer multicast in the data centre and balance load across senders in many-to-one communication, while performing at least as well as the state of the art with respect to goodput and flow completion time for long and short unicast flows, respectively. Supporting one-to-many and many-to-one application workloads is very important given how extremely common they are in modern data centres \cite{ElmoMulticast}. SCDP achieves this remarkable combination by integrating systematic rateless coding with receiver-driven flow control, packet trimming and in-network priority scheduling. 

RaptorQ codes incur some minimal network overhead, only when loss occurs in the network, but our experimental evaluation showed that this is negligible compared to the significant performance benefits of supporting one-to-many and many-to-one workloads. RaptorQ codes also incur computational overhead and associated latency when when loss occurs. However, we showed that this is rare for short flows because of MLFQ. For long flows, block pipelining alleviates the problem by splitting large blocks into smaller ones and decoding each of these smaller blocks while retrieving the next one. As a result, latency is incurred only for the last smaller block. RaptorQ codes have been shown to perform at line speeds even on a single core; we expect that with hardware offloading the overall overhead will not be significant. 

\black{As part of our future work, we aim at developing an SCDP prototype (in-kernel and/or using user-space network stack) and exploring its performance with real application workloads. We will also explore machine learning-based approaches for setting the initial window on a per-flow basis. More specifically, we will investigate the applicability of Reinforcement Learning in updating the initial window value for new or existing flows based on the (partially) observable state of the network (e.g. as \cite{sigcomm-rl-2020} performs congestion control). A key argument in this paper was that RaptorQ coding should be the centrepiece of the data transport mechanism, in order to enable a unified approach for efficiently dealing with all supported communication modes. As part of our future work, we will investigate this argument further by developing extensions of existing unicast data centre protocols (e.g. \cite{NDP}) that can handle one-to-many and many-to-one data transport and compare their performance with SCDP.}

\bibliographystyle{IEEEtran}
\bibliography{reference}

\end{document}